# Applications of integrated optical microcombs

Yang Sun[1, †], Jiayang Wu,[1, †, 5] Mengxi Tan[2], Xingyuan Xu[3], Yang Li[1], Roberto Morandotti[4], Arnan Mitchell[2], and David Moss[1, 6]

[1]*Optical Sciences Centre, Swinburne University of Technology, Hawthorn, VIC 3122, Australia*

[2]*School of Engineering, RMIT University, Melbourne, VIC 3001, Australia*

[3]*State Key Laboratory of Information Photonics and Optical Communications, Beijing University of Posts and Telecommunications, Beijing 100876, China.*

[4]*INRS –Énergie, Matériaux et Télécommunications, Varennes, QC J3X 1S2, Canada*

[5]*e-mail: jiayangwu@swin.edu.au*

[6]*e-mail: dmoss@swin.edu.au*

[†]*These authors contribute equally.*

**Keywords:** Integrated photonics, microcombs, microwave photonics, optical communications, precision measurements, neuromorphic computing, quantum optics.

**Abstract**

Optical microcombs represent a new paradigm for generating laser frequency combs based on compact chip-scale devices, which have underpinned many modern technological advances for both fundamental science and industrial applications. Along with the surge in activity related to optical micro-combs in the past decade, their applications have also experienced rapid progress – not only in traditional fields such as frequency synthesis, signal processing, and optical communications, but also in new interdisciplinary fields spanning the frontiers of light detection and ranging (LiDAR), astronomical detection, neuromorphic computing, and quantum optics. This paper reviews the applications of optical microcombs. First, an overview of the devices and methods for generating optical microcombs is provided, which are categorized into material platforms, device architectures, soliton classes, and driving mechanisms. Second, the broad applications of optical microcombs are systematically reviewed, which are categorized into microwave photonics, optical communications, precision measurements, neuromorphic computing, and quantum optics. Finally, the current challenges and future perspectives are discussed.

## 1. Introduction

High-speed information processing drives the information age. Along with the ever-increasing demand for data capacity and processing speed, photonic technologies have become a fast-evolving frontier with growing interdisciplinary interest [1], facilitating advances in a variety of fields such as optical communications [2], ultrafast computing [3], precision measurements [4], artificial intelligence [5], and quantum optics [6]. With a sequence of equidistant laser frequencies that are able to link the optical and electrical domain, laser frequency combs (LFCs) provide a key technology to, for example, achieve the most accurate time references to date [7, 8], underpinning many technical breakthroughs in science and industry [9].

Since their first experimental demonstration in 2007 [10], LFCs generated by compact microresonators, or so-called "optical microcombs", have become an exceptionally active research field drawing enormous attention [11-13]. Compared to conventional LFCs generated by mode-locked solid-state or fiber lasers [14-16], optical microcombs feature a small device volume and high compatibility for on-chip integration, which make them powerful alternatives of LFCs with reduced size, weight, and power consumption (SWaP). In addition, the repetition rates of optical microcombs, which typically range from gigahertz to terahertz and are well beyond what conventional LFCs can possibly offer [17], intrinsically enable optical microcombs to address a range of applications that deal with high-bandwidth electrical signals, which have close connections with many core industries such as telecommunications, information processing, precision measurements, and artificial intelligence.

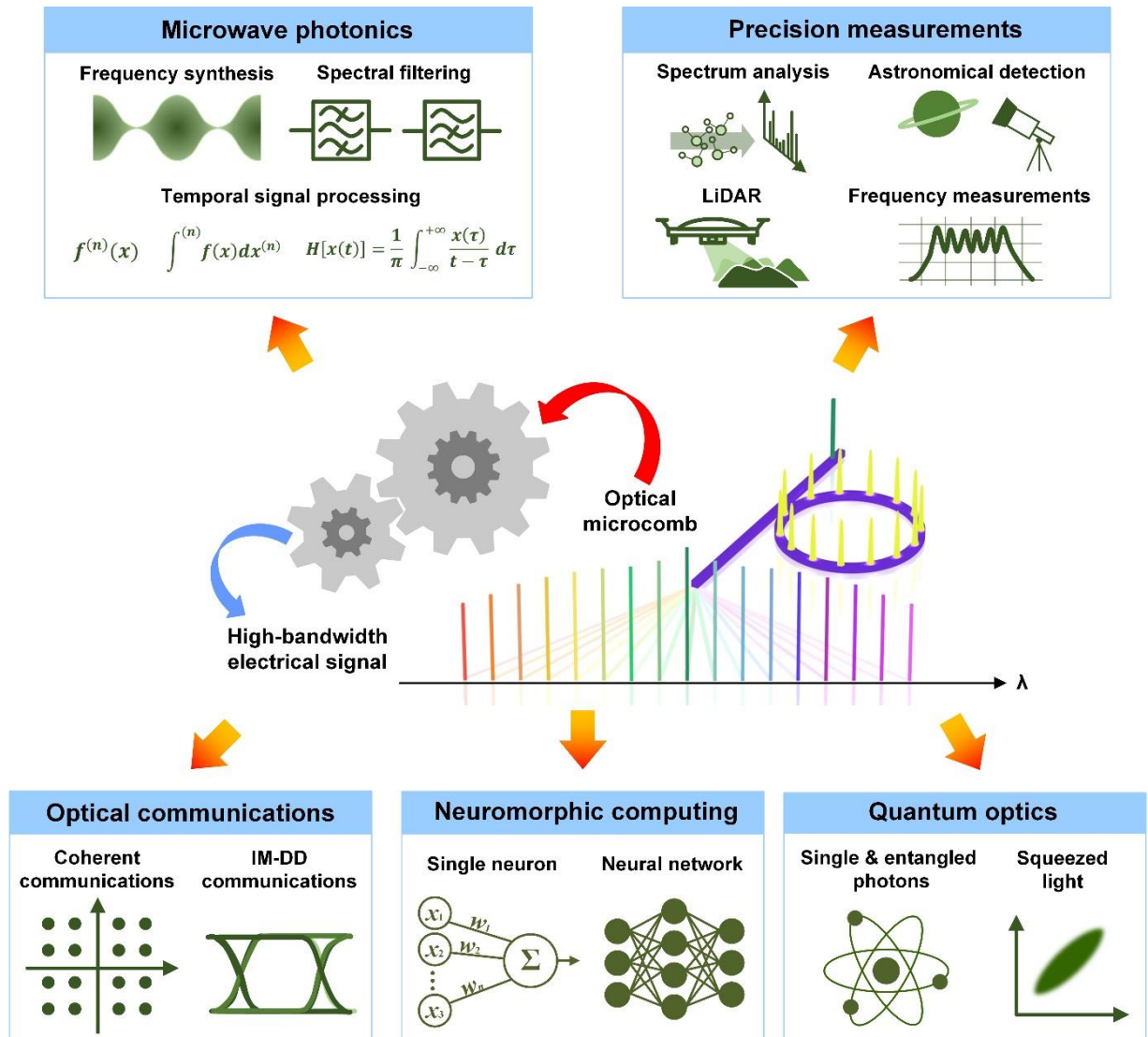

**Figure 1.** Broad applications of optical microcombs. LiDAR: light detection and ranging. IM-DD: intensity modulation - direct detection.

In tandem with the significant research on optical microcombs themselves over the past decade, there have been enormous advances in their applications in terms of both classical and quantum optics, with new applications being continually reported. As shown in **Figure 1**, the applications of optical microcombs can be divided into five main categories – microwave photonics, optical communications, precision measurements, neuromorphic computing, and quantum optics. The applications to microwave photonics refers to the use of optical microcombs to achieve signal handling in microwave photonic systems [18], where microwave signals are up-converted into the optical domain, and then pass through optical modules before being down-converted back to the microwave domain. The optical communications applications represent the realization of high-speed data transmission based on optical microcombs, with high-speed electrical signals being modulated onto optical microcombs that serve as sources of multiple optical carriers. Precision measurements include the use of optical microcombs to achieve measurement functions such as spectrum analysis, ranging, and frequency measurements. The neuromorphic computing applications include the use of optical microcombs to realize different computing functions in neuromorphic systems, which take inspiration from biological visual cortex systems to create power-efficient computing hardware that is capable of handling sophisticated tasks [19]. Quantum optics based on microcombs provide novel quantum optical sources for quantum information science, mainly involving the generation of single photons, entangled photons, and squeezed light [6].

The chronology for the applications of optical microcombs is shown in **Figure 2(a)**. **Figure 2(b)** shows the number of relevant publications in Science Citation Index (SCI) journals versus year since 2012. Both of these highlight the rapid growth in the use of optical microcombs for different applications, which has been driven mainly by three factors. First, the performance of optical microcombs has been continually improving in terms of its spectral bandwidth, energy efficiency, phase coherence, and stability, and this has significantly enhanced their performance in these applications. Secondly, optical microcombs have become more widely used for traditional and demanding applications, such as frequency synthesis [20-28], spectral filtering [29-33], temporal signal processing [34-40], and optical communications [33, 41-45]. Finally, many new and exciting interdisciplinary applications have emerged, spanning the frontiers of light detection and ranging (LiDAR) [46-49], astronomical detection [50-53], neuromorphic computing [54-56], and quantum optics [11, 57, 58], which bring significant new opportunities.

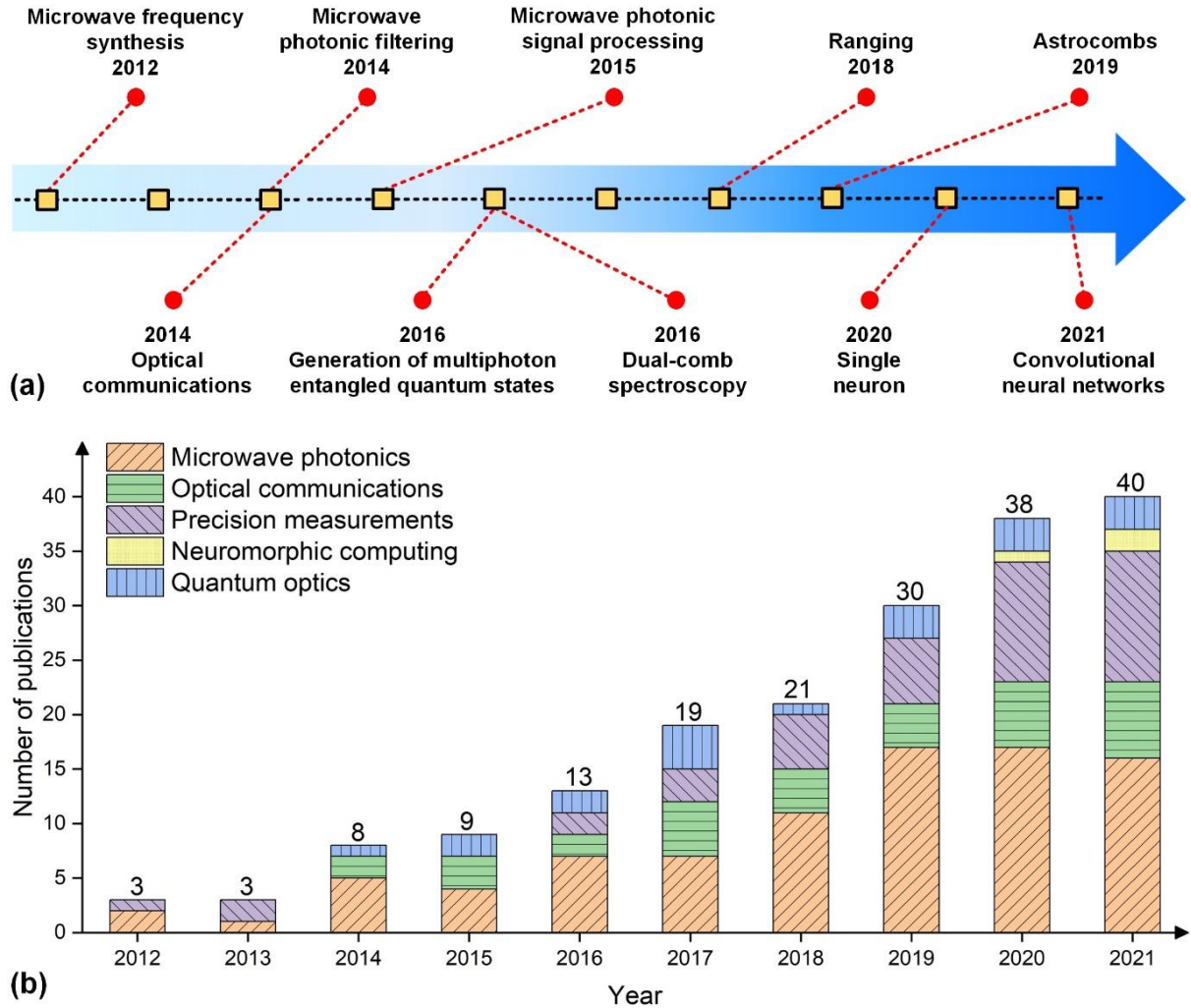

**Figure 2.** (a) Chronology for different applications of optical microcombs, including the first demonstrations of microwave frequency synthesis [20], microwave photonic filtering [29], optical communications [41], microwave photonic signal processing [34], generation of multiphoton entangled quantum states [59], dual-comb spectroscopy [60], ranging [46, 47], single neuron [54], astrocombs [50, 51], and convolutional neural networks [55, 56]. (b) Number of publications on different applications of optical microcombs in Science Citation Index journals versus year since 2012. Data were taken from ISI Web of Science.

In this paper, we review the applications of optical microcombs, including microwave photonics, optical communications, precision measurements, neuromorphic computing, and quantum optics. While optical microcombs have been the subject of recent reviews [9, 11-13, 61, 62], they have predominantly focused on the physics or dynamics of the optical microcombs themselves, rather than their applications. In addition to reviewing the state-of-the-art of these fields, we provide quantitative analysis for the key performance parameters for some of the most important applications. We also highlight the open challenges and future directions for both conventional and emerging applications.

This paper is structured as follows. In **Section 2**, we review the advances in generating optical microcombs, being categorized into material platforms, device architectures, soliton classes, and driving mechanisms. Next, the applications of optical microcombs to microwave photonics are discussed in **Section 3**, including frequency synthesizers, microwave photonic filters, and microwave photonic signal processors. In **Section 4**, the use of optical microcombs for optical communications in both coherent and intensity modulation - direct detection (IM - DD) transmission systems is reviewed. In **Section 5**, precision measurements based on optical microcombs are reviewed, including dual-comb spectroscopy, ranging, astrocombs, frequency measurements, and spectrum channelizers. In **Section 6**, neuromorphic computing applications of optical microcombs are reviewed, including those in both single neurons and neural networks. The applications of quantum microcombs are summarized in **Section 7**, including the generation of single / entangled photons and squeezed light. The current challenges and future perspectives are discussed in **Section 8**. Finally, conclusions are given in **Section 9**.

## 2. Generation of optical microcombs

In this section, we review the devices and methods used for generating high-performance optical microcombs, which are a prerequisite for their practical applications. The advances in development of material platforms, modification of device architectures, and investigation of different soliton states and driving mechanisms, not only improve the microcombs' performance such as spectral bandwidth [63], phase coherence [64], operation stability [65], and energy efficiency [66], but also reduce the fabrication cost, system size, and manipulation complexity.

This section is organized into four categories: material platforms, device architectures, soliton classes, and driving mechanisms. The first category reviews the progress in development of new material platforms and towards the goal of achieving high-volume manufacturing. The second discusses novel device designs for improving microcomb performance. The third summarizes the different classes of soliton states, and the last discusses the driving mechanisms for generating microcombs with high coherence, wide bandwidths, high stability and efficiency, and simple and reliable initiation and operation.

**2.1 Device platforms**

*2.1.1 Material platforms*

Following the first report of optical microcombs based on a silica toroid microcavity [10], they have been realized in many material platforms [67], highlighted by the first report of microcombs in photonic integrated platforms, which were also compatible with the complementary metal-oxide-semiconductor (CMOS) fabrication technology for silicon electronic chips [68, 69]. The two initial CMOS compatible nonlinear platforms – silicon nitride and high-index doped silica glass (Hydex), arguably opened up the field of integrated microcomb sources, when the community realized that they could ultimately be achieved in a fully integrated form.

**Figure 3** shows a development roadmap of optical microcombs based on different material platforms, including both bulk-optic resonators and integrated platforms. A range of material platforms including silica ($SiO_2$) [10, 20, 63, 70, 71], magnesium fluorides ($MgF_2$) [72-74], calcium fluorides ($CaF_2$) [75, 76], barium fluorides ($BaF_2$) [77], strontium fluorides ($SrF_2$) [78], silicon (Si) [79, 80], silicon nitride ($Si_3N_4$) [69, 81, 82], Hydex glass [68, 83], aluminum nitride (AlN) [84], and diamond [85] have been reviewed previously [83]. More recently, optical microcombs have been realized in new material platforms, with an emphasis towards integrated devices, including aluminum gallium arsenide (AlGaAs) [86-88], deuterated $Si_3N_4$ (D-SiN) [89], silicon carbide (SiC) [90], tantalum pentoxide ($Ta_2O_5$) [91], gallium phosphide (GaP) [92], and lithium niobate ($LiNbO_3$) [93-95].

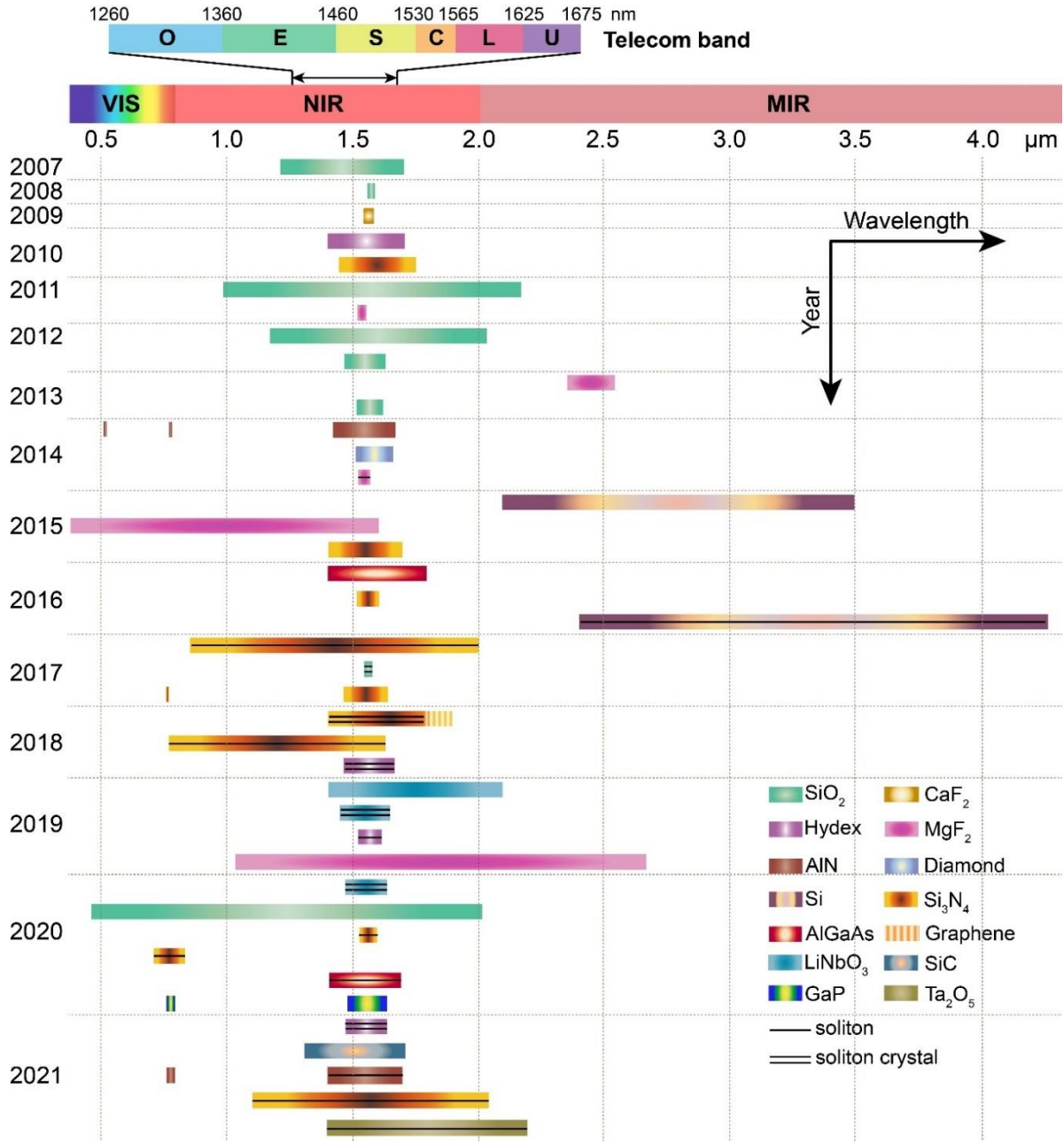

**Figure 3.** Development roadmap of optical microcombs based on various material platforms. Values of wavelength ranges are taken from: Ref. [10] for 2007, Ref. [96] for 2008, Ref. [75] for 2009, Ref. [68] for 2010 (Hydex), Ref. [69] for 2010 ($Si_3N_4$), Ref. [63] for 2011 ($SiO_2$), Ref. [72] for 2011 ($MgF_2$), Refs. [20, 97] for 2012, Ref. [73] for 2013 ($MgF_2$), Ref. [71] for 2013 ($SiO_2$), Ref. [98] for 2014 (AlN), Ref. [85] for 2014 (diamond), Ref. [65] for 2014 ($MgF_2$), Ref. [79] for 2015 (Si), Ref. [74] for 2015 ($MgF_2$), Ref. [99] for 2015 ($Si_3N_4$), Ref. [86] for 2016 (AlGaAs), Ref. [82] for 2016 ($Si_3N_4$), Ref. [80] for 2016 (Si), Refs. [100, 101] for 2017 ($Si_3N_4$), Ref. [102] for 2017 ($SiO_2$), Refs. [103, 104] for 2018 ($Si_3N_4$), Ref. [105] for 2018 (Hydex), Refs. [94, 95] for 2019 ($LiNbO_3$), Ref. [106] for 2019 (Hydex), Ref. [107] for 2019 ($MgF_2$), Ref. [108] for 2020 ($LiNbO_3$), Ref. [109] for 2020 ($SiO_2$), Refs. [110, 111] for 2020 ($Si_3N_4$), Ref. [88] for 2020 (AlGaAs), Ref. [92] for 2020 (GaP), Ref. [112] for 2021 (Hydex), Ref. [90] for 2021 (SiC), Ref. [113] for 2021 (AlN), Ref. [114] for 2021 ($Si_3N_4$), and Ref. [91] for 2021 ($Ta_2O_5$).

The search for new material platforms has been motivated by many factors, not just to expand the wavelength range beyond the telecom band around 1550 nm where most state-of-the-art applications of optical microcombs are centered, but even for devices operating within this same wavelength range in order to improve their performance and efficiency by enhancing the optical nonlinearity. The predominant integrated platforms for optical microcombs have been $Si_3N_4$ [69] and Hydex [68] and while these offer many advantages such as negligible nonlinear absorption and very low linear loss, their Kerr nonlinearity is quite low compared to silicon [115]. Hence, there is an on-going search for new material platforms that feature both high nonlinearity and low (linear and nonlinear) loss, in order to improve microcomb performance. In terms of expanding the wavelength range of microcombs outside the telecom band, while this may not be necessary for mainstream traditional applications, many emerging applications such as dual-comb spectroscopy, astrocombs, LiDAR, and visible light communication require microcombs operating at new wavelengths.

AlGaAs has been widely used for laser diodes and photodetectors (PDs) that operate at photon energies around its bandgap near 800 nm [116]. Because this is roughly twice the photon energy as 1550 nm, it was one of the first platforms to be investigated as a nonlinear optical platform for the telecom band [117]. AlGaAs also has a very high Kerr coefficient ($n_2$) of about $10^{-17}$ $m^2\,W^{-1}$ and a large transparency window from around 800 nm in the near-infrared to the mid-infrared region [117]. This allows the engineering of very high nonlinear figure of merit (FOM) [118] to achieve excellent device performance [119]. Recently, it has become a new and interesting focus for optical microcomb generation [86]. The key to this has been the successful development of the very challenging integrated AlGaAs-on-insulator platform to achieve low-loss, high-index-contrast AlGaAs waveguides. **Figure 4(a-i)** shows an AlGaAs-on-insulator microring resonator (MRR) used for optical microcomb generation [87]. The aluminum fraction was engineered to be 20% to alleviate the two-photon absorption (TPA) in the C-band. The waveguide had a low propagation loss of ~0.4 dB/cm, yielding a high-quality (Q) factor of ~$1.53 \times 10^6$ for a fabricated MRR with a radius of ~12 µm. **Figure 4(a-ii)** shows the generated comb spectrum, which had a free spectral range (FSR) of ~1 THz and covered a wavelength range from ~1450 nm to ~1700 nm.

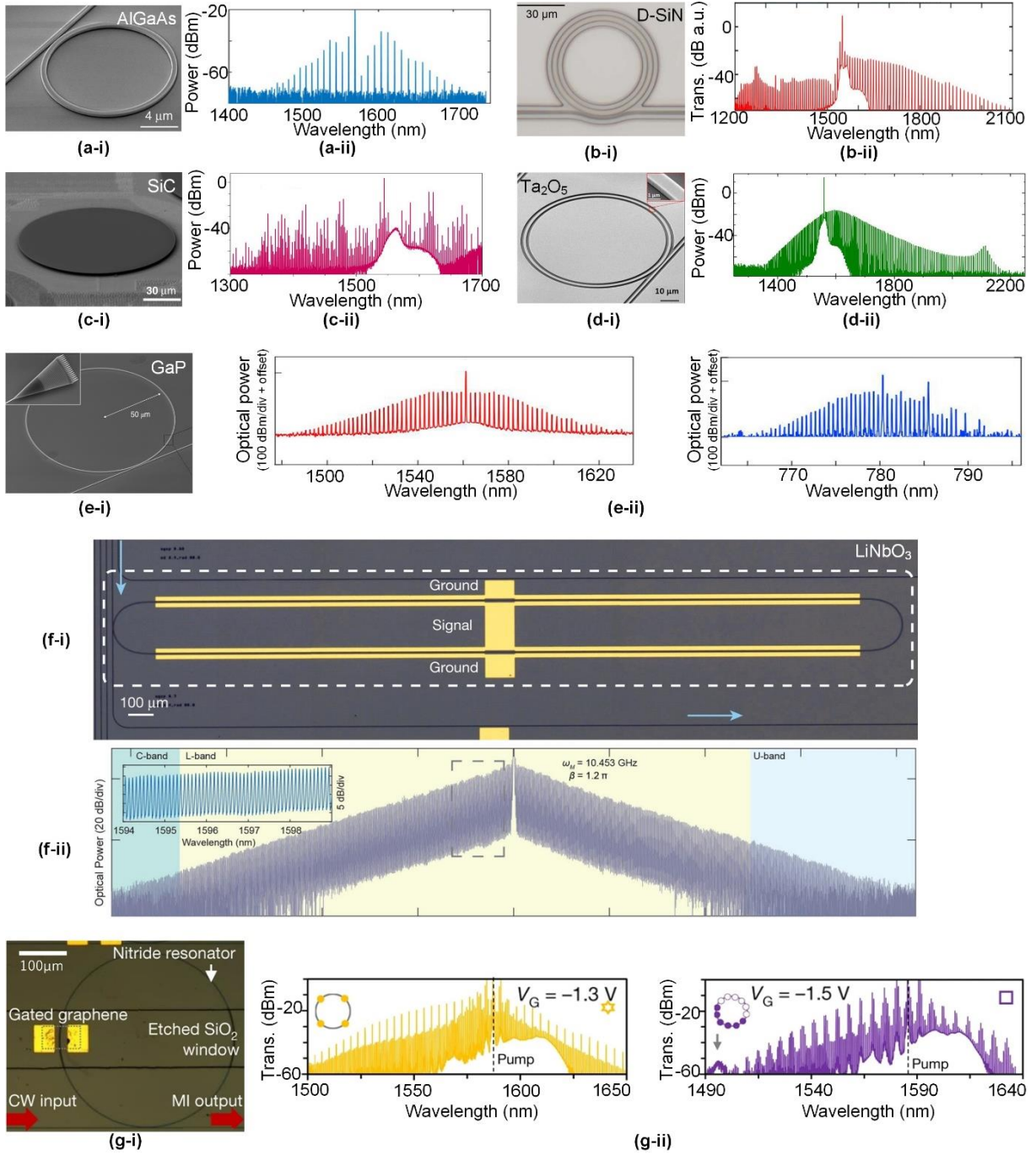

**Figure 4.** Recent progress in optical microcomb generation based on new material platforms. (a) Aluminum gallium arsenide (AlGaAs). (b) Deuterated silicon nitride (D-SiN). (c) Silicon carbide (SiC). (d) Tantalum pentoxide ($Ta_2O_5$). (e) Gallium phosphide (GaP). (f) Lithium niobate ($LiNbO_3$). (g) Silicon nitride ($Si_3N_4$) with gated graphene. In (a) – (g), (i) shows the microresonators used for generating microcombs and (ii) shows the generated comb spectra. (a) Reprinted with permission from [*Nat. Commun.*, 11, 1 (2020)].[87] (b) Reprinted with permission from [*Opt. Lett.*, 43, 1527 (2018)].[89] (c) Reprinted with permission from [*Light: Sci. Appl.*, 10, 139 (2021)].[90] (d) Reprinted with permission from [*Optica.*, 8, 811 (2021)].[91] (e) Reprinted with permission from [*Nat. Photonics.*, 14, 57, (2020)].[92] (f) Reprinted with permission from [*Nature.*, 568, 373 (2019)].[93] (g) Reprinted with permission from [*Nature.*, 558, 410 (2018)].[103]

Deuterated SiN (D-SiN) has been developed to reduce the linear loss of $Si_3N_4$ waveguides for efficient optical microcomb generation [89]. The use of deuterium rather than hydrogen in the silane precursor effectively eliminates the N-H bonds, the source of high optical absorption in the telecom band, without needing to anneal the devices at high temperatures (> 1000 °C). A waveguide propagation loss of ~0.31 dB/cm in the range of 1.5 µm – 1.6 µm was achieved by using isotopically substituted precursors in the plasma-enhanced chemical vapor deposition (PECVD) process to shift the N-H overtone absorption band from ~1.5 µm to ~2 µm [120]. **Figure 4(b-i)** shows a D-SiN MRR used for optical microcomb generation, which had a radius of ~23 µm and a Q factor of ~$5.6 \times 10^5$. **Figure 4(b-ii)** shows the generated comb spectrum featuring an FSR of ~1 THz over a 900-nm wavelength range from ~1200 nm to ~2100 nm.

SiC has been developed as a CMOS-compatible integrated photonic platform for linear optics since 2011 [121] and nonlinear optics since 2013 [122]. It has a wide transparency window from ~0.37 µm to ~5.6 µm and a large bandgap (*e.g.*, ~3.26 eV for 4H poly-types) that yields negligible TPA at near-infrared wavelengths [122]. These, together with its reasonably large Kerr nonlinearity (~$8 \times 10^{-19}$ $m^2\,W^{-1}$ [90]), make SiC attractive for microcomb operation. Recent progress in fabricating ultralow loss SiC films [123, 124] has resulted in integrated SiC devices capable of generating optical microcombs. **Figure 4(c-i)** shows a 4H-SiC microdisk resonator with a radius of ~100 µm and a Q factor of ~$7.1 \times 10^6$ [90]. **Figure 4(c-ii)** shows the comb spectrum generated at a low pump power of ~13 mW, which had an FSR of ~260 GHz covering a wide wavelength range from ~1300 nm to ~1700 nm. More recently, octave-spanning (1100 nm − 2400 nm) microcomb generation in a 4H-SiC MRR with a radius of ~36 µm and a Q factor above $10^6$ has also been demonstrated by optimizing device fabrication and engineering mode dispersion [125].

$Ta_2O_5$ is a CMOS-compatible dielectric that has traditionally been used for high reflectivity mirror coatings [126], and has been recently studied as a platform for nonlinear optics [127], including optical microcomb generation [91]. It has a typical bandgap > 3.8 eV [128, 129], with an $n_2$ of about 3 times that of $Si_3N_4$ (at 1550 nm). **Figure 4(d-i)** shows a $Ta_2O_5$ MRR fabricated via ion-beam sputtering to deposit a crack-free tantalum film on a thermally oxidized silicon wafer [91]. It has a radius of ~23 µm and a Q factor of ~$3.8 \times 10^6$. **Figure 4(d-ii)** shows the comb spectrum generated at a low pump power of ~20 mW, which covers a wide wavelength range from ~1400 nm to ~2200 nm.

GaP has been of interest as an optoelectronic material for about 50 years [130], mainly because of its relatively large bandgap (~2.25 eV) which makes it attractive for solid-state light-emitting devices in the visible wavelength range. It has been studied as a nonlinear optical platform motivated by its high third-order nonlinearity ($n_2$ = ~$1.1 \times 10^{-17}$ $m^2 W^{-1}$), together with its negligible TPA (for wavelengths longer than half of the bandgap, at ~1.1 µm), as well as its large refractive index ($n > 3$ at 1550 nm) [131]. Furthermore, since it is a zincblende crystal it has a nonzero second-order nonlinearity which opens up new nonlinear processes beyond what the third-order Kerr nonlinearity can provide. Recently, GaP has been developed as an integrated material platform for optical microcomb generation [92]. **Figure 4(e-i)** shows a GaP MRR with a radius of ~50 µm and a Q factor of ~$2.5 \times 10^5$ that generated optical microcombs with a spacing of ~250 GHz in the near-infrared region (1500 nm – 1620 nm) [92]. Because of its large second-order nonlinearity, the comb was able to be simultaneously frequency doubled into the visible region, as shown in **Figure 4(e-ii)**.

For noncentrosymmetric materials, the second-order optical nonlinearity enables the ability to not only frequency double microcombs but also to employ novel processes in generating microcombs based on the Pockels electro-optic (EO) effect [132]. Recently, $LiNbO_3$ with a large Pockels coefficient has been integrated on chip for generating optical microcombs [93, 94, 133]. **Figure 4(f-i)** shows an integrated $LiNbO_3$ MRR, which had a waveguide cross section of ~1.4 µm × 0.6 µm and a Q factor of ~$1.5 \times 10^6$ [93]. The spectrum of microcombs generated by the LiNbO$_3$ MRR is shown in **Figure 4(f-ii)**, which had a low FSR of ~10.453 GHz and more than 900 comb lines spanning from 1560 nm to 1640 nm. LiNbO$_3$ also has a large third-order nonlinearity, which enables the more conventional generation of Kerr optical microcombs [95]. The LiNbO$_3$ integrated platform with both second and third-order nonlinearities can achieve both microcomb generation and modulation simultaneously, thus avoiding the need for heterogeneous integration of different materials.

Recent advances in graphene optoelectronics [134] have motivated its use for engineering optical micro-combs. By integrating a two-dimensional (2D) graphene film on a Si$_3$N$_4$ MRR, gate-tunable optical microcombs have been demonstrated [103]. **Figure 4(g-i)** shows a hybrid graphene-Si$_3$N$_4$ MRR, where the bare Si$_3$N$_4$ MRR had a Q factor of ~$1.6 \times 10^6$. After gating of the graphene film, the hybrid MRR could still preserve a Q factor of up to ~$1.0 \times 10^6$. The chromatic dispersion of the hybrid MRR was modulated by gate tuning the Fermi level of graphene, thus allowing tunable transitions of the comb states. **Figure 4(g-ii)** shows the measured comb spectra taken at different gate voltages.

*2.1.2 High-volume manufacturing*

The reliable and high yield manufacturing of on-chip microcomb sources with a high degree of integration will be critical for the practical implementation of microcombs outside the laboratory. Although the generation of optical microcombs based on CMOS-compatible integrated MRRs made from SiN and Hydex was demonstrated in 2010 [68, 69], these optical microcombs were excited via continuous-wave (CW) pump from off-chip laser sources. In recent years, there has been significant progress in the fabrication of on-chip optical microcomb sources with co-integrated CW laser sources [33, 135, 136]. On the other hand, despite being an important class of microresonators for generating optical microcombs, whispering-gallery-mode (WGM) microcavities mainly rely on discrete devices implemented based on bulk optics – *i.e.*, non-integrated platforms. Recently, however, there have been exciting advances in developing fabrication methods for on-chip integration of the WGM microcavities [21, 60, 137, 138]. All of these advances leverage the wafer-scale mass-production in foundries and have underpinned many new breakthroughs in system-level applications.

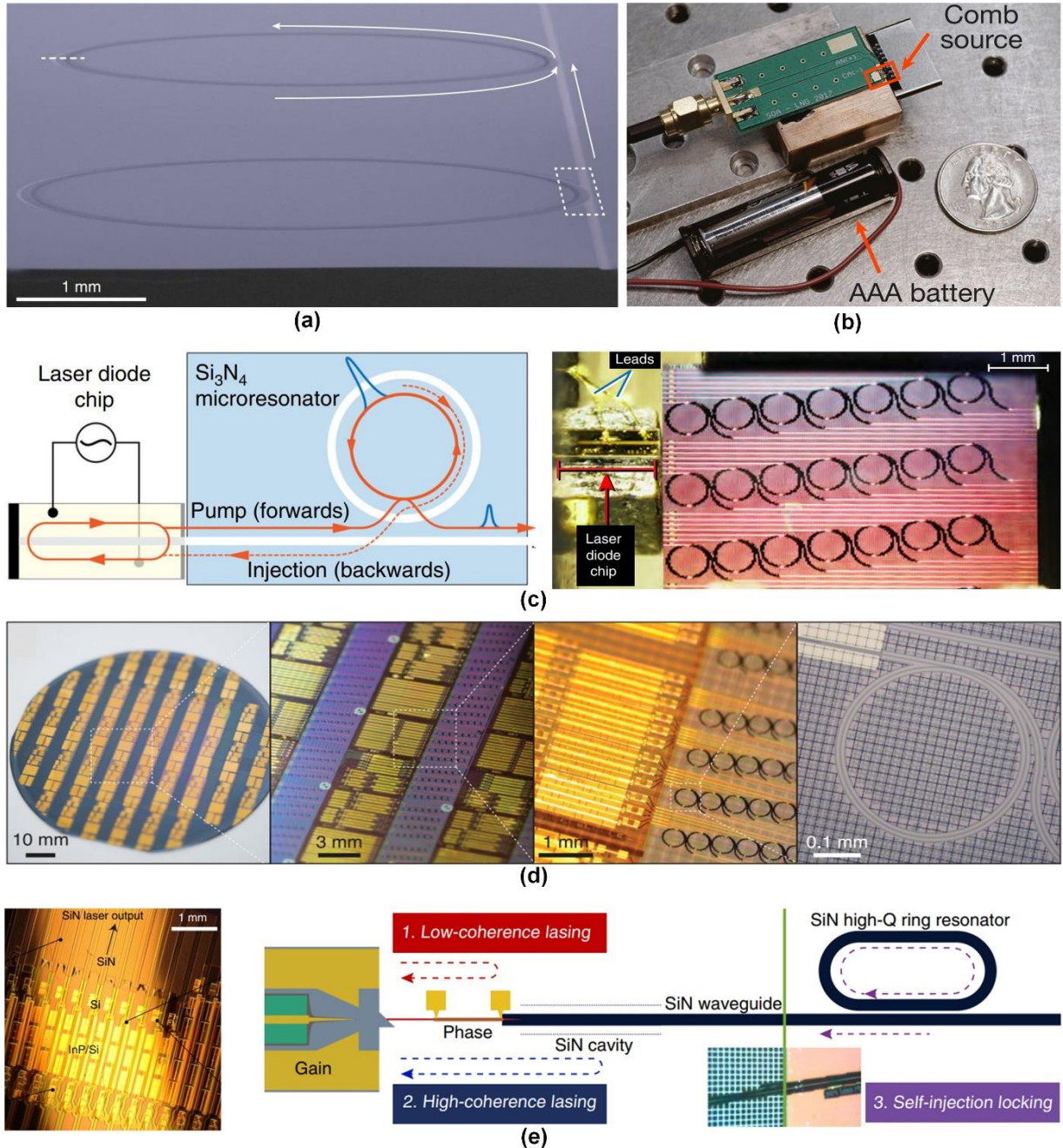

**Figure 5**. Recent progress in high-volume manufacturing of optical microcomb devices. (a) A microcomb generator based on integrated silica ridge resonators with ultrahigh Q factors. (b) A battery-operated microcomb generator. (c) Electrically pumped microcomb generation by using a commercial InP laser diode. (d) Heterogeneously integrated microcomb generators on a silicon substrate. (e) A microcomb generator based on a high-power, low-noise laser fully integrated with a $Si_3N_4$ MRR. (a) Reprinted with permission from [*Nat. Photonics.*, 12, 297 (2018)].[138] (b) Reprinted with permission from [*Nature.*, 562, 401 (2018)].[139] (c) Reprinted with permission from [*Nat. Commun.*, 11, 6384 (2020)].[140] (d) Reprinted with permission from [*Science.*, 373, 99 (2021)].[135] (e) Reprinted with permission from [*Nat. Commun.*, 12, 6650 (2021)].[136]

The on-chip integration of discrete silica WGM microcavities with record high Q factors above $2 \times 10^8$ has been realized (**Figure 5(a)**) [138], where specially designed $Si_3N_4$ waveguides enabled efficient energy coupling to other co-integrated devices. By pumping an integrated silica WGM resonator with a CW power of ~25 mW, soliton microcombs with repetition rates of ~15 GHz have been successfully generated.

A portable battery-operated microcomb source has been realized by co-integrating a III-V reflective semiconductor optical amplifier (RSOA), a $Si_3N_4$ laser cavity, and a $Si_3N_4$ MRR (**Figure 5(b)**) [139], where soliton combs were generated with a low pump power of ~98 mW supplied by a standard AAA battery.

Another electrically pumped microcomb generator has been demonstrated by using a commercial indium phosphide (InP) laser diode to pump a $Si_3N_4$ MRR (**Figure 5(c)**) [140], where soliton combs were generated with a pump power < 1 W and transitions of comb states were observed by tuning the current of the laser diode.

Heterogeneously integrated microcomb sources on a silicon substrate have also been reported [135], where InP/Si semiconductor lasers and $Si_3N_4$ MRRs were monolithically integrated via wafer bonding (**Figure 5(d)**), generating DKS states with a comb spacing of ~100 GHz. Subsequently, InP/Si/$Si_3N_4$ lasers with both high output power (> 10 mW) and narrow linewidth (< 1 kHz) have been integrated (**Figure 5(e)**) [136], and successful optical microcomb generation was demonstrated using these lasers heterogeneously integrated with high-Q $Si_3N_4$ MRRs.

**2.2 Device architectures**

In addition to developing new material platforms and fabrication technologies, significant attention has been devoted to the modification of device architectures. **Table 1** summarizes typical architectures of resonators for generating optical microcombs, including toroid [10, 63, 72-76, 109, 141], wedge [47, 142-144], ridge [138], sphere [70], disk [21, 51, 60, 90], rod [22-24, 71, 145-147], single ring [68, 93, 135, 148], and multiple rings [149-151]. The resonators with different architectures show differences in the Q factors as well as the repetition rates for the generated microcombs. The device structures are also closely related to the material platforms, with different device architectures showing different compatibility for on-chip integration. In this section, we review and discuss the modification of device architectures for microcomb generation, including dispersion engineering, coupling control, and multi-mode interaction. Since the different device architectures for generating optical microcombs have already been reviewed in previous literature [11, 12, 83], here we focus on recent progress after 2018, which has improved the microcomb performance in terms of spectral bandwidth, power consumption, conversion efficiency, and stability.

**Table 1. Comparison of different device architectures for generating optical microcombs.**

| Resonator Structure | Q factor | Repetition rate (GHz) | On-chip integration | Material platforms |
|---|---|---|---|---|
| Toroid | $10^8 - 10^{10}$ | $10^1 - 10^2$ | No | $SiO_2$ [10, 63, 109, 141], $MgF_2$ [72-74], $CaF_2$ [75, 76] |
| Wedge | $10^8$ | $10^0 - 10^1$ | Yes | $SiO_2$ [47, 142-144] |
| Ridge | $10^8$ | $10^1$ | Yes | $SiO_2$ [138] |
| Sphere | $10^7$ | $10^2$ | No | $SiO_2$ [70] |
| Disk | $10^6 - 10^8$ | $10^0 - 10^2$ | Yes | $SiO_2$ [21, 51, 60], SiC [90] |
| Rod | $10^8 - 10^9$ | $10^1 - 10^2$ | No | $SiO_2$ [71, 145, 146], $MgF_2$ [22-24, 147] |
| Single ring | $10^5 - 10^7$ | $10^1 - 10^3$ | Yes | $Si_3N_4$ [26, 55, 89, 104, 114, 135, 139, 140, 150, 152], Hydex [56, 68, 83, 112], Si [148], $LiNbO_3$ [93, 95, 108], AlGaAs [33, 88], AlN [98, 113], Diamond [85], GaP [92], $Ta_2O_5$ [91] |
| Multiple rings | $10^5 - 10^6$ | $10^2$ | Yes | $Si_3N_4$ [149-151] |

Waveguide dispersion in particular is an important parameter that determines the spectral bandwidth of optical microcombs. For optical resonators, their resonance frequencies can be expanded in a Taylor series around the pumped mode as follows [11, 12]

$$\omega_l = \omega_{l_0} + \sum_{n=1}^{N} \frac{D_n}{n!}(l-l_0)^n \qquad (1)$$

where $l$ (= 0, 1, 2, …) are the labelled mode numbers of the mode spectrum, $l_0$ is the eigennumber of the pump mode, $\omega_l$ are eigenfrequencies of the resonator, and $N$ is the order of truncation for the expansion. In **Eq. (1)**, the Taylor expansion coefficients $D_n$ are related to the dispersion coefficients of the propagation constant $\beta_n$ as below [11]

$$D_n = \left(\frac{2\pi}{L}\right)^n \frac{d^{n-1}}{d\omega^{n-1}}\left[\frac{\omega-\omega_0}{\beta(\omega)-\beta(\omega_0)}\right]\bigg|_{\omega=\omega_0} \approx -v_g \left(2\pi\Delta f\right)^n \beta_n \qquad (2)$$

where $L$ is the round-trip length, $\omega_0$ is the pump frequency, $v_g = 1/\beta_1$ is the group velocity, and $\Delta f$ is the FSR. In **Eq. (2)**, $D_1 = 2\pi\Delta f$ is related to the FSR, $D_2$ is the dispersion parameter widely used for characterizing the waveguide dispersion. Another frequently used parameter is the group velocity dispersion $\beta_2$, and the relation between $D_2$ and $\beta_2$ can be given by

$$D_2 = -\frac{2\pi c}{\lambda^2}\beta_2 \qquad (3)$$

where $c$ and $\lambda$ are the light speed in vacuum and light wavelength, respectively. Waveguides with normal dispersion correspond to $D_2 < 0$ or $\beta_2 > 0$, whereas those with anomalous dispersion correspond to $D_2 > 0$ or $\beta_2 < 0$. For materials with positive Kerr coefficients $n_2$ (*e.g.*, silicon, silicon nitride, and silica), waveguides with anomalous dispersion are preferable since they can mitigate the phase mismatch for nonlinear optical processes such as four-wave mixing (FWM), self-phase modulation (SPM), and cross-phase modulation (XPM) [153, 154]. It should also be noted that the high-order ($n \geq 3$) dispersion items in **Eq. (2)** should be taken into account for optical microcombs with large spectral bandwidths.

To obtain a large comb spectral bandwidth, the waveguide dispersion needs to be engineered to achieve broadband phase matching. Octave-spanning dissipative Kerr soliton (DKS) states covering the wavelength window for biological imaging (700 nm – 1400 nm) have been generated by engineering the waveguide dispersion of a $Si_3N_4$ MRR (**Figure 6(a)**), which had a cross section of ~1300 nm × 740 nm, resulting in anomalous dispersion within the wavelength range of interest [104]. Based on precise control of dispersive waves via dispersion engineering (**Figure 6(b)**) [100], DKS states with a spectral bandwidth (from ~150 THz to ~350 THz) exceeding a full octave have also been generated. By engineering the dispersion of a Ge-on-Si MRR to achieve both dispersion flattening and dispersion hybridization, two-octave spanning mid-infrared microcombs from ~2.3 μm to ~10.2 μm have been generated at a low pump power of ~180 mW [155]. Increased comb flatness and reduced pump power can be achieved by engineering the waveguide zero-dispersion wavelengths. Octave spanning microcombs with increased spectral flatness and decreased pump power enabled by engineering the zero dispersion wavelengths of strip / slot hybrid MRRs have been theoretically investigated in Refs. [156, 157].

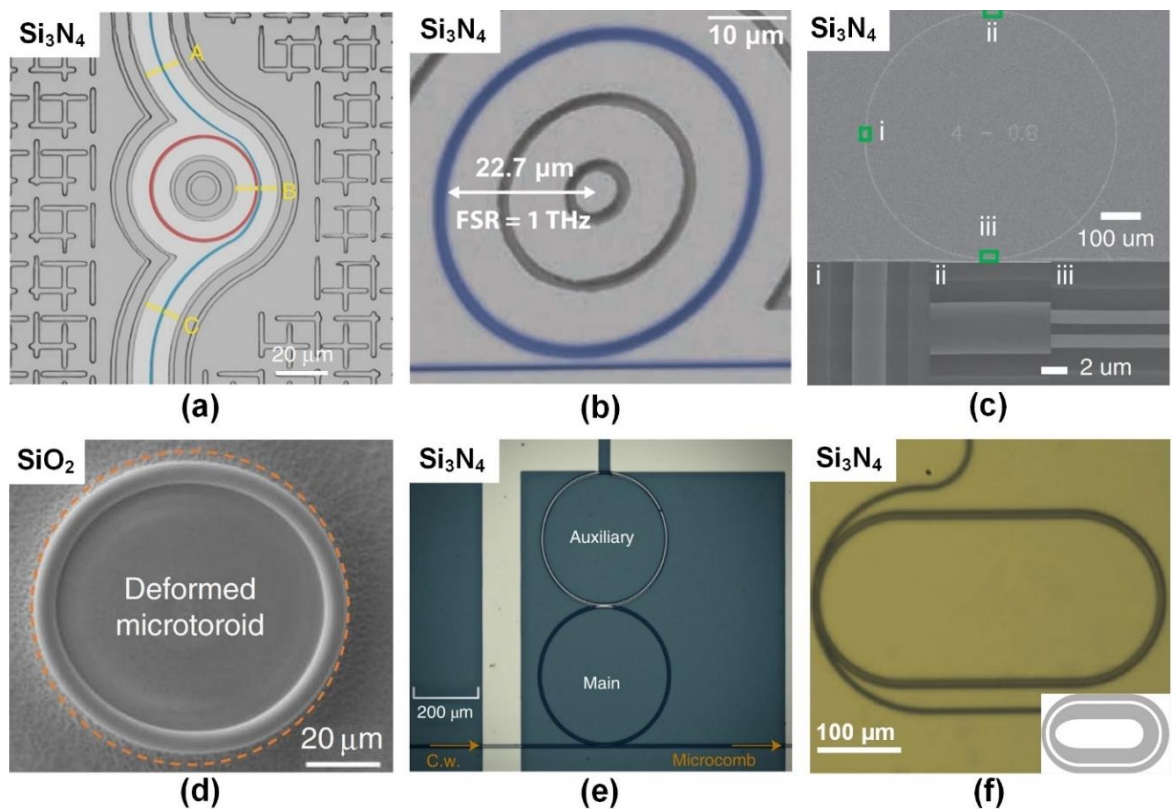

**Figure 6**. Recent progress in new device design for generating optical microcombs. (a) Waveguide geometry engineering in a $Si_3N_4$ MRR. (b) Dispersive waves engineering in a $Si_3N_4$ MRR. (c) Dispersion oscillation engineering in a tapered $Si_3N_4$ MRR with varied waveguide width. (d) Coupling engineering based on a deformed $SiO_2$ microtoroid resonator. (e) Multi-mode interaction engineering in two coupled $Si_3N_4$ MRRs. (f) Multi-mode interaction engineering in a $Si_3N_4$ concentric MRR. (a) Reprinted with permission from [*Nat. Commun.*, 9, 1 (2018)].[104] (b) Reprinted with permission from [*Optica.*, 4, 684 (2017)].[100] (c) Reprinted with permission from [*Light: Sci. Appl.*, 9, 52 (2020)].[152] (d) Reprinted with permission from [*Nat. Commun.*, 11, 2336 (2020)].[109] (e) Reprinted with permission from [*Nat. Photonics.*, 15, 305 (2021)].[151] (f) Reprinted with permission from [*Nat.*

*Commun.*, 8, 8 (2017)].[150]

Engineering the waveguide dispersion can also improve comb stability. By tapering the waveguide width of a $Si_3N_4$ MRR (**Figure 6(c)**), thus creating group velocity dispersion (GVD) oscillation, the frequency detuning stability zone has been improved by over an order of magnitude relative to microcombs generated by comparable MRRs but with a uniform waveguide width [152].

Engineering the coupling between microresonators and the access, or bus, waveguides to achieve broadband light extraction with high efficiency is also needed for realizing optical microcombs with large spectral bandwidths.

By using a pulley coupling configuration, up to 20-dB enhancement of light extraction at short wavelengths has been achieved compared to devices using directional couplers [158]. By slightly deforming a $SiO_2$ microtoroid resonator (**Figure 6(d)**) to create a chaotic tunneling channel [109], broadband collection of the generated optical microcombs via a nanofiber has been achieved, yielding a comb spectral bandwidth exceeding two octaves (from ~450 nm to ~2008 nm).

Amongst all the resonator structures, ring resonators have shown the highest compatibility with integrated platforms as well as the highest accuracy in tailoring the resonance modes. This makes it possible to generate localized anomalous dispersion for microcomb generation by engineering the interaction between different ring resonators that feature normal dispersion.

Microcomb generation based on two coupled $Si_3N_4$ MRRs with normal dispersion was first demonstrated in Ref. [149], where different comb states were achieved by using a micro-heater to adjust the mode interactions between the two MRRs. Recently, DKS has been generated in photonic diatomic molecules formed by coupled $Si_3N_4$ MRRs with normal dispersion (**Figure 6(e)**) [151], showing attractive advantages in achieving high conversion efficiency, uniform power distribution, and low-power operation. Coherent microcombs have also been generated by using a 300-nm-thick $Si_3N_4$ concentric MRR (**Figure 6(f)**), where a tapered concentric racetrack structure was chosen to selectively excite the resonant mode having anomalous dispersion [150].

**2.3 Soliton classes**

According to the classical theory of microcombs [11, 83], for microresonators with low cavity loss and considering a single mode and monochromatic driving, the evolution of the intracavity optical complex field envelope can be expressed by the mean-field Lugiato-Lefever (LLE) equation as below [11]

$$t_R \frac{\partial E(t,\tau)}{\partial t} = \left[ -\alpha - i\delta_0 + iL \sum_{n \geq 2} \frac{\beta_n}{n!} \left( i \frac{\partial}{\partial \tau} \right)^n + i\gamma |E|^2 L \right] E + \sqrt{1-\kappa} E_{in} \qquad (4)$$

where $t_R$ is the round-trip time, $E$ is the envelope of intracavity field, $E_{in}$ is the external driving field, $\tau$ is time expressed in a reference frame − moving with the group velocity of light at the pump frequency, $t$ is the slow time of the microresonator related to the number of round-trips m as $E(t = mt_R, \tau)$, $\alpha$ describes the total losses of the microresonator, $\delta_0$ is the phase detuning of the pump field, $L$ is the round-trip length of the microresonator, $\kappa$ is the power coupling coefficient of the input/output coupler, $\gamma$ is the nonlinear parameter, and $\beta_n$ are the dispersion coefficients of the propagation constant in **Eq. (2)**.

Different types of solutions of **Eq. (4)** correspond to different microcomb operation regimes, such as primary combs, chaotic states, and DKS states. Amongst them, DKS states, with high coherence and low noise resulting from the dual balance between the nonlinearity and dispersion as well as loss and gain, are the most widely used for practical applications [42, 46, 47]. In **Table 2**, we compare different types of soliton states generated by optical microcombs, including bright solitons, dark solitons, breather solitons, Dirac solitons, Stokes solitons, Brillouin-Kerr solitons, soliton crystals, soliton molecules, and laser cavity-solitons.

**Table 2. Comparison of different soliton classes of optical microcombs.**

| Soliton classes | Characteristics | Dispersion | Platform | Refs. |
|---|---|---|---|---|
| Bright solitons | Generated by using a red-detuned pump to excite a microresonator with anomalous dispersion. | Anomalous | $MgF_2$<br>$Si_3N_4$<br>$SiO_2$<br>Hydex<br>AlN<br>AlGaAs<br>Si | [65]<br>[110, 159]<br>[109, 160]<br>[106, 161]<br>[113, 162]<br>[88]<br>[80] |
| Dark solitons | Generated by using a blue-detuned pump to excite a microresonator with normal dispersion. | Normal | $Si_3N_4$ | [43, 99] |
| Breather solitons | Generated in presence of dynamical instabilities, where bright or dark solitons experience periodic variation in amplitude and duration. | Both | $Si_3N_4$<br>$MgF_2$<br>Si | [163-165]<br>[164]<br>[163] |
| Dirac solitons | Generated by engineering dispersion induced by nonlinear coupling between TE and TM modes on the basis of DKS. | Both | $SiO_2$ | [166] |
| Stokes solitons | Generated through Kerr-effect trapping and Raman amplification created by initially formed DKS. | Anomalous | $SiO_2$ | [167] |
| Brillouin-Kerr solitons | Generated by using a blue-detuned pump to excite red-detuned Brillouin lasing in a microresonator. | Anomalous (Brillouin mode) | $SiO_2$ | [168] |
| Soliton crystals | Self-organized ensembles of multiple co-propagating solitons with a crystal-like profile in the angular domain. | Anomalous | $SiO_2$<br>$Si_3N_4$<br>Hydex<br>$LiNbO_3$ | [102]<br>[103, 169]<br>[105, 112, 170]<br>[94, 108] |
| Soliton molecules | Bound states of solitons are achieved when there is a balance between attractive and repulsive effects. | Anomalous | $MgF_2$ | [171] |
| Laser cavity solitons | Generated based on filter-driven four-wave mixing in a gain fiber loop with a nested microresonator. | Anomalous | Hydex | [106] |

Bright and dark solitons are two basic soliton types determined by the dispersion of microresonators. Bright solitons are generated in microresonators with anomalous dispersion and excited by a red-detuned pump, in contrast to requirement for microresonators to have normal dispersion and a blue-detuned pump in the case of dark solitons. Unlike the stochastic nature of the number of peaks generated for bright solitons, there is a deterministic pathway towards mode-locking in the normal-dispersion region, which allows for the repeatable generation of dark solitons [99]. Dark solitons also display much higher power-conversion efficiency [43, 172], which is compelling for coherent optical communications. One reason for these advantages lies in the fact that dark solitons, being in some sense the inverse of single soliton states, have a naturally high power punctuated by dark solitons – and so the overall intracavity power of these states is very similar to the chaotic background. Hence there is very little photo-induced resonance shift when generating these states, which is the chief challenge for bright single solitons. For $Si_3N_4$ MRRs, the challenge of achieving anomalous dispersion to generate bright solitons has confronted the field since its beginning since this requires very thick (typically > 700 nm) $Si_3N_4$ layers which, because $Si_3N_4$ needs to be annealed at high temperatures to reduce the loss, often induces cracking [173, 174]. In contrast, the generation of dark solitons based on microresonators with normal dispersion avoids the need to grow thick $Si_3N_4$ layers [99].

On the basis of bright and dark solitons, derivative soliton types can be realized by exploiting other effects such as dynamical instabilities, mode-coupling induced dispersion, stimulated Raman scattering, and stimulated Brillouin scattering. These include breather solitons, Dirac solitons, Stokes solitons, and Brillouin-Kerr solitons.

Breather solitons are generated in presence of dynamic instabilities, where bright or dark solitons experience a periodic evolution in amplitude and duration [163]. The study of breather solitons is useful for understanding the soliton dynamics and provides guidance for generating stable soliton microcombs [163]. The typical operation regime of breather solitons is between the modulation instability regime and the steady soliton regime. Inter-mode breather solitons, which can be triggered by avoided mode crossings, have been demonstrated in the conventionally stable soliton regime [175].

Dirac solitons are generated by engineering dispersion induced by nonlinear mode-coupling in the visible band [166]. They normally have asymmetric soliton comb spectra due to different mode compositions on the different sides of the spectra.

Stokes solitons are derived from the Raman effect when the primary soliton is generated. They have frequency separation with primary solitons, which has significant potential for mid-infrared comb generation [167].

Brillouin-Kerr solitons are formed by exciting red-detuned Brillouin lasing in a microresonator via a blue-detuned pump. They are easier to generate due to long soliton steps and stable access to single soliton states. They have shown advantages in achieving narrow-linewidth comb lines and stable repetition rates, which are beneficial for generating low-noise microwave signals [168].

Interactions amongst co-existing solitons within the same resonant cavity can also be engineered to form solution groups with new capabilities, such as soliton crystals and soliton molecules.

Soliton crystals are self-organized ensembles of multiple co-propagating solitons with a crystal-like profile in the angular domain [102]. They are easy to generate, even with slow manual pump wavelength tuning, and are naturally robust [102]. As in dark solitons, the reason for this lies in the fact that the intracavity power for soliton crystals is much higher than single soliton states and so when forming them out of the chaotic background, there is very little photo or thermal induced resonance shift, meaning that rapid pump detuning is not required. Although anomalous dispersion and mode crossings are needed, they are not challenging to realize via dispersion engineering [44]. Soliton crystals feature scalloped shaped spectra which, although has been considered a disadvantage for practical applications, has in fact been shown not to present a significant drawback [44], partly due to the much higher efficiency soliton crystals achieve compared to single soliton states.

Soliton molecules are the bound states of solitons achieved when there is a balance between attractive and repulsive effects. They can be generated by engineering the interaction amongst co-existing solitons in the same resonant cavity and using a discrete pump in the red-detuned regime. Soliton molecules are reproducible with high coherence and high conversion efficiency [171], making them appealing for optical communications and precision measurements.

Laser cavity-soliton (LCS) microcombs are a new class of microcomb-based solitons realized by nesting a microresonator in a fiber loop with gain [106]. They are generated based on filter-driven FWM [176] and offer many highly attractive features, such as a high mode efficiency (defined as the fraction of optical power residing in the comb modes relative to the most powerful line) and a reduced average pump power. The LCSs have a theoretical mode efficiency that can approach 100%. A key reason for this high efficiency is the fact that LCSs do not require a CW background state for the solitons to operate, which frees up all of the optical energy to be able to reside in the optical pulses and hence in the comb wavelengths. Another important advantage of LCSs is the flexibility in adjusting their repetition rates, which can be tuned by more than a megahertz through simply adjusting the overall fiber loop laser cavity length.

Featuring high coherence and low noise, soliton microcombs have underpinned many recent breakthroughs in microwave photonics [83, 177], optical communications [42, 44], precision measurements [46, 60], neuromorphic computing [55, 56], and quantum optics [57, 178]. Moreover, the rich variety of different types of solitons, each having its own distinctive characteristics, significantly increases the range of performance for soliton microcombs that can be used for practical applications. For applications in microwave photonics and optical communications, two widely used families of soliton states are dark solitons [43, 99] and soliton crystals [44, 102], mainly because they are intrinsically easier to generate than bright single soliton states and are more energy-efficient and stable. As mentioned above, the reasons for all of these advantages stem from the same issue – the fact that the intracavity energy for these states is much higher than for single solitons. Hence there is very little power drop when they are formed, making them much easier to generate and stabilize, and also yielding much higher power levels in the cavity that allows for higher energy efficiency.

**2.4 Driving mechanisms**

The exact mechanism employed to drive or generate microcombs is crucial for engineering optical microcombs for different purposes, particularly to generate soliton states with high coherence, low noise, and high efficiency. In **Table 3**, we compare different driving mechanisms for generating optical microcombs. Previously, typical driving mechanisms, including frequency scanning [10, 65, 179, 180], power kicking [181-183], forward and backward tuning [184], two-colour pumping [185-187], EO modulation [97, 188, 189], self-injection locking [74, 190, 191], filter-driven FWM [106, 176, 192], integrated heaters [99, 149, 193], and self-referencing [194-196], have been reviewed in many articles [12, 83, 177]. Here we review and discuss new and innovative mechanisms introduced after 2018 [88, 106, 110, 141, 145, 147, 159, 197-200], generally with the aim of achieving simple and natural generation of high-performance microcombs with high stability and efficiency. All of these factors are crucial to enhance the performance and reduce the SWaP and cost of devices for practical applications. Many of these new mechanisms are aimed at solving some key issues that bright single solitons face, including the need for complex pump dynamics [65] to generate the soliton states and external feedback control to stabilize them. Improving the efficiency is another key aim since DKS states require a CW background that does not contribute to the comb power, and which takes up a significant part of the optical energy. This accounts for their generally low efficiency, typically < 5%.

The generation of single DKS based on cryogenic cooling was first demonstrated in a Si$_3$N$_4$ MRR [201], where the significantly reduced thermorefractive coefficient at low temperatures below 60 K enabled direct generation of bright DKS via adiabatic pump frequency tuning. Recently, this method has also been employed for generating stable bright DKS in an AlGaAs MRR [88], where the MRR's thermo-refractive coefficient was reduced more than 100 times compared to room temperature by cryogenic cooling it to 4K – 20K.

By launching a second CW pump into a Si$_3$N$_4$ MRR in the opposite direction [159], the intra-cavity thermal dragging dynamics have been effectively suppressed, resulting in stable bright DKS generation and soliton bursts. Similarly, by employing an auxiliary laser to compensate thermal shifts in a silica microrod resonator (**Figure 7(a)**), the length of soliton steps of the generated soliton microcombs has been significantly extended by 2 orders of magnitude [145]. Significant spectral bandwidth broadening of soliton microcombs generated by a silica microtoroid resonator has also been achieved by injecting a second pump in the normal dispersion regime (**Figure 7(b)**) [141], which compensated thermal shifts and generated another synchronized frequency comb via cross-phase modulation (XPM), thus extending the comb spectra into normal dispersion wavelengths.

In order to address the challenge of rapid and dynamic pump tuning that bright DKS states require, piezoelectric control of soliton microcombs generated in Si$_3$N$_4$ MRRs has been demonstrated (**Figure 7(c)**) [198]. By using integrated AlN actuators, a large actuation bandwidth of ~1 MHz was achieved – well surpassing those of integrated heaters (typically < 10 kHz) [198].

**Table 3. Driving mechanisms for generating optical microcombs.**

| Mechanisms | Methods | Year [a] | Refs. |
|---|---|---|---|
| Frequency scanning | Sweeping the pump from blue to the red-detuned regime before the microresonator is heated up by the thermo-optic effect. | 2007 | $SiO_2$ [10], $MgF_2$ [65], $Si_3N_4$ [179], AlN [180] |
| Two-colour pumping | Using two pumps at different wavelengths to generate microcombs based on cascaded FWM. | 2009 | $MgF_2$ [185, 186], Hydex [187] |
| Filter-driven FWM | Embedding a four-port MRR in a gain fiber laser loop to generate LCS microcombs based on passive laser mode locking. | 2012 | Hydex [106, 176] |
| EO modulation | Stabilizing a microcomb by locking the comb spacing to the spacing of EO modulated pump sidebands. | 2012 | $SiO_2$ [97, 188, 189] |
| Self-injection locking | Locking a microcomb by narrowing the pump linewidth via the self-injection locking effect and keeping light within the resonance. | 2014 | $MgF_2$ [74, 191] |
| Power kicking | Stabilizing the pump in short soliton steps by using modulators to accurately control the pump power and timing sequence. | 2015 | $Si_3N_4$ [182, 183], $SiO_2$ [181] |
| Integrated heaters | Using integrated heaters to assist mode selection, control of mode interactions, or passive thermal locking in microcomb generation. | 2015 | $Si_3N_4$ [99, 149, 193] |
| Self-referencing | Stabilizing a microcomb based on phase locking of the carrier-envelope offset frequency via self-referencing. | 2016 | $SiO_2$ [194], $Si_3N_4$ [195], AlN [196] |
| Forward and backward tuning | First sweeping the pump forward and then backward at a slow speed to generate deterministic single soliton microcombs. | 2017 | $MgF_2$ [184], $Si_3N_4$ [184] |
| Auxiliary laser | Using an auxiliary laser located at blue-detuned regime to compensate the decrease of intracavity heat during the generation of soliton microcombs. | 2019 | $Si_3N_4$ [159], $SiO_2$ [141, 145], Hydex [202] |
| Cryogenic cooling | Stabilizing a microcomb by cryogenic cooling the microresonator to quench the thermorefractive effects. | 2020 | AlGaAs [88] |
| Piezoelectric control | Using integrated actuators to achieve a large actuation bandwidth and realize rapid and dynamic pump tuning. | 2020 | $Si_3N_4$ [198] |
| Turnkey operation | Achieving turnkey mode locking of a microcomb by exploiting backscattered light from the microresonator to the pump laser cavity. | 2020 | $Si_3N_4$ [110] |
| Pulse pump | Using optical pulses as the pump for microcomb generation to reduce the power consumption and improve the conversion efficiency. | 2021 | $MgF_2$ [147], $Si_3N_4$ [147, 199] |
| Self-starting oscillation | Generating robust self-emergence microcombs in a microresonator-filtered fiber laser system by tailoring the slow nonlinearities in such system. | 2022 | Hydex [192] |

[a] The year corresponds to that when the first demonstration was reported.

Simple and robust mode locking in a $Si_3N_4$ MRR, occurring immediately after turning on a co-integrated pump laser (*i.e.*, so-called turnkey soliton microcombs), has been realized based on self- and cross- phase modulation induced by backscattered light from the MRR to the pump laser cavity (**Figure 7(d)**) [110], thus avoiding complex startup protocols and feedback control circuitry.

Self-starting soliton microcomb generation has been demonstrated based on the photorefractive effect in a $LiNbO_3$ MRR [94], where the photorefractive effect showed a more significant influence on the refractive index change than the thermo-optic effect, allowing the red-detuned regime to become a thermal stable region for generating self-starting soliton microcombs. Self-emergence (*i.e.*, spontaneous generation) of robust micro-cavity solitons has also been achieved in a microresonator-filtered fiber laser system (**Figure 7(e)**) [192]. By tailoring the slow nonlinearities arising from the fiber amplifier and the thermal response of a Hydex MRR, the temporal cavity-solitons were transformed into the dominant attractor of the system, which enabled consistent generation of the chosen soliton state by simply turning the system on, as well as displaying spontaneous recovery after complete disruption. Moreover, the generated soliton state was highly stable over long timeframes without external control, and the startup parameters only had to be set once.

In addition to CW pumping, pulse pumping can be used for soliton microcomb generation. Pulse pumped soliton generation was first demonstrated based on a fiber-based Fabry–Pérot microresonator [203]. Recently, pulse pumping has also been used for generating bright DKSs in both a $MgF_2$ crystalline resonator and a $Si_3N_4$ MRR [147]. Up to a ~10-fold reduction in the required average pump power was achieved by pumping these resonators with picosecond optical pulses generated by an injection-locked III-V gain-switched laser (GSL) (**Figure 7(f)**). By using picosecond pulses to drive a dispersion-engineered $Si_3N_4$ MRR, successful generation of soliton-based resonant supercontinuum combs with over 2200 comb lines at a repetition rate of ~28 GHz has also been realized (**Figure 7(g)**) [199], establishing a critical connection between supercontinuum generation and microcomb generation.

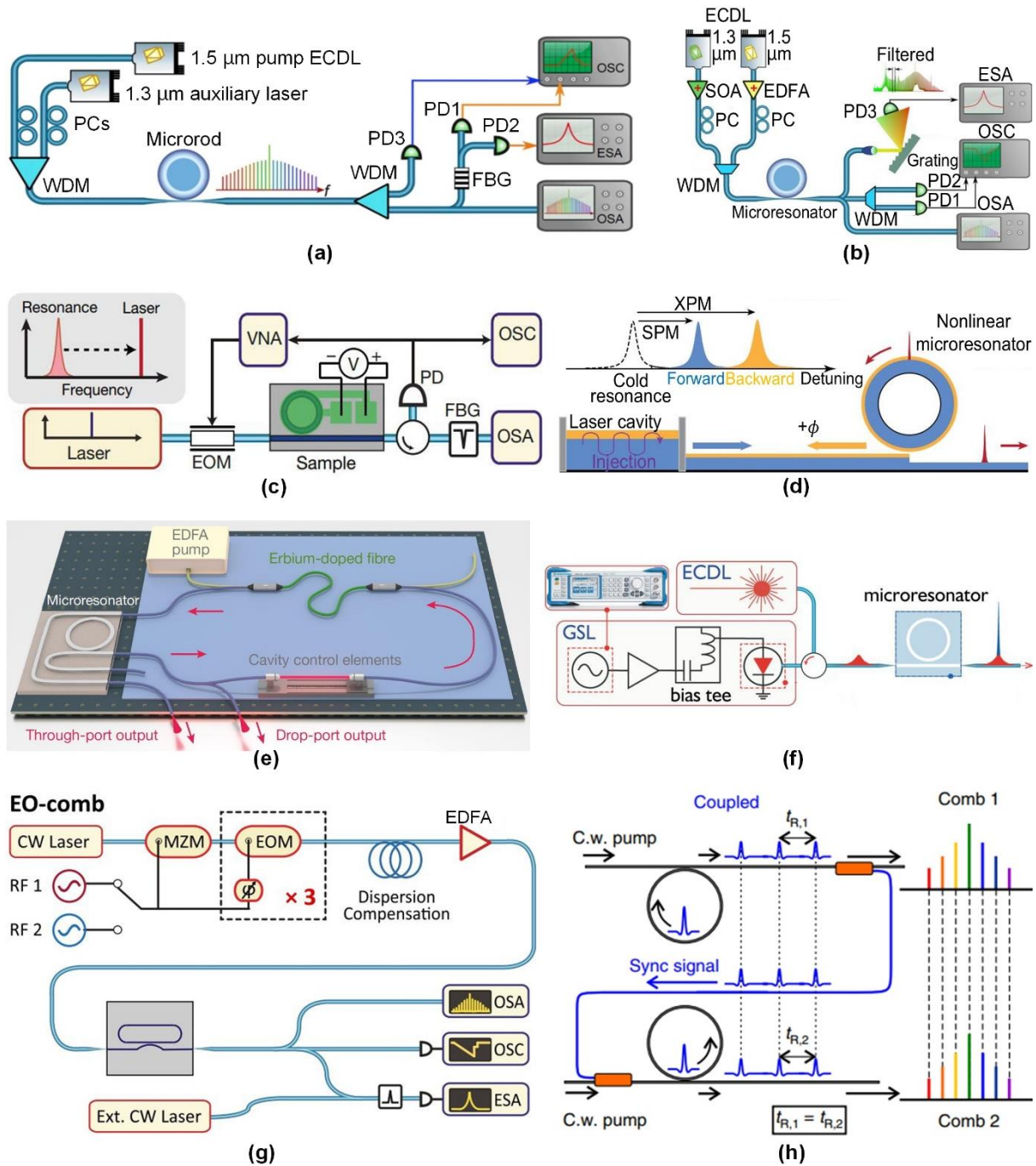

**Figure 7**. Recent progress in microcomb new driving mechanisms. (a) Access range extension of soliton microcombs generated by a silica microrod resonator by using an auxiliary laser. (b) Spectral bandwidth broadening of soliton microcombs generated by a silica microtoroid resonator via bichromatic pumping. (c) Piezoelectric-controlled DKS generation in a $Si_3N_4$ MRR. (d) Turnkey soliton microcomb generation in a $Si_3N_4$ MRR. (e) Self-emergence robust soliton microcomb generation in a Hydex MRR. (f) DKS generation in both a $Si_3N_4$ MRR and a $MgF_2$ crystalline resonator pumped by optical pulses generated from a gain-switched laser (GSL). (g) Soliton-based resonant supercontinuum generation in a $Si_3N_4$ microresonator pumped by optical pulses. (h) Synchronization of two soliton microcombs in two $Si_3N_4$ MRRs. (a) Reprinted with permission from [*Optica.*, 6, 206 (2019)].[145] (b) Reprinted with permission from [*Nat. Commun.*, 11, 6384 (2020)].[141] (c) Reprinted with permission from [*Nature.*, 583, 385 (2020)].[198] (d) Reprinted with permission from [*Nature.*, 582, 365 (2020)].[110] (e) Reprinted with permission from [*Nature.*, 608, 303 (2022)].[192] (f) Reprinted with permission from [*Nat. Commun.*, 12, 9 (2021)].[147] (g) Reprinted with permission from [*Optica.*, 8, 771 (2021)].[199] (h) Reprinted with permission from [*Nat. Photonics.*, 12, 688 (2018)].[197]

In addition to driving a single microresonator for optical microcomb generation, synchronization of two soliton microcombs generated by different MRRs has also been demonstrated (**Figure 7(h)**) [197], where synchronization enabled by the coupling between two $Si_3N_4$ MRRs on different chips was established by transmitting a fraction of one's output to the other's input via an optical fiber with a length of ~20 m. The fundamental power limit of microresonator-based combs was also overcome by coherent combination of the synchronized microcombs.

## 3. Microwave photonics based on optical microcombs

Microwave photonics, which utilizes photonic technologies to process microwave signals and realize functions that are difficult to achieve in conventional microwave systems, has a long history dating back to the 1970s [204]. The rise of microwave photonics has been paralleled by advances in fiber optics, and so experienced significant progress during the fiber optic boom in the late 1990s, greatly benefiting from the growth of integrated photonics after the millennium. For the most demanding applications of microwave photonics, soliton microcombs display a very high degree of coherence, and so can be used for synthesizing high-quality microwave frequencies after photodetection. In addition, the multiple comb lines of optical microcombs can be used as discrete taps for the microwave transversal filter systems, which enable many spectral filtering and temporal signal processing applications that are not particularly demanding in terms of comb coherence. Since the previous review of microwave photonic applications of optical microcombs [83], there have been rapid advances in this field, with many new breakthroughs in research and practical applications. In this section, we review and highlight the recent progress in this field, which is categorized into microwave frequency synthesizers, microwave photonic filters, and microwave photonic signal processors.

## 3.1 Frequency synthesizers

The synthesis of microwave or optical signals with accurate control of their frequencies is a fundamental requirement for microwave photonic applications. Optical microcombs provide a new solution to realize broadband frequency synthesizers on a chip scale. Photodetection of mode-locked microcombs to obtain microwave beat notes corresponding to the microcombs' repetition rates provides a method to directly generate microwave signals [26]. Another widely used method for microwave frequency synthesis is optical-to-microwave frequency division, which uses a microcomb locking to a stable reference light source to divide optical frequency down to microwave frequency [205]. The reverse process of the optical-to-microwave frequency division enables up-conversion of microwave frequency to optical frequency, which can be used for optical frequency synthesis [21].

Generally, frequency stability and spectral purity are two basic performance characteristics of a frequency synthesizer. Allan deviation has been widely used for the quantitative analysis of the frequency stability [21, 22, 24, 26-28, 144, 191]. **Figure 8(a)** compares the Allan deviations of frequency synthesizers based on optical microcombs as well as LFCs generated by solid-state lasers and mode-locked fiber lasers. As can be seen, although the frequency synthesizers based on optical microcombs show a relatively low stability (*i.e.*, high values of Allan deviations) for the synthesized microwave signals, they are capable of generating signals at high frequencies above 10 GHz [24, 27, 28]. On the other hand, the spectral purity of the generated signals is affected by a number of factors such as relative-intensity noise of pump lasers, shot noise, Q factors of microresonators, thermal fluctuations, quantum vacuum fluctuations, pump-resonance detunings, and microcomb driving mechanisms [11, 191], and is normally characterized by the single-sideband (SSB) phase noise [191]. **Figure 8(b)** compares the SSB phase noise at different offset frequencies (1 kHz, 10 kHz, and 10 MHz) of frequency synthesizers based on optical microcombs as well as LFCs generated by solid-state lasers and mode-locked fiber lasers. As can be seen, the frequency synthesizers based on optical microcombs have already achieved high spectral purities that are comparable with their bulky counterparts.

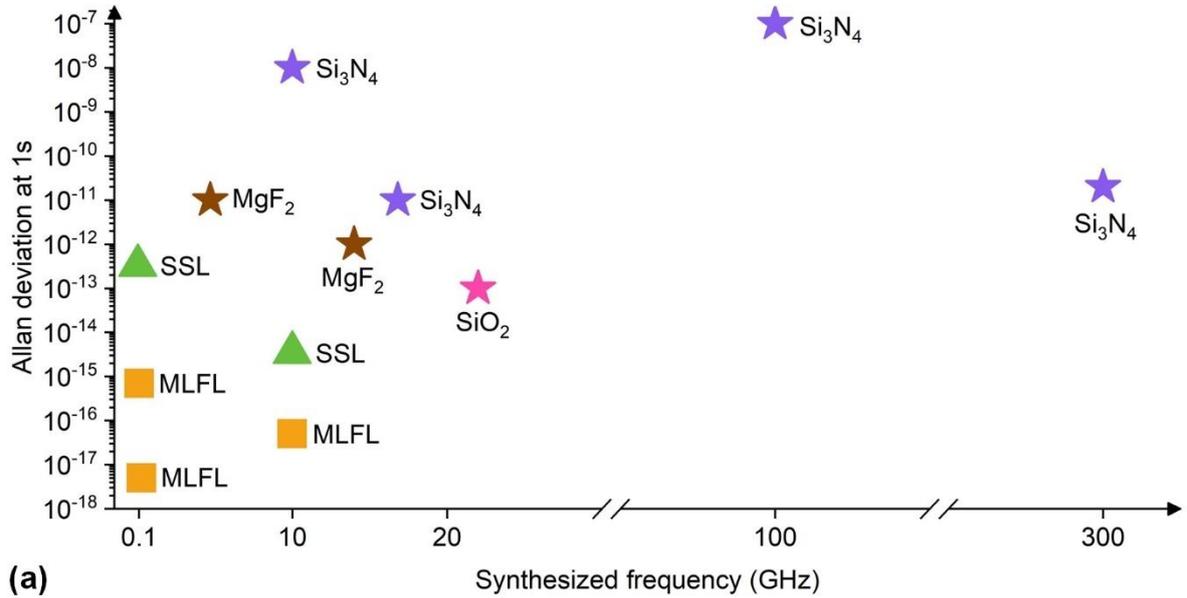

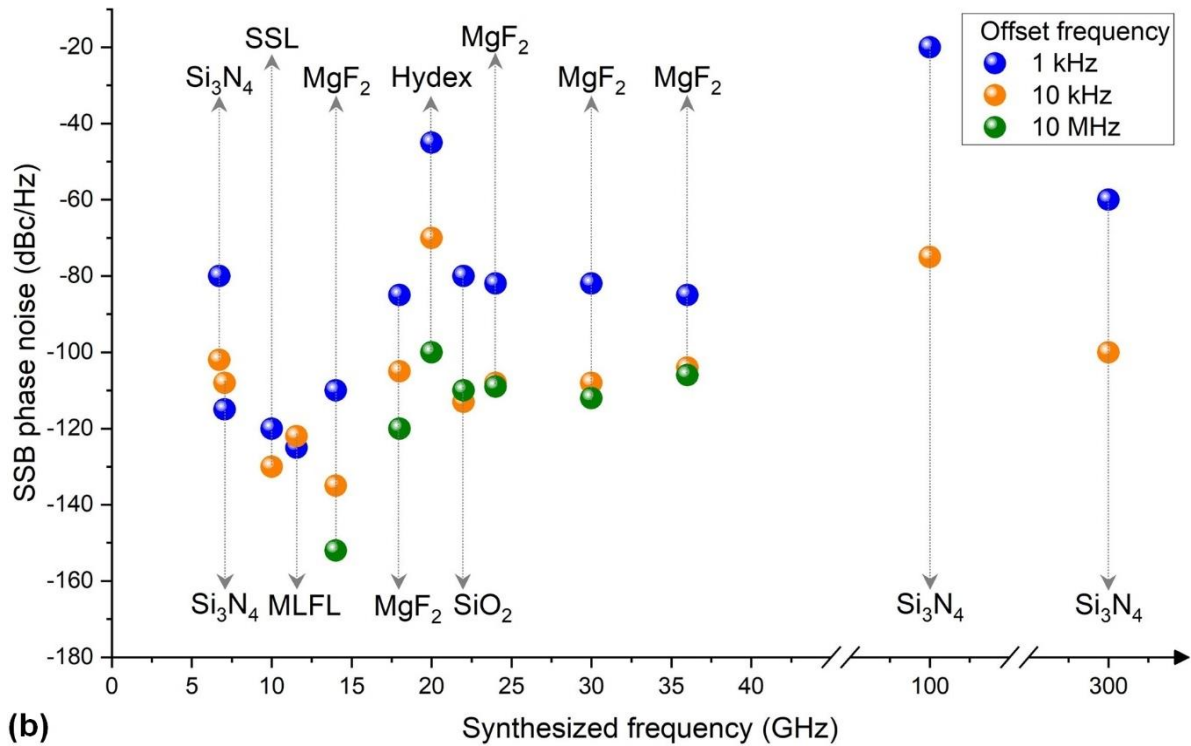

**Figure 8**. (a) Allan deviations (at an integration time of 1 s) and (b) single-sideband (SSB) phase noise (at 1-kHz, 10-kHz, and 10-MHz offset frequencies) versus synthesized frequencies for frequency synthesizers based on different microcombs as well as LFCs generated by solid-state lasers (SSLs) and mode-locked fiber lasers (MLFLs). Values of Allan deviations and SSB phase noise are taken from Ref. [24] for the 16.82-GHz, 6.73-GHz, and 7.05-GHz signals synthesized based on microcombs generated by a $Si_3N_4$ MRR, Ref. [27] for the 100-GHz signal synthesized by microcombs generated by a $Si_3N_4$ MRR, and Ref. [28] for the 300-GHz signal synthesized by microcombs generated by a $Si_3N_4$ MRR. Values of Allan deviations are taken from Ref. [24] for the 4.7-GHz

signal synthesized based on microcombs generated by a MgF$_2$ whispering-gallery-mode (WGM) resonator, Ref. [16] for the 0.01-GHz signal synthesized based on LFCs generated by a solid-state laser, Ref. [206] for the 10-GHz signal synthesized based on LFCs generated by two SSLs, Ref. [207] for the 10-GHz signal synthesized based on LFCs generated by two MLFLs, Ref. [26] for the 10-GHz and 20-GHz signals synthesized based on microcombs generated by two Si$_3$N$_4$ MRRs, Ref. [22] for the 14-GHz signal synthesized based on microcombs generated by a MgF$_2$ WGM resonator, Ref. [144] for the 22-GHz signal synthesized by microcombs generated by a SiO$_2$ wedge resonator, values are taken from Thorlabs, Inc. for the 0.1-GHz [208] and the 0.25-GHz [209] signals synthesized by commercial LFCs generated by mode-locked fiber lasers. Values of SSB phase noise are taken from Ref. [210] for the 11.55-GHz signal synthesized based on LFCs generated by two MLFLs, Ref. [23] for the 14-GHz signal synthesized based on microcombs generated by a MgF$_2$ WGM resonator, Ref. [211] for 18-GHz, 24-GHz, 30-GHz, and 36-GHz signal synthesized based on microcombs generated by a MgF$_2$ WGM resonator, Ref. [25] for 20-GHz signal synthesized based on microcombs generated by a Hydex MRR, Ref. [20] for 22-GHz signal synthesized based on microcombs generated by a SiO$_2$ microdisk resonator.

Theoretically, the phase noise floor after photodetection can be given by [191]

$$L = \frac{q\,Z\,R\,P_{DC}}{P_{RF}} \tag{5}$$

where $q$ is the elementary charge, $Z$ is the resistance of the PD load, $R$ is the responsivity of the PD, $P_{DC}$ is the DC optical power reaching the PD, and $P_{RF}$ is the power of the generated radio frequency (RF) signal. The SSB phase noise at different offset frequencies is affected by different factors. For the offset frequency below 1 kHz, SSB phase noise is mainly induced by microresonators' frequency fluctuations [11]. For the offset frequency above 10 MHz, the SSB phase noise is dominated by the shot noise, which can be reduced by employing a narrow-band RF filter after the PD [191]. The SSB phase noise at offset frequencies between 1 kHz and 10 MHz is mainly due to the transfer of laser relative-intensity noise (RIN) to microwave phase modulation by microcomb dynamics [191]. The phase noise induced by the RIN of the pump laser can be expressed as [23, 92]

$$S(f) = \left( \frac{D_1}{2\pi} \frac{4\eta c n_2}{\kappa V_{eff} n_0^2} \frac{1}{f} P_{in} \right)^2 S_{RIN}(f) \tag{6}$$

where $D_1/2\pi$ is the FSR, $\kappa/2\pi$ is the cavity energy decay rate, $\eta$ is the coupling impedance of the resonator, $c$ is the speed of light, $V_{eff}$ is the mode volume, $n_2$ is the Kerr nonlinearity, $n_0$ is the refractive index, $f$ is the pump frequency, $P_{in}$ is the pump power, and $S_{RIN}(f)$ is the RIN of the pump laser.

Recent advances in microcomb-based frequency synthesizers have shown significant increase in performance, with the frequency range extending from the microwave to the optical domain. **Table 4** summarizes the performance metrics of state-of-the-art frequency synthesizers based on optical microcombs.

Note that for different works some of the performance parameters (*e.g.*, phase noise and Allan deviation) cannot be directly compared, since their experimental configurations are very different, with significant associated differences in complexity, volume, and cost.

A key challenge has been to reduce the frequency spacing, or FSR, of the microcombs into mainstream microwave bands, well below the 100 – 600 GHz range where they were first demonstrated [11]. A recent breakthrough achieved an FSR of ~ 49 GHz [212], which enabled microwave frequency conversion from the L band (1–2 GHz) to the U band (40–60 GHz) [25], achieving an SSB phase noise of −90 dBc/Hz at $10^4$ Hz. Even lower FSRs have since been reported, including a microwave frequency synthesizer generating signals in the X (8–12 GHz) and K (18–27 GHz) bands (**Figure 9(a)**) [26], produced by photodetection of DKS states formed in $Si_3N_4$ MRRs with centimeter-scale circumferences and an ultralow propagation loss of ~1.4 dB/m. The relative Allan deviation was on the order of $10^{-8}$ at an integration time of 1s, together with a low SSB phase noise of ~-110 dBc/Hz at $10^4$ Hz.

A 100-GHz microwave frequency synthesizer with improved signal-to-noise ratio as well as microwave signal linewidth has been demonstrated (**Figure 9(b)**) [27], The constructive interference among the RF beat notes generated by a mode-locked soliton microcombs from a $Si_3N_4$ MRR was exploited to achieve a 5.8-dB improvement of the microwave signal power compared to that using two-laser heterodyne detection. A 100-fold reduction in the microwave signal linewidth over the pump laser was also achieved after the transfer of the pump frequency variation to the comb repetition rate.

**Table 4. Performance metrics of microcomb-based frequency synthesizers. WGM: whispering gallery mode.**

| Resonator for generating microcombs | Frequency range | Phase noise (dBc/Hz) [a] | Allan deviation [b] | Year | Refs. |
|---|---|---|---|---|---|
| SiO$_2$ microdisk resonator | 2.6 – 220 GHz | -113 at $10^4$ Hz | – | 2012 | [20] |
| MgF$_2$ WGM resonator | ~9.9 GHz | -170 at $10^7$ Hz | ~$10^{-11}$ | 2015 | [191] |
| MgF$_2$ WGM resonator | 6 GHz, 12 GHz, 18 GHz, 24 GHz, 30 GHz, and 36 GHz | <-100 at $10^4$ Hz | – | 2016 | [211] |
| Si$_3$N$_4$ MRR and SiO$_2$ microdisk resonator | 4 THz around 1550 nm | – | $9.2 \times 10^{-14}$ | 2018 | [21] |
| MgF$_2$ crystalline resonator | ~14 GHz | -130 at $10^4$ Hz | ~$10^{-12}$ | 2019 | [22] |
| MgF$_2$ crystalline resonator | ~14 GHz | <-135 at $10^4$ Hz | – | 2020 | [23] |
| MgF$_2$ crystalline resonator  Si$_3$N$_4$ MRR | 4.7 GHz, 7.05 GHz, 6.73 GHz, 10.09 GHz, 16.82 GHz | <-115 at $10^3$ Hz  <-60 at $10^3$ Hz | ~$10^{-11}$ | 2020 | [24] |
| Si$_3$N$_4$ MRRs | ~10 GHz  ~20 GHz | -110 at $10^4$ Hz | ~$10^{-8}$ | 2020 | [26] |
| Hydex MRR | 8.9 – 25.9 GHz | -70 at $10^4$ Hz | – | 2020 | [25] |
| Si$_3$N$_4$ MRR | ~100 GHz | <-60 at $10^4$ Hz | ~$10^{-7}$ | 2021 | [27] |
| Si$_3$N$_4$ MRR | ~300 GHz | -100 at $10^4$ Hz | $2 \times 10^{-11}$ | 2021 | [28] |
| SiO$_2$ wedge resonator | ~22 GHz | -110 at $10^3$ Hz  -88 at $10^2$ Hz | ~$10^{-13}$ | 2022 | [144] |
| SiO$_2$ microrod resonator | 11.4 GHz | -107 at $10^3$ Hz  -133 at $10^4$ Hz | – | 2022 | [146] |

[a] Here we show the single sideband (SSB) phase noise at a given frequency offset from the carrier.
[b] Here we show the Allan deviation at an integration time of 1s.

Optical-to-microwave frequency division via LFCs has proven to be an effective way to generate microwave signals with extremely high purity and frequency stability [207, 213]. Recently, a ~300-GHz frequency synthesizer has been demonstrated based on frequency division using a soliton microcomb [28], where photomixing between a ~3.6-THz optically carried reference generated by stimulated Brillouin scattering in a fiber ring cavity and a ~300-GHz soliton microcomb generated by a Si$_3$N$_4$ MRR was achieved in a broadband uni-travelling-carrier photodiode, achieving a phase noise of ~-100 dBc/Hz at $10^4$ Hz.

Another microwave frequency synthesizer based on optical-to-microwave frequency division has been demonstrated using a soliton microcomb with a repetition rate of ~14 GHz generated by a $MgF_2$ crystalline resonator (**Figure 9(c)**) [23]. The frequency division was achieved by using a transfer oscillator method based on electronic division and mixing, yielding a low phase noise below ~-135 dBc/Hz at $10^4$ Hz for the generated ~14-GHz microwave signal.

Based on a new frequency division method that transfers the spectral purity of a cavity soliton oscillator into the subharmonic frequencies of the microcomb's repetition rate, a low-noise microwave frequency synthesizer has been demonstrated (**Figure 9(d)**) [24], where a soliton microcomb was injected into a GSL driven by a sinusoidal current with a frequency equaling to the subharmonic frequency of the microcomb's repetition rate, leading to the generation of a dense GSL comb dividing the microcomb's line spacing. Successful microwave signal generation was demonstrated by dividing the repetition rates of soliton microcombs generated by a ~14.09-GHz-FSR $MgF_2$ crystalline resonator and a ~100.93-GHz-FSR $Si_3N_4$ MRR.

A microwave frequency synthesizer with reduced phase noise has been demonstrated by creating intracavity potential gradient via injection locking to trap soliton microcombs and control their repetition rate [22], achieving a low SSB phase noise of ~-130 dBc/Hz at $10^4$ Hz for ~14-GHz microwave signals generated by a $MgF_2$ crystalline resonator.

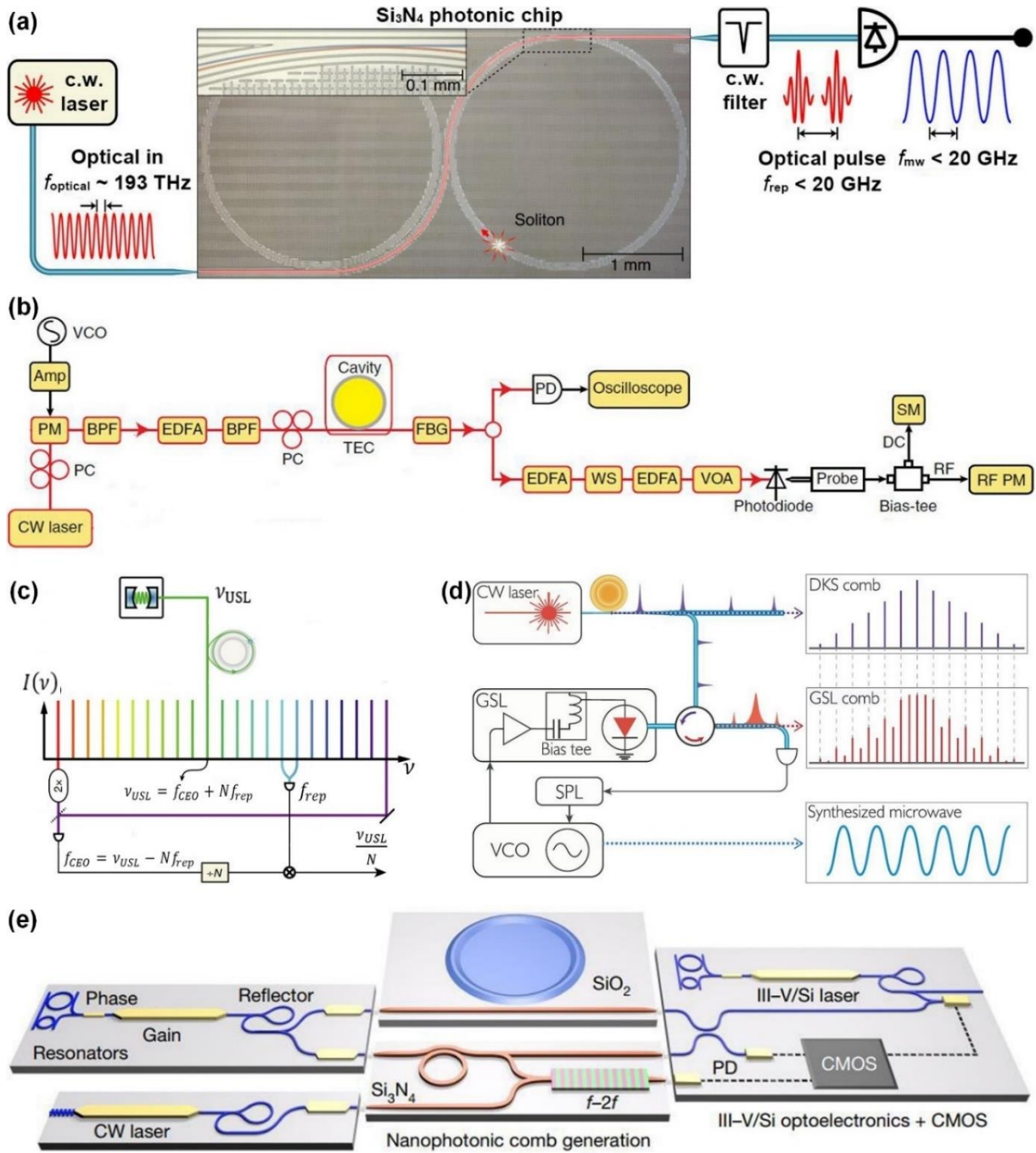

**Figure 9.** Recent progress in microcomb-based frequency synthesizers. (a) Microwave frequency synthesizers in the X and K bands based on soliton microcombs generated by $Si_3N_4$ MRRs. (b) A 100-GHz microwave frequency synthesizer based on soliton microcombs generated by a $Si_3N_4$ MRR. (c) A microwave frequency synthesizer using a transfer oscillator approach to divide the repetition rate of a soliton microcomb. (d) A microwave frequency synthesizer realized by injecting a soliton microcomb into a gain-switched laser (GSL). (e) An optical frequency synthesizer with *f*-2*f* stabilization based on two soliton microcombs generated by a $Si_3N_4$ MRR and a $SiO_2$ microdisk. (a) Reprinted with permission from [*Nat. Photonics.*, 14, 486 (2020)].[26] (b) Reprinted with permission from [*Light: Sci. Appl.*, 10, 4 (2021)].[27] (c) Reprinted with permission from [*Nat. Commun.*, 11, 374 (2020)].[23] (d) Reprinted with permission from [*Sci. Adv.*, 6, 6 (2020)].[24] (e) Reprinted with permission from [*Nature.*, 557, 81 (2018)].[21]

Recently, a palm-sized microwave frequency synthesizer has been reported [144], where a fiber delay line spool with a centimeter-scale diameter was employed as a time reference to stabilize the repetition rate of a microcomb generated by a silica wedge resonator. The generation of ~22-GHz microwave signals was experimentally demonstrated, achieving a phase noise of ~-110 dBc/Hz at $10^3$ Hz and ~-88 dBc/Hz at $10^2$ Hz. The measured Allan deviation was on the level of $10^{-13}$ at 1 s. By employing a voltage-controlled oscillator and an acousto-optic frequency shifter, a tunable repetition rate over 100 kHz bandwidth was achieved.

A low-noise microwave frequency synthesizer has been realized through photodetection of soliton microcombs generated by a ~6-mm-diameter silica microrod resonator with a spherical cross section [146]. High-purity fiber preform was lathe machined to form the large-mode-volume microrod resonator with a Q factor exceeding $4 \times 10^9$, enabling an extremely low power of ~110 µW for comb initiation. The synthesized microwave signal in the X band (~11.4 GHz) had a phase noise of ~-107 dBc/Hz and ~-133 dBc/Hz at $10^3$ Hz and $10^4$ Hz, respectively.

In addition to microwave frequency synthesizers, an optical frequency synthesizer has also been demonstrated (**Figure 9(e)**) [21], where two soliton combs generated by a $Si_3N_4$ MRR and a $SiO_2$ microdisk created a phase-coherent link between microwave and optical frequencies, enabling programmable optical frequency output with an ultrahigh resolution of ~1 Hz across 4 THz band around 1550 nm. The output optical frequency was highly stable, with a synthesis error $< 7.7 \times 10^{-15}$. The measured Allan deviation was $(9.2 \pm 1.4) \times 10^{-14}$ at 1 s – about two orders of magnitude lower than previous microcomb-based frequency synthesizers [191].

**3.2 Microwave photonic filters**

Microwave photonic filters are indispensable components in microwave photonic systems. Microcomb-based microwave photonic filters are typically implemented in the transversal filter structure based on discrete-time processing of microwave signals [83], where the microcomb serves as a multi-wavelength source to provide discrete-time channels or taps. The spectral transfer function of the transversal filter can be engineered via the design of the tap coefficients for different channels, thus allowing for highly reconfigurable filter shapes. In the following, we provide quantitative analysis and review of recent advances of microcomb-based microwave photonic filters.

For a microwave filter based on discrete-time processing, the system function can be given by [214]

$$H\left(z^{-1}\right) = \frac{\sum_{n=0}^{R} a_n z^{-n}}{1 - \sum_{k=1}^{M} b_k z^{-k}} \quad (7)$$

where $z^{-1}$ is the basic delay between the sampled signals, and $a_n$ and $b_k$ are the filter coefficients. In **Eq. (7)**, the numerator and the denominator correspond to the finite and infinite impulse parts, respectively, with $R$ and $M$ denoting their orders. When $b_k = 0$ for all $k$ (=1, 2, 3, … , $M$), the filter is a transversal filter, and the corresponding impulse response can be expressed as [214]

$$h(t) = \sum_{n=0}^{R} a_n \delta(t - n\Delta t) \quad (8)$$

where $\Delta t$ is the time delay between two adjacent taps. The time difference between different channels is a featured characteristic of transversal filters, and they are also termed "delay-line filters" in many literatures [83, 212, 215]. The progressive time delays can be engineered to implement true time delay lines [216], which are fundamental building blocks in phased array antennas [212, 217] and microwave beamformers [218, 219]. After Fourier transformation from **Eq. (8)**, the spectral transfer function of a transversal filter can be given by [214]

$$H(\omega) = \sum_{n=0}^{N-1} a_n e^{-j\omega n \Delta t} \quad (9)$$

where $N = R + 1$ is the tap number. Due to the discrete-time processing nature of the transversal filter, the filter shape corresponding to **Eq. (9)** will appear periodically in the RF transmission spectrum with an FSR given by

$$FSR_{RF} = 1/\Delta t \quad (10)$$

**Eq. (9)** forms the basis for the implementation of microcomb-based microwave photonic filters with arbitrary filter shapes through designing the tap coefficients $a_n$ ($n$ = 0, 1, 2, 3, …, $N$−1) for different wavelength channels. For example, a low-pass (sinc) filter can be realized by shaping microcombs to have the same tap coefficient for each tap as below [30]

$$a_{sinc,n} = 1, \ n = 0, 1, 2, 3, …, N-1 \quad (11)$$

On the basis of the low-pass filter in **Eq. (11)**, Gaussian apodization can be applied to improve the main-to-secondary sidelobe ratio (MSSR), where the tap coefficients $a_{sinc,\,n}$ ($n$ = 0, 1, 2, 3, …, $N$−1) need to be modified to the product of itself and a Gaussian function as below [30]

$$a_{gau,n} = a_{sinc,n} \cdot e^{-\frac{(n-b)^2}{2\sigma^2}}, \ n = 0, 1, 2, 3, …, N-1 \quad (12)$$

where $\sigma$ and $b$ are the root mean square width and the peak position of the Gaussian function, respectively. A band-pass filter can be realized by multiplying $a_{gau,\,n}$ in **Eq. (12)** by a sinc function as below [30]

$$a_{BPF,n} = a_{gau,n} \cdot \cos\frac{f_{center} \pi n}{FSR_{RF}}, \ n = 0, 1, 2, 3, …, N-1 \quad (13)$$

where $f_{center}$ is the center frequency of the band-pass filter. In principle, for $N \rightarrow \infty$ in **Eq. (9)**, reconfigurable filters with arbitrary filter shapes can be realized by applying corresponding tap coefficients to each wavelength channel. For practical systems with limited numbers of available channels, negligible deviations between the practical and the ideal filter shapes can be achieved when $N$ is sufficiently large.

**Figure 10** provides a quantitative analysis for the influence of the number of comb lines (*i.e.*, tap number of the transversal filter system) on the performance of the microcomb-based microwave photonic filters, using low-pass and band-pass filters designed based on **Eqs. (11)** and **(13)** as examples. **Figure 10(a-i)** shows the simulated transmission spectra of the low-pass filters with the same cut-off frequency of 1 GHz but different tap numbers ranging from 10 to 80. The filtering resolution (*i.e.*, 3-dB bandwidth) and the MSSR extracted from **Figure 10(a-** **i)** are shown in **Figures 10(a-ii)** and **(a-iii)**, respectively. **Figure 10(b)** shows the corresponding results for the band-pass filters with the same center frequency of 10 GHz but different tap numbers. As can be seen, both the filtering resolution and the MSSR of the low-pass and band-pass filters can be improved by increasing the tap number. This reflects the advantage of implementing transversal filter systems based on optical microcombs, which can simultaneously provide a large number of discrete taps using a compact device in chip scale. Another important advantage for the transversal filter systems is their high degree of reconfigurability. By changing the tap coefficients, a diverse range of spectral responses can be realized based on the same system. This is very challenging to realize for microwave photonic filters based on passive optical filters [220, 221].

In one of the first microwave applications of low FSR (< 100GHz) microcombs, an 80-tap microwave photonic transversal filter was demonstrated based on soliton crystal microcombs generated by a 49-GHz-FSR Hydex MRR (**Figure 11(a)**) [30]. The increased number of taps, – about 4 times that of previous work based on a similar structure [217], yielded a filtering resolution that was 4 times higher. Many different filter shapes, together with tunable center frequency and bandwidth, have also been demonstrated.

To improve the frequency resolution of the microcomb-based transversal filter systems, bandwidth scaling was introduced [31], where another MRR was used as a passive Vernier comb filter to slice and sample the shaped comb before photodetection, yielding a high resolution of ~117 MHz and a large microwave instantaneous bandwidth of ~4.64 GHz (**Figure 11(b)**).

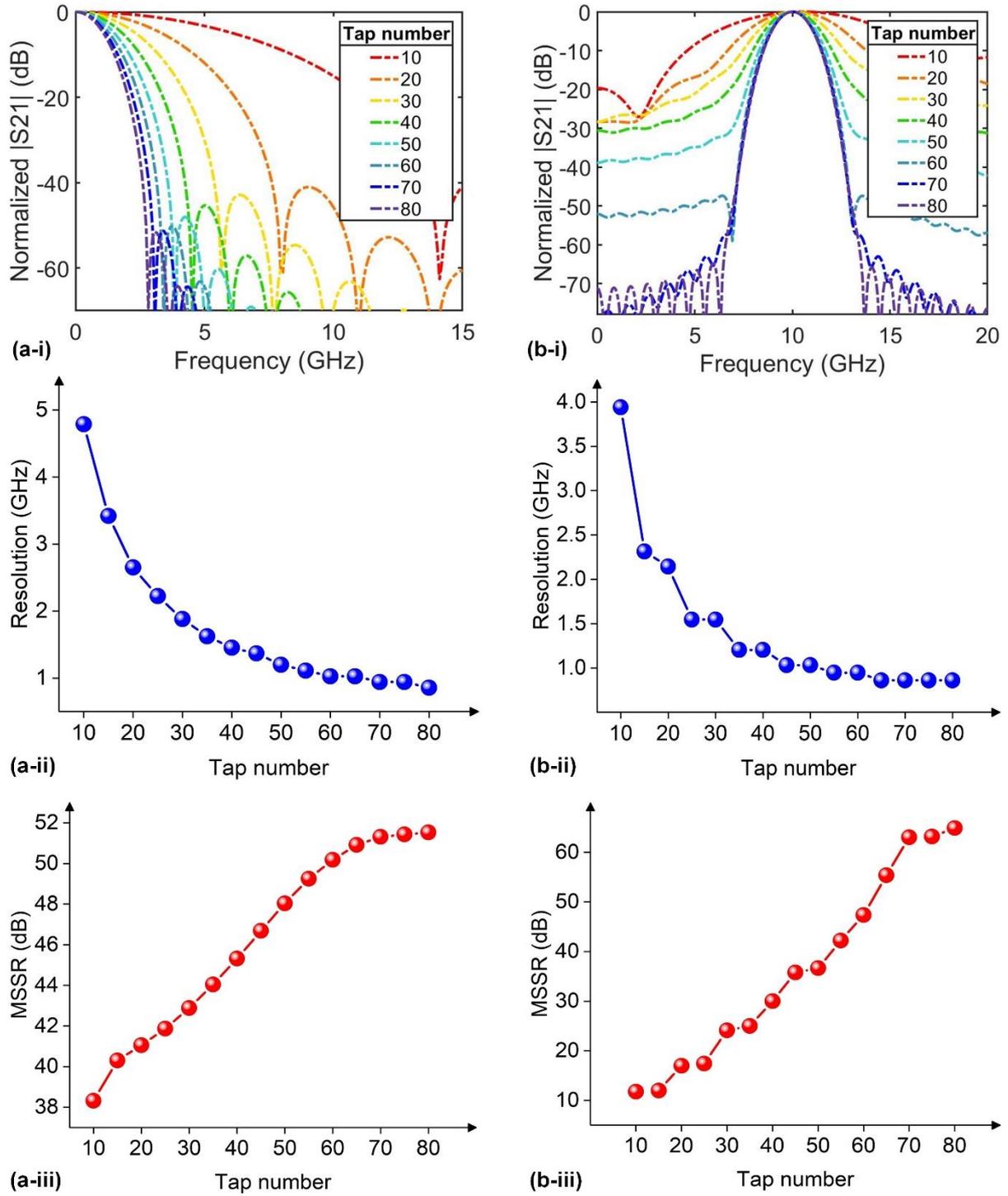

**Figure 10.** Influence of tap number on the performance of microcomb-based microwave photonic filters. (a) Low-pass filters with the same cut-off frequency of 1 GHz but different tap numbers ranging from 10 to 80. (b) Band-pass filters with the same center frequency of 10 GHz but different tap numbers ranging from 10 to 80. In (a) and (b), (i), (ii), and (iii) show the simulated transmission spectra, the filtering resolutions (*i.e.*, 3 dB bandwidth) versus tap numbers, and the main-to-secondary sidelobe ratios (MSSR) versus tap numbers, respectively.

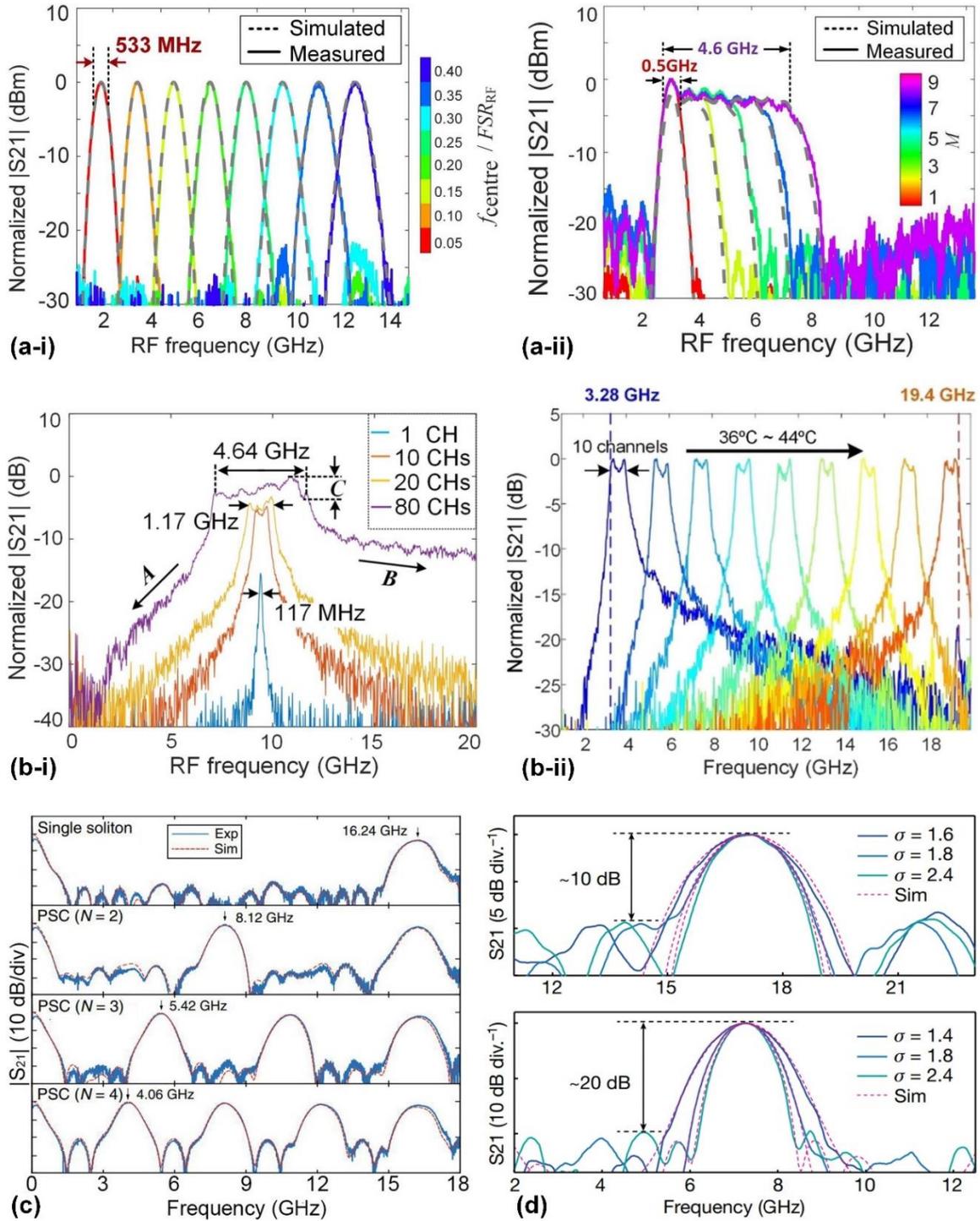

**Figure 11**. Recent progress in microcomb-based microwave photonic filters. (a) An 80-tap microwave photonic transversal filter based on soliton crystal microcombs generated by a Hydex MRR. (i) and (ii) show the measured spectral responses of bandpass filters with a tunable center frequency and a tunable bandwidth, respectively. (b) A microwave photonic transversal filter based on bandwidth scaling. (i) and (ii) show the measured spectral responses of a bandpass filter with various tap numbers and a bandpass filter with tunable center frequency, respectively. (c) Reconfigurable microwave photonic filters without external pulse shaping based on rich soliton states formed in a $Si_3N_4$ MRR. (d) A reconfigurable microwave photonic filter using soliton microcombs generated by an AlGaAs-on-insulator source chip to drive a silicon-in-insulator (SOI) shaping chip. (a) Reprinted with permission from [*J. Lightwave Technol.*, 37, 1288 (2019)].[30] (b) Reprinted with permission from [*APL Photonics.*, 4, 8 (2019)].[31] (c) Reprinted with permission from [*Nat. Commun.*, 11, 4377 (2020)].[32] (d) Reprinted with permission from [*Nature.*, 605, 457 (2022)].[33]

Reconfigurable microwave photonic filters without the need for external pulse shaping have been demonstrated (**Figure 11(c)**) [32], where the filter shape was reconfigured all-optically by exploiting the rich family of soliton states formed in a $Si_3N_4$ MRR. Perfect soliton crystals were triggered at different resonances to multiply the comb spacing, thus dividing the center frequency of bandpass filters. A bandpass filter with tunable center frequency from ~0.85 GHz to ~7.51 GHz was also demonstrated based on spectral interference between two solitons in the same cavity.

Recently, a microcomb-based microwave photonic filter with a significantly improved degree of integration has also been demonstrated (**Figure 11(d)**) [33], where soliton microcombs generated by a heterogeneously integrated source chip were employed to drive a silicon-on-insulator (SOI) shaping chip. The source chip, which contained an InP distributed feedback (DFB) laser and an AlGaAs MRR, generated turnkey dark soliton microcombs with a spacing of ~180 GHz. The SOI shaping chip was composed of a Mach-Zehnder modulator, an MRR shaping array, and spiral delay lines. Reconfigurable bandpass filtering with tunable 3-dB bandwidth from ~1.97 GHz to ~2.42 GHz was demonstrated. The filtering performances of using on-chip and off-chip delay lines were also compared.

**3.3 Microwave photonic signal processors**

Microcomb-based transversal filter systems can be used not only for spectral filtering but also for temporal signal processing. Different temporal signal processing functions can be realized by engineering the discrete impulse response of the transversal filter system, *i.e.*, the inverse Fourier transform of the spectral transfer function in **Eq. (9)**. Although implemented with photonic hardware, the transversal filter systems essentially perform digital processing, which can not only maintain the high processing speed of optical processing but also achieve improved processing accuracy compared to optical analog processing [83, 222, 223].

In **Table 5**, we summarize and compare the different functions that have been demonstrated using microcomb-based microwave photonic signal processors. On the basis of previously demonstrated Hilbert transform [34] and temporal differentiation [224], many new signal processing functions have been demonstrated after 2018, including fractional-order differentiation [37], fractional-order Hilbert transform [35], integration [38], phase encoding [40], and arbitrary waveform generation [39].

**Table 5. Performance comparison of microcomb-based microwave photonic signal processors. CS: comb spacing. WC: wavelength channel. RMSE: root-mean-square error.**

| Functions | CS (GHz) | No. of WC | Operation bandwidth or processing speed | RMSE | Year | Refs. |
|---|---|---|---|---|---|---|
| Integral-order Hilbert transform | ~200 | 20 | 0.3 − 16.9 GHz | – | 2015 | [34] |
| Integral-order differentiation | ~200 | 8 | ~8.5 GHz | ~7.24% | 2017 | [224] |
| Square root differentiation | ~200 | 7 | ~8.5 GHz | ~4.02% | 2018 | [83] |
| Fractional-order Hilbert transform | ~49 | 17 | 0.48 − 16.45 GHz | ~2.92% | 2019 | [35] |
| Fractional-order differentiation | ~49 | 27 | ~15.49 GHz | – | 2020 | [37] |
| Integration | ~49 | 81 | ~11.9 GHz | – | 2020 | [38] |
| Phase-encoding | ~49 | 60 | ~6 Gbit/s [d] | – | 2020 | [40] |
| Arbitrary waveform generation | ~49 | 81 | ~5.6 GHz | – | 2020 | [39] |
| Arbitrary waveform generation | ~100 [a] | 21 [b] | < 4 GHz [c] | – | 2022 | [225] |

[a] This value is taken from Fig. 4(a) in Ref. [225].
[b] This value is taken from Figs. 2 and 3 in Ref. [225].
[c] The demonstrated operation bandwidth was limited by the oscilloscope.

Differentiation and integration are basic signal processing functions, and more complex functions such as phase encoding and arbitrary waveform generation can be realized based on them. The spectral transfer function of $N$th-order temporal differentiation is given by [224]

$$H_{diff}(\omega) = (j\omega)^N \quad (14)$$

where $j = \sqrt{-1}$, $\omega$ is the angular frequency, and $N$ is the differentiation order that can be either an integer or a fraction. The spectral transfer function of $N$th-order temporal integration is the inverse of the $N$th-order differentiation, which can be given by [38]

$$H_{int}(\omega) = \left(\frac{1}{j\omega}\right)^N \tag{15}$$

Hilbert transform has been widely used for signal processing in radar systems, measurements, speech processing, and image processing [226]. The transfer function of a Hilbert transformer is given by [35]

$$H(\omega) = \begin{cases} e^{-j\phi}, & 0 \leq \omega < \pi \\ e^{j\phi}, & -\pi \leq \omega < 0 \end{cases} \tag{16}$$

where $\phi = P \times \pi/2$ is the phase shift, with $P$ denoting the order of the Hilbert transformer. When $P = 1$, **Eq. (16)** corresponds to a classical integral Hilbert transformer. According to **Eq. (16)**, a Hilbert transformer can be regarded as a phase shifter with phase shifts of $\pm\phi$ around a center frequency. The corresponding impulse response can be given by [34]

$$h(t) = \begin{cases} 1/\pi t, & t \neq 0 \\ \cot(\phi), & t = 0 \end{cases} \tag{17}$$

For microcomb-based transversal filter systems, differentiation, integration, and Hilbert transform can be realized by designing filters in **Eqs. (14) – (16)** according to **Eq. (9)**. In principle, the operation bandwidth of a microcomb-based transversal filter system is limited by the Nyquist zone, *i.e.*, half of the comb spacing [214]. Microcombs provide a large Nyquist zone typically on the order of 10's or 100's of GHz, which is well beyond the processing bandwidth of electronic devices and challenging to achieve for the LFCs generated by solid-state lasers and mode-locked fiber lasers. For practical systems, the operation bandwidth can be expressed as

$$OB = \frac{1}{\Delta t} = \frac{1}{\beta \cdot \Delta\lambda \cdot L} \tag{18}$$

where $\Delta t$ is the time delay between adjacent taps, $\Delta\lambda$ is the comb spacing, $\beta$ and $L$ are the dispersion coefficient and length of the dispersive medium (*e.g.*, single-mode fibers), respectively. Note that the *OB* in **Eq. (18)** is the same as the $FSR_{RF}$ in **Eq. (10)**. This is because both the spectral filter and the temporal signal processor are based on the same transversal filter system. According to **Eq. (18)**, a broader operation bandwidth can be achieved by decreasing the comb spacing or the length of the dispersive medium. It is also worth mentioning that such broadening of the operation bandwidth is not limited, and it is still subject to the limitation of the Nyquist zone.

For signal processors, the root-mean-square error (RMSE) is widely used to characterize the processing accuracy, which can be expressed as [227]

$$\text{RMSE} = \sqrt{\sum_{i=1}^{n}\frac{(Y_i - y_i)^2}{n}} \qquad (19)$$

where $Y_1, Y_2, \ldots, Y_n$ are the values of ideal signal processing result, $y_1, y_2, \ldots, y_n$ are values of practical signal processing result, and $n$ is the number of sampled points. **Figure 12** provides a quantitative analysis for the influence of the tap numbers on the RMSEs of microcomb-based microwave photonic signal processors, using integral and fractional differentiation as examples. Note that the RMSEs here are the best values that can be theoretically achieved. **Figure 12 (a-i)** and **(a-ii)** show the simulated amplitude and phase responses of first-order differentiators with different tap numbers ranging from 10 to 80. **Figure 12 (a-iii)** shows the corresponding output waveforms for the Gaussian input signal. The calculated RMSEs between the output waveforms and the ideal differentiation result are shown in **Figure 12 (a-iv)**. **Figure 12 (b)** shows the corresponding results for 0.5-order differentiators. As can be seen, the processing accuracy increases with tap number for both the integral and fractional differentiators. Similar to that shown in **Figure 10,** this reflects the fact that the increase of the tap number can also improve the performance of the transversal filter systems for temporal signal processing.

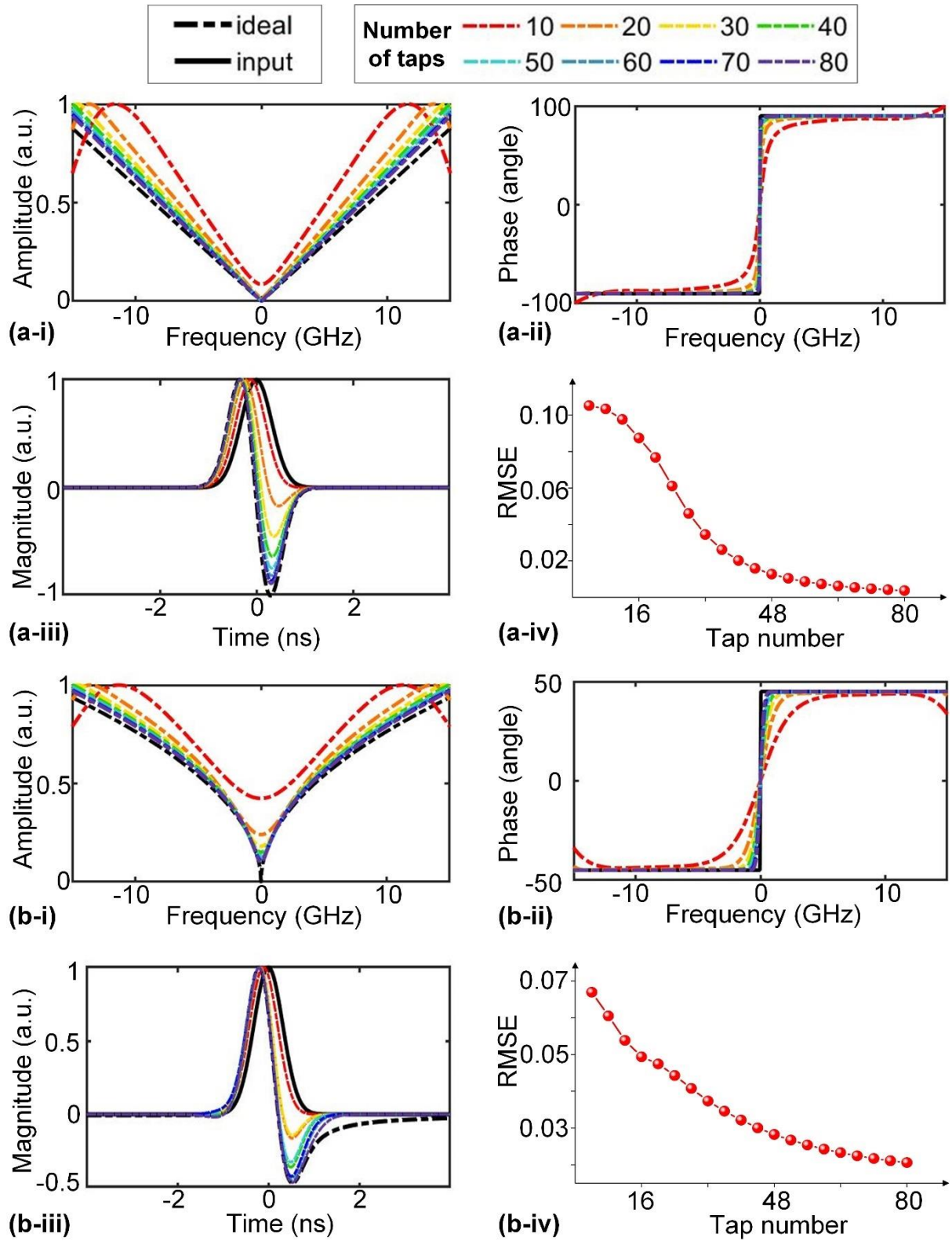

**Figure 12.** Influence of tap number on the performance of microcomb-based microwave photonic signal processors. (a) First-order differentiators with different tap numbers ranging from 10 to 80. (b) 0.5-order differentiators with different tap numbers ranging from 10 to 80. In (a) and (b), (i), (ii), (iii), and (iv) show the simulated amplitude responses, phase responses, the output waveforms for Gaussian input signal, and the root-mean-square errors (RMSEs) between the output waveforms and the ideal results versus tap numbers, respectively.

The high reconfigurability of the transversal filter system also enables the realization of versatile processing functions based on a single system – simply through comb shaping according to specific tap coefficients and without any changes of the hardware [34, 35, 37-40, 83, 224]. This is usually challenging for signal processing based on spatial light devices [228-231] and passive integrated photonic devices [232-235], making the microcomb-based microwave photonic signal processors attractive for practical applications with diverse processing requirements.

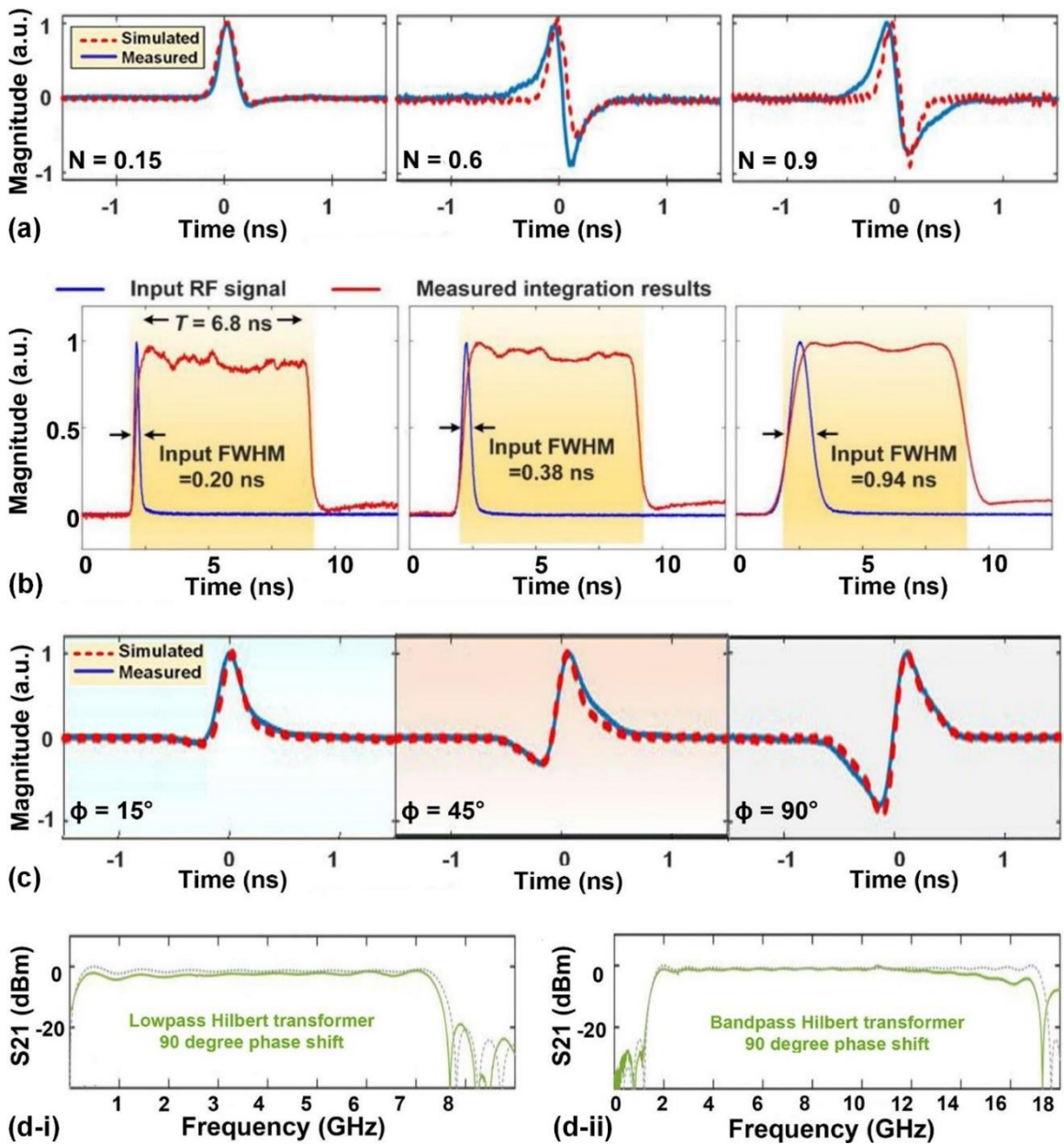

**Figure 13**. Recent progress in realizing new basic computing functions using microcomb-based microwave photonic signal processors. (a) Theoretical and measured output temporal waveforms of a fractional-order

differentiator. (b) Theoretical and measured output temporal waveforms of an integrator. (c) Theoretical and measured output temporal waveforms of a fractional-order Hilbert transformer. (d) Theoretical and measured spectral response of a fractional-order Hilbert transformer with tunable (i) bandwidth and (ii) center frequency. (a) Reprinted with permission from [*IEEE Trans. Circuits Syst. II-Express Briefs.*, 67, 2767 (2020)].[37] (b) Reprinted with permission from [*IEEE Trans. Circuits Syst. II-Express Briefs.*, 67, 3582 (2020)].[38] (c) Reprinted with permission from [*J. Lightwave Technol.*, 37, 6097 (2019)].[35] (d) Reprinted with permission from [*J. Lightwave Technol.*, 39, 7581 (2021)].[36]

A fractional-order differentiator with reconfigurable orders from 0.15 to 0.9 has been demonstrated (**Figure 13(a)**) [37], where 27 wavelength channels of soliton crystal microcombs generated by a ~49-GHz-FSR Hydex MRR were employed as discrete taps for the transversal filter system, resulting in an operation bandwidth of ~15.49 GHz. By using 81 wavelength channels generated by the same comb source, an integrator has also been demonstrated (**Figure 13(b)**) [38], which had an integration time window of ~6.8 ns and an operation bandwidth of ~11. 9 GHz, yielding a large time-bandwidth product of ~81.

A fractional-order Hilbert transformer with reconfigurable orders from 0.17 to 0.83 has been demonstrated (**Figure 13(c)**) [35], where 17 taps were used, achieving an operation bandwidth up to 9 octaves, in contrast to 5 octaves achieved for previous integral-order Hilbert transformer [34]. By changing the tap coefficients, tunable bandwidth (from ~1.2 GHz to ~15.3 GHz) and center frequency (from baseband to ~9.5 GHz) of the fractional-order Hilbert transformer were also demonstrated (**Figure 13(d)**) [36].

Microwave phase-encoded signals, with low power densities and random phases designed to avoid their hosts being tracked, have been widely used for secure communications in radar systems. Recently, high-speed microwave phase encoding has been realized using a microcomb-based transversal filter system (**Figure 14(a)**) [40], showing reduced system size, complexity, and instability compared to previous methods based on polarization modulators [236, 237], dual parallel modulators [238, 239], and Sagnac loops [240, 241]. The operation principle of microwave phase encoding was the same as that of the microwave photonic integrator in **Figure 13(b)**, except that phase flipping of delayed replicas was introduced via differential photodetection to assemble a phase-encoded sequence in the time domain. By using 60 wavelength channels of a soliton crystal microcomb generated by a Hydex MRR, microwave phase encoding with a high encoding rate of ~6 Gbit/s and a high pulse compression ratio of ~29.6 was achieved. A reconfigurable encoding rate was also demonstrated by changing the length of each phase code.

Microcomb-based transversal filter systems have also been used for broadband arbitrary waveform generation (AWG), with wide applications to radar, remote sensing, and communication systems [242-244]. Similar to phase encoding, the waveform generation was achieved based on temporal integration, except that the flattened comb lines were replaced by shaped comb lines that correspond to different temporal waveforms [39]. By changing the tap coefficients, the generation of square waveforms with different duty ratios, sawtooth waveforms with different slope ratios, and various chirp waveforms was demonstrated using the same system with 81 wavelength channels (**Figure 14(b)**).

A microwave AWG based on coherent dual-microcomb sampling has also been demonstrated [225], which avoids the need for long optical delay lines to generate sufficient delays between the comb lines. Microwave frequencies were generated by beating a signal soliton microcomb with a local soliton microcomb, both of which were generated by $Si_3N_4$ MRRs with a repetition rate offset of ~150 MHz. Amplitude and phase shaping of the signal soliton microcomb was mapped onto the generated microwave frequency comb, enabling the synthesis of triangle, square, and "UVA"-like waveforms in the experimental demonstration. A tunable repetition rate of the microwave frequency comb was also achieved by adjusting the repetition rate of the local soliton microcomb (**Figure 14(c)**).

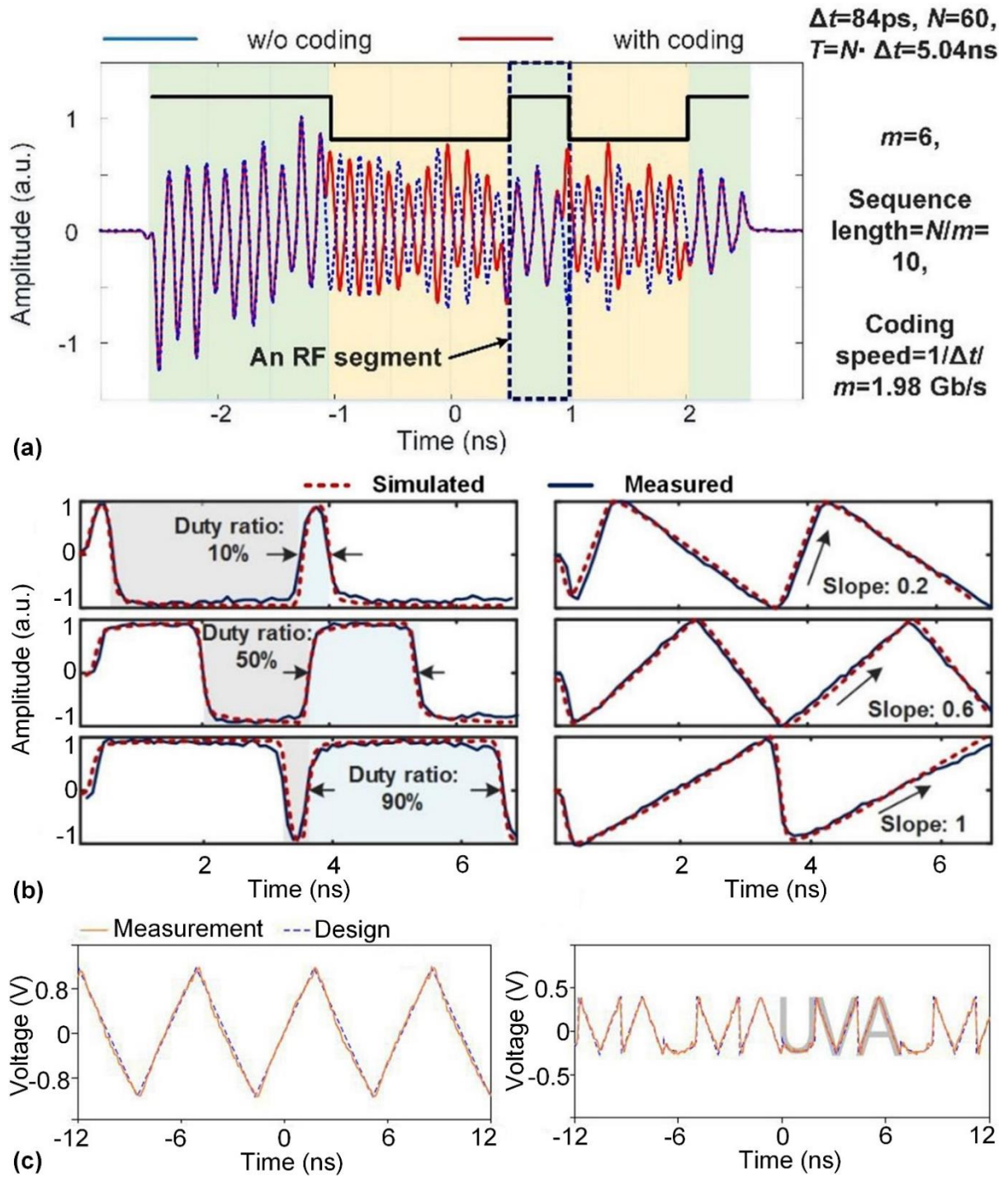

**Figure 14.** Recent progress in realizing other functions using microcomb-based microwave photonic signal processors. (a) Microwave photonic phase encoding. (b) Arbitrary waveform generation (AWG): theoretical and measured tunable RF square and sawtooth waveforms. (c) AWG: theoretical and measured triangle and "UVA"-like waveforms. (a) Reprinted with permission from [*J. Lightwave Technol.*, 38, 1722 (2020)].[40] (b) Reprinted with permission from [*J. Lightwave Technol.*, 38, 6221 (2020)].[39] (c) Reprinted with permission from [*Photonics Res.*, 10, 932 (2022)].[225]

# 4. Optical communications based on optical microcombs

Optical communications represent another important class of applications that has played a critical role in the information age. In optical communication systems, high-speed electrical signals in certain modulation formats are modulated onto optical carriers, and the optical links provide enormous bandwidth for achieving massively parallel data transmission via wavelength division multiplexing (WDM). In contrast to the use of bulky and power-hungry discrete laser arrays to generate multiple optical carriers in conventional WDM optical communication systems [245], optical microcombs can provide a number of equally spaced optical carriers by using a single device with a compact footprint, with the spacing between them (on the order of $10^1$ or $10^2$ GHz) matching state-of-the-art International Telecommunication Union (ITU) WDM spectral grid [246]. This offers a competitive edge for implementing WDM communication systems with greatly reduced SWaP.

The high achievable phase coherence between the comb lines of mode-locked microcombs also enables the realization of coherent optical communications that have stringent requirements for the spectral purity. State-of-the-art microcomb-based coherent optical communication systems have achieved data rates up to ~55 Tbit/s [42], which is well beyond those of conventional coherent optical communication systems based on LFCs generated by mode-locked lasers [247] or single-CW-seeded cavity-less parametric combs [248]. In terms of noise performance, although the bit error ratios (BERs) of microcomb-based coherent optical communications have not reached the level as high as those based on the bulk instrumentation [249, 250], a high BER on the order of $10^{-5}$ has been achieved [42].

In this section, we review and discuss the use of optical microcombs for optical communications, including both coherent optical communications and intensity modulation and direct detection (IM-DD) optical communications. In **Table 6**, we summarize and compare the performance of state-of-the-art microcomb-based optical communication systems.

**Table 6. Performance comparison of microcomb-based optical communication systems. WC: wavelength channel. MF: modulation format. QAM: quadrature amplitude modulation. PAM4: pulse-amplitude four-level modulation. SE: spectral efficiency. TD: transmission distance. BER: bit error ratio. NRZ: non return to zero.**

| No. of WCs | Microcomb characteristic | MF | Maximum data rate | Maximum SE | TD | BER | Year | Refs. |
|---|---|---|---|---|---|---|---|---|
| 20 | A low-noise comb | 16-QAM | 1.44 Tbit/s | 6 bits/s/Hz | 300 km | $7.5 \times 10^{-4}$ | 2014 | [41] |
| 179 | Two single soliton combs | 16-QAM | 55.0 Tbit/s | 5.2 bits/s/Hz | 75 km | $6.7 \times 10^{-5}$ | 2017 | [42] |
| 20 | A mode-locked dark-pulse comb | 64-QAM | 4.4 Tbit/s | – | 80 km | $2.5 \times 10^{-3}$ | 2018 | [43] |
| 80 | A soliton crystal comb | 64-QAM | 44.2 Tbit/s | 10.4 bits/s/Hz | 76.6 km | $2.6 \times 10^{-2}$ | 2020 | [44] |
| 52 | A single soliton comb | 256-QAM | 12 Tbit/s | 10.4 bits/s/Hz | 82 km | $\sim 10^{-2}$ | 2021 | [251] |
| 52 | A single soliton comb | 16-QAM | 8 Tbit/s | 6.7 bits/s/Hz | 2100 km | $\sim 10^{-2}$ | 2021 | [251] |
| 20 | Two single soliton combs | 16-QAM | 1.68 Tbit/s | – | 50 km | $\sim 10^{-4}$ | 2022 | [252] |
| 145 | A single soliton comb | NRZ | 1.45 Tbit/s | 1 bits/s/Hz | 40 km | $< 10^{-9}$ | 2022 | [45] |
| 20 | A dark soliton microcomb | PAM4 | 2 Tbit/s | – | 2 km | $\sim 10^{-2}$ | 2022 | [33] |

## 4.1 Coherent optical communications

Coherent optical communications have been the mainstream application of microcombs in optical communication systems, where coherent optical microcombs with comb spacings matching the ITU WDM spectral grid standards act as both optical carriers at the transmitters and local oscillators (LOs) at the receivers [253, 254]. Compared to the IM-DD optical communications, coherent optical communications provide an attractive advantage in achieving high communication capacities, resulting from the fact that the amplitude, phase, and polarization of optical carriers are all employed to carry information [42]. The spectral efficiencies in coherent optical communications can be significantly increased by adopting advanced modulation formats [41, 42], which enables ultra-dense optical data transmissions. Another important advantage of coherent optical communication systems is the improvement of the receiver sensitivity by using heterodyne or homodyne detection [2, 255, 256]. The noise in coherent receivers mainly comes from the LO-induced shot noise, as compared to the dark current and thermal noise for the IM-DD systems [256, 257].

The data rate of a microcomb-based WDM optical communication system can be expressed as [41, 42]

$$DR = SR \times BPS \times N \times M \qquad (20)$$

where $SR$ is the symbol rate of the modulated electrical signal, $BPS$ is the number of bits per symbol, $N$ is the number of WDM channels that equals to the number of available comb lines, and $M$ is the number of polarization states.

In addition to the raw data transmission rate, another important performance parameter is spectral efficiency (SE), or the number of bits/s transmitted for a given bandwidth (bits/s/Hz), which can be given by [41, 42]

$$SE = \frac{SR \times BPS \times M}{CS} \qquad (21)$$

where $CS$ is the comb spacing. The SE is critical since the optical bandwidth of the telecommunications band is fixed, and so this determines the ultimate achievable transmission data rate. According to **Eq. (21)**, improved $SE$ can be achieved by employing advanced modulation formats to increase the $BPS$. In practical communications, error correction coding technologies such as forward error correction (FEC) are usually employed to control the errors occurred during the communication process, which results in a net SE that is slightly lower than that calculated based on **Eq. (21)**.

The bit error ratio (BER), which is defined as the number of error bits divided by the total number of bits transferred within a specific time interval, is widely used for characterizing the quality of data transmission [258, 259]. For *M*-ary quadrature amplitude modulation (QAM) signals that are widely used in coherent optical communication systems, assuming the optical additive white Gaussian noise is the dominant source of errors and the reception is data-aided, the BER can be given by [42, 258]

$$\text{BER} = \frac{1-M^{-\frac{1}{2}}}{\frac{1}{2}\log_2 M} erfc\sqrt{\frac{\text{OSNR}}{10}} \qquad (22)$$

where *M* is the number of symbols in the complex constellation plane, OSNR is the measured optical signal to noise ratio, and *erfc* is the complementary error function [258]. For bright soliton microcombs used for coherent optical communications, the powers of the comb lines can be approximately given by [42]

$$P(\mu) = \frac{\kappa_{ex} D_2 \hbar \omega_0}{4g} \text{sech}^2\left(\frac{\kappa \mu}{2}\sqrt{\frac{D_2 \hbar \omega}{\kappa_{ex} g P_{in}}}\right) \qquad (23)$$

where $\mu = (\omega - \omega_0)/D_1$ is the comb mode index relative to the pump mode, $\omega$ and $\omega_0$ are the angular frequencies of the comb and pump modes, respectively, and $D_1/2\pi$ is the FSR of the microresonator. The total cavity loss-rate is given by $\kappa = \kappa_0 + \kappa_{ex}$, which includes the internal loss rate $\kappa_0$ and the coupling rate $\kappa_{ex}$, $D_2$ is the dispersion parameter, $\hbar$ is the reduced Planck constant, $g/2\pi$ is the nonlinear coupling constant, and $P_{in}$ is the input pump power. According to **Eq. (23)**, the power degradation at comb lines far away from the pump would lead to reduced OSNRs and hence increased BER. Therefore, in order to achieve massively parallel data transmission in broad bandwidths, the dispersion and coupling of the microresonators used for generating microcombs need to be properly engineered.

The first demonstration of coherent data transmission using a multi-wavelength microcomb source was reported in 2014 [41], where 20 comb lines of a low-noise microcomb generated by a $Si_3N_4$ MRR were modulated with quadrature phase-shift keying (QPSK) and 16-QAM signals, achieving a maximum data rate of ~1.44 Tbit/s and a SE of ~6 bits/s/Hz.

Massively parallel coherent optical communications have been demonstrated using two interleaved soliton microcombs generated by $Si_3N_4$ MRRs (**Figure 15(a)**) [42], where the microcombs with channel spacings of ~100 GHz served as both a multi-carrier source at the transmitter and a LO at the receiver, achieving a maximum data rate of ~55.0 Tbit/s for 16-QAM signals modulated on 179 parallel optical carriers spanning over the C and L bands.

Mode-locked dark-pulse microcombs generated by a $Si_3N_4$ MRR with normal dispersion have been used for ~4.4 Tbit/s coherent optical data communication (**Figure 15(b)**) [43]. The high internal power conversion efficiency of the dark-pulse microcombs ($> 30\%$) yielded a low pump power $< 400$ mW and a high transmitted OSNR $> 33$ dB for 64-QAM signals modulated on 20 optical carriers in the C band.

Recently, coherent optical communication with a record high SE has been demonstrated using a soliton crystal microcomb generated by a Hydex MRR (**Figure 15(c)**) [44], where the soliton crystal microcomb with a small comb spacing (~49 GHz), high conversion efficiency (~40%), and high stability enabled a maximum data rate of ~44.2 Tbit/s and a maximum SE of ~10.4 bit/s/Hz, enabled by the use of a higher modulation format of 64-QAM across 80 comb lines in the C band. This work also featured a live field trial demonstration over an installed optical fiber network in the metropolitan Melbourne area.

With no need for filtering or guard bands within the channel, the use of broadband optical superchannels consisting of multiple densely spaced subchannels (with an overall bandwidth larger than that of the PDs) in coherent communications can improve the SE. Coherent superchannel communication based on soliton microcombs generated by a silica wedge resonator has been demonstrated [251], where 52 comb lines with a spacing of ~22.1 GHz were employed for each superchannel. Experimental demonstrations of ~12 Tbit/s transmission of a 256-QAM signal over ~82 km and ~8 Tbit/s transmission of 16-QAM signal over ~2100 km were carried out, achieving spectral efficiencies of ~10.4 and ~6.7 bits/s/Hz, respectively.

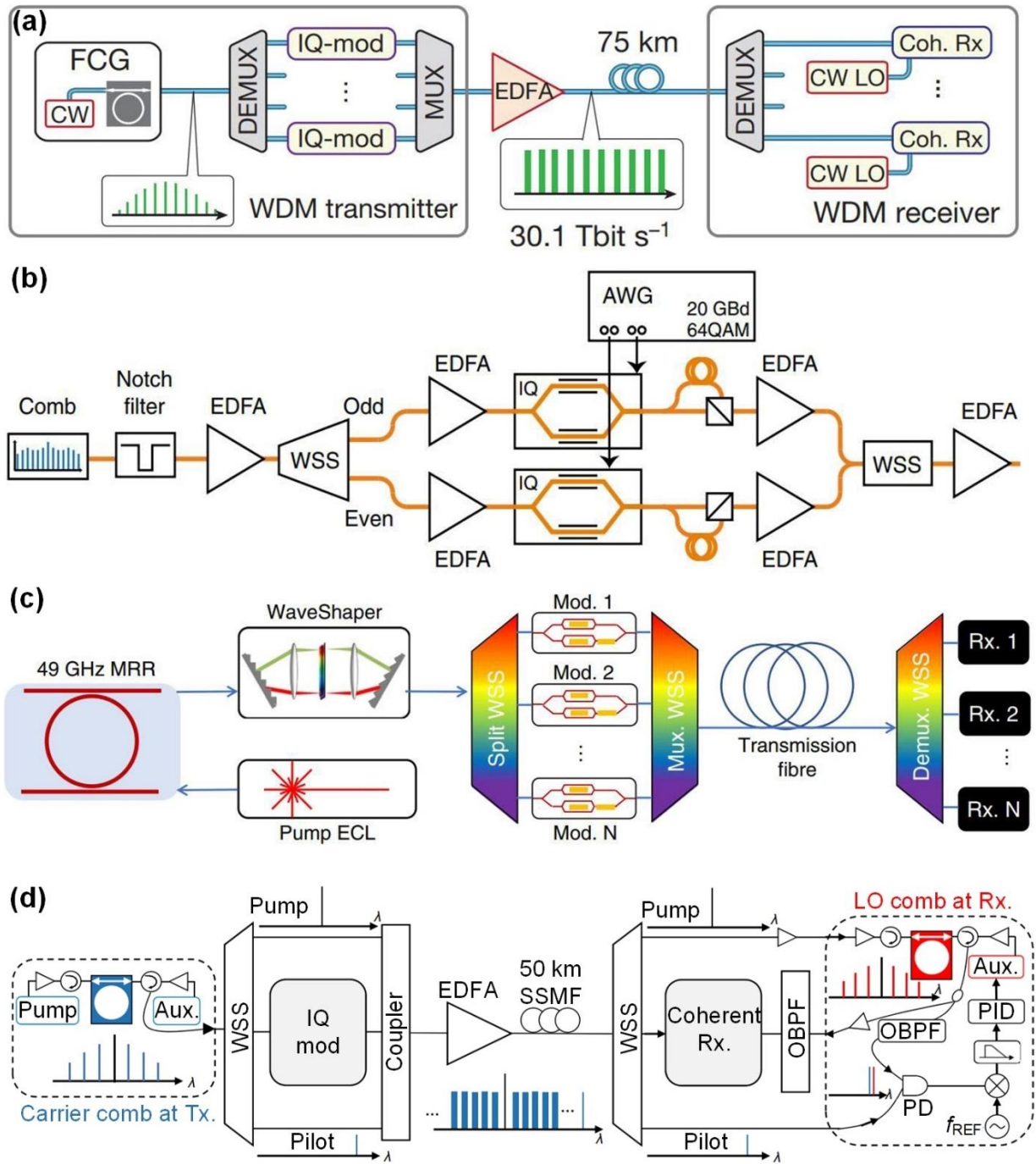

**Figure 15.** Coherent optical communications based on optical microcombs. (a) ~55.0 Tbit/s coherent communication over 75 km based on two interleaved soliton microcombs generated by $Si_3N_4$ MRRs. (b) ~4.4 Tbit/s coherent communication over 80 km based on a mode-locked dark-pulse microcomb generated by a normal-dispersion $Si_3N_4$ MRR. (c) ~44.2 Tbit/s coherent communication over 76.6 km based on a soliton crystal microcomb generated by a Hydex MRR. (d) ~1.68 Tbit/s coherent communication over 50 km based on coherence-cloned soliton microcombs generated by $Si_3N_4$ MRRs. (a) Reprinted with permission from [*Nature.*, 546, 274 (2017)].[42] (b) Reprinted with permission from [*Nat. Commun.*, 9, 1 (2018)].[43] (c) Reprinted with permission from [*Nat. Commun.*, 11, 1 (2020)].[44] (d) Reprinted with permission from [*Nat. Commun.*, 13, 1070 (2022)].[252]

Coherent communication with reduced power consumption and system complexity has also been demonstrated based on coherence-cloned soliton microcomb regeneration over a long distance (**Figure 15(d)**) [252], where the frequency and phase of 20 comb lines of a soliton microcomb generated by a $Si_3N_4$ MRR at transmitter were regenerated at the receiver 50 km away through sharing pump laser and two-point locking, yielding totally saved frequency offset estimation and substantially reduced carrier phase estimation in coherent detection. The demonstration of 1.68 Tbit/s transmission of 16-QAM signal over 50 km was reported, achieving a BER on the order of $10^{-4}$.

**4.2 Intensity modulation - direct detection optical communications**

In addition to coherent optical communications, optical microcombs have been used as multiwavelength sources for IM-DD optical communications [33, 45]. In contrast to coherent optical communications that achieve high sensitivity needed for long-distance communications, the IM-DD optical communication systems have a reduced complexity that that is advantageous for short-distance communications in local area networks where low latency and low cost are important [45].

In an alternative approach based on bulk-optic microcombs, IM-DD optical communications over 40 km has been demonstrated based on optical microcombs generated by a $MgF_2$ crystalline microresonator with an FSR of ~10 GHz (**Figure 16(a)**) [45], where 145 comb lines covering the entire C-band were employed as the optical carriers of the WDM channels. A maximum data rate of ~1.45 Tbit/s and a maximum SE of ~1 bit/s/Hz were experimentally achieved, together with a BER $< 10^{-9}$ for an operation without using the FEC.

In terms of direct detection systems, a microcomb-based PAM4 IM-DD communication system with a high integration level has also been demonstrated (**Figure 16(b)**) [33], where 20 comb lines of a dark soliton microcomb generated by an AlGaAs-on-insulator MRR were used as the WDM optical carriers and launched into an SOI transmitting-receiving chip consisting of silicon EO modulators and germanium PDs. A maximum data transmission rate of ~2 Tbit/s was achieved by using an external cavity laser (ECL) to pump the MRR. When the ECL was replaced by an on-chip DFB laser, the maximum achieved data transmission rate was ~448 Gbit/s. After direct detection by using on-chip Ge PDs, a BER on the order of $10^{-2}$ was achieved.

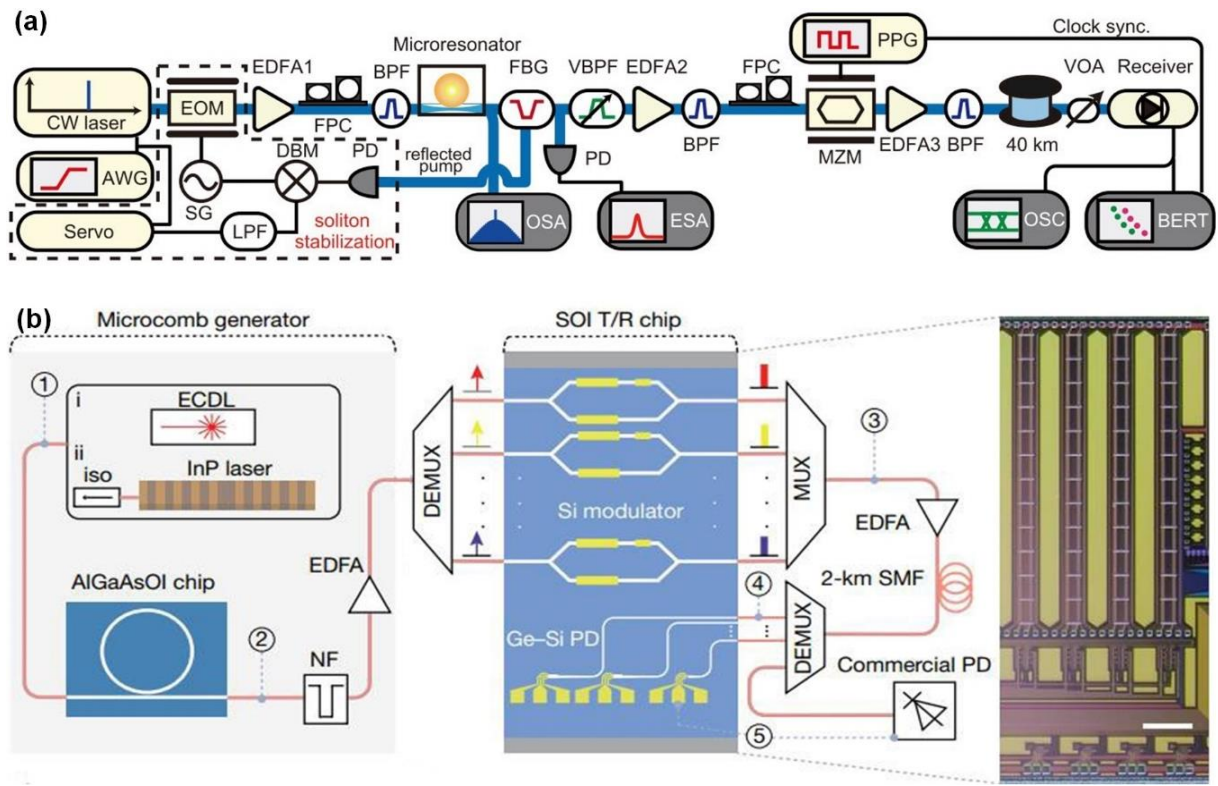

**Figure 16.** Intensity modulation and direct detection (IM-DD) optical communications based on optical microcombs. (a) ~1.45 Tbit/s data transmission over 40 km based on a single soliton microcomb generated by a $MgF_2$ crystalline microresonator. (b) ~2 Tbit/s PAM4 data transmission over 2 km based on an AlGaAs-on-insulator comb source and an SOI transmitting-receiving chip. (a) Reprinted with permission from [*Opt. Express.*, 30, 1351 (2022)].[45] (b) Reprinted with permission from [*Nature.*, 605, 457 (2022)].[33]

# 5. Precision measurements based on optical microcombs

With a large instantaneous bandwidth, low loss, and strong immunity to electromagnetic interference, precision measurements based on photonic technologies have wide applications in many fields, such as ranging, spectra analysis, frequency measurement, radar, and astronomy [46-51, 143, 260, 261]. Compared with measurement systems based on bulky light sources such as CW lasers, mode-locked pulsed lasers, and amplified spontaneous emission sources [262-264], optical microcombs provide powerful new solutions to realize ultrafast and highly precision measurements with chip-scale devices. The overlap between the repetition rates of optical microcombs and the microwave frequency band also provides an intrinsic link connecting microwave and optical domain [9], which has been utilized to realize diverse measurement functions. In this section, we review and discuss the use of optical microcombs for precision measurements, which is categorized into dual-comb spectroscopy, ranging, astrocombs, frequency measurements, and spectrum channelizers. In **Table 7**, we compare the performance metrics of different measurements applications based on optical microcombs.

**Table 7. Performance comparison of different measurement functions based on optical microcombs. CS: comb spacing. Freq.: frequency.**

| | Microresonators | CS (GHz) | Wavelengths range (nm) | Spectral resolution (GHz) | Acquisition time (µs) | Year | Refs. |
|---|---|---|---|---|---|---|---|
| Dual-comb spectroscopy | Two silica microdisk resonators | ~22 | 1538 – 1562 | ~22 | ~20 | 2016 | [60] |
| | Two $Si_3N_4$ MRRs | ~450 | near 1561 | ~450 | ~20 | 2018 | [260] |
| | Two silicon MRRs | ~127 | 2900 – 3100 | ~127 | ~2 | 2018 | [148] |
| | A silica wedge resonator | ~22 | Near 3300 | ~1.4 | – | 2020 | [261] |
| | A silica wedge resonator | ~22 | Near 3300 | ~1.4 | ~2 × $10^5$ | 2021 | [265] |
| | Microresonators | CS (GHz) | Wavelength range (nm) | Spatial resolution (µm) | Acquisition time (µs) | Year | Refs. |
| Ranging | Two $Si_3N_4$ MRRs | ~100 | 1529 – 1620 | ~2 | ~0.01 | 2018 | [46] |
| | A silica wedge resonator | ~9.36 | 1530 – 1570 | ~0.2 | ~176 | 2018 | [47] |
| | A $Si_3N_4$ MRR | ~99 | 1550 – 1630 | ~1 × $10^4$ | – | 2020 | [48] |
| | A Hydex MRR | ~48.97 | 1530 – 1595 | ~0.028 | ~28 | 2020 | [49] |
| | A tapered $Si_3N_4$ MRR | ~88.5 | 1575 – 1615 | ~0.003 | – | 2021 | [266] |
| | Microresonators | CS (GHz) | Wavelength range (nm) | Calibration precision (m/s) | Instability (mHz) | Year | Refs. |
| Astrocombs | A silica disk resonator | ~20 | 1530 – 1580 | ~3.4 | ~10 at $10^3$ s | 2019 | [51] |
| | A $Si_3N_4$ MRR | ~23.7 | 1486 – 1685 | ~0.25 | ~1000 at 1 s | 2019 | [50] |
| | Microresonators | CS (GHz) | Wavelength range (nm) | Parameters and values | Precision (GHz) | Year | Refs. |
| Frequency measurements | A silica disk resonator | ~15 | 700 – 2100 | Freq. drift 180 mHz/s | – | 2018 | [267] |
| | A silica wedge resonator | ~22 | 1545 – 1560 | Optical freq. ~192.79 THz | ~0.0025 | 2019 | [143] |
| | Two $Si_3N_4$ MRRs | ~197 ~216 | 1430 – 1660 | Microwave freq. ~197 GHz | ~0.995 | 2020 | [268] |
| | A Hydex MRR | ~49 | 1500 – 1620 | freq. spectrum 200 – 2400 GHz | 0.075 | 2021 | [269] |
| | Microresonators | CS (GHz) | Wavelength range (nm) | Slice resolution (GHz) | Channel numbers | Year | Refs. |
| Spectrum channelizers | Two Hydex MRRs | ~200 | 1535 – 1565 | ~1.04 | 20 | 2018 | [270] |
| | Two Hydex MRRs | ~49 | 1535 – 1570 | ~0.12 | 92 | 2020 | [271] |

## 5.1 Dual-comb spectroscopy

Dual-comb spectroscopy (DCS) based on multi-heterodyne detection of two Vernier LFCs is a key approach that can achieve broadband sampling of optical spectra. In DCS, acquisition time, *i.e.*, the time to acquire the optical spectra, is an important parameter determined by the repetition rates of the LFCs. Optical microcombs with large repetition rates provide a powerful solution for realizing broadband DCS with low acquisition times and fast measurements [60, 148, 260]. In the spectral domain, the output of a DCS system is an RF comb consisting of a series of heterodyne beats corresponding to different pairs of Vernier comb teeth. The minimum time $\tau$ to resolve the RF comb lines and acquire a single spectrum can be given by [264]

$$\tau = 1/\Delta f_{rep} \tag{24}$$

where $\Delta f_{rep}$ is the difference between the repetition rates of the two combs. For practical DCS systems, the acquisition time $T$ should satisfy

$$T \geq \tau \tag{25}$$

According to **Eqs. (24)** and **(25)**, a larger comb spacing allows for a larger $\Delta f_{rep}$ and hence a smaller acquisition time $T$. This reflects the advantage of realizing DCS based on optical microcombs compared to conventional LFCs [262, 272, 273].

In DCS, the wavelength ranges of microcombs are a critical parameter since they need to match the absorption bands of the materials being sensed. Hence, this motivates the need for micro-combs at both shorter and particularly longer (*e.g.*, mid-infrared) optical wavelengths compared to more traditional ones that operate in the telecom band.

In 2016, DCS based on microcombs generated by two compact silica disk resonators was first reported for measuring molecular absorption spectra in the telecom band [60], where the large comb spacing of the microcombs (~22 GHz) yielded significantly improved acquisition rates as compared with mode-locked fiber lasers.

By using thermo-optic microheaters, the simultaneous generation of two microcombs from two Si$_3$N$_4$ MRRs on the same chip was realized with a single pump laser at ~1561 nm (**Figure 17(a)**) [260], which significantly reduced the system complexity. Real-time DCS measurement of dichloromethane's absorption across a range of 170 nm near the pump wavelength was demonstrated, achieving an acquisition time as short as ~20 μs.

Subsequently, mid-infrared DCS based on optical microcombs generated by silicon MRRs has been demonstrated (**Figure 17(b)**) [148]. A single CW light at ~3 μm was employed to pump two silicon MRRs, generating two mutually coherent mode-locked frequency combs across a wavelength range of 2.6 μm – 4.1 μm. The silicon MRRs had p-i-n junctions to sweep out the free carriers and were fabricated using an etchless method. DCS measurement of acetone in the liquid phase covering 2.9 μm – 3.1 μm was performed, achieving a spectral resolution of 127 GHz and an acquisition time of ~2 μs.

To improve the spectral resolution of mid-infrared DCS based on optical microcombs, a technique termed interleaved difference-frequency generation (iDFG) was employed (**Figure 17(c)**) [261], where mid-infrared combs were generated from two near-infrared combs. The microwave signal derived from a 1.5-μm soliton microcomb was used to drive a 1.0-μm electro-optic comb, enabling a densified comb spacing of the generated mid-infrared combs for finer spectral sampling. A spectral resolution of ~1.4 GHz was achieved, representing a 100-fold improvement compared with previous reports [148].

The iDFG technique has also been used for mid-infrared DCS based on 1.55-μm counter-propagating soliton microcombs in the same silica wedge resonator (**Figure 17(d)**) [265]. Here, the system complexity was significantly reduced as a result of the mutual coherence of the counter-propagating soliton microcombs in the same microcavity. DCS measurements of ethane and methane around 3.3 μm were experimentally demonstrated, achieving a high normalized precision of ~1.0 ppm·m·s$^{1/2}$.

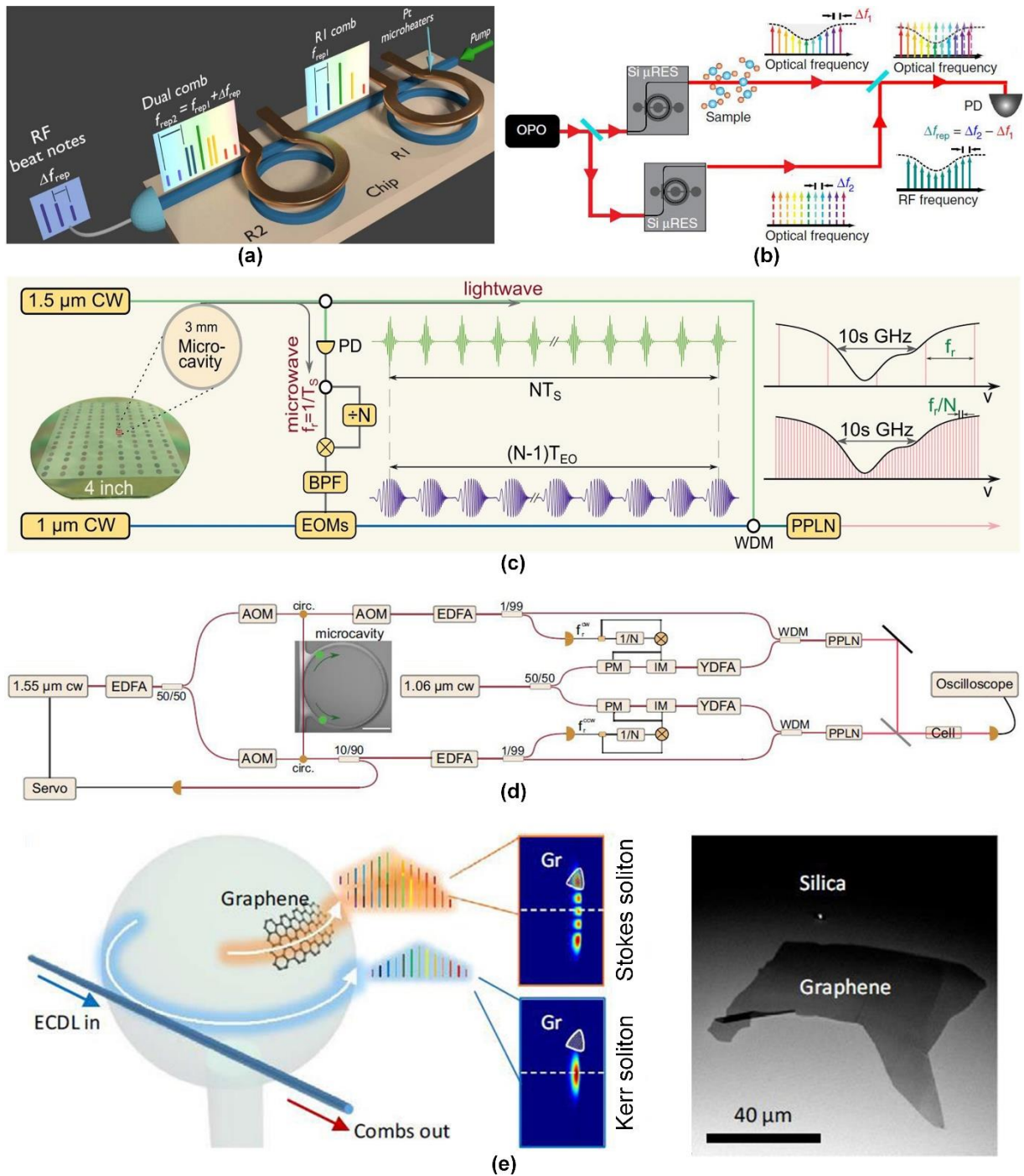

**Figure 17.** Dual-comb spectroscopy (DCS) based on optical microcombs. (a) DCS measurement of dichloromethane's absorption by using two microcombs on a single chip with a single pump. (b) DCS measurement of acetone in the liquid phase based on mid-infrared microcombs generated by two silicon MRRs. (c) DCS measurement of methane gas's absorption based on mid-infrared combs generated by combining a 1.5-μm soliton microcomb and a 1.0-μm electro-optic frequency comb to pump a periodically poled lithium niobate (PPLN) crystal. (d) DCS measurement of a mixture of methane and ethane gas's absorption based on mid-infrared combs generated by 1.55-μm counter-propagating soliton microcombs. (e) Gas molecule detection based on three Stokes solitons generated by a graphene-deposited silica microsphere resonator. (a) Reprinted with permission from [*Sci. Adv.*, 4, 9 (2018)].[260] (b) Reprinted with permission from [*Nat. Commun.*, 9, 6 (2018)].[148] (c) Reprinted with permission from [*Optica.*, 7, 309 (2020)].[261] (d) Reprinted with permission from [*Nat. Commun.*, 12, 6573 (2021)].[265] (e) Reprinted with permission from [*Nat. Commun.*, 12, 6716 (2021)].[274]

Gas molecule detection based on multiple Stokes solitons generated by a silica microsphere asymmetrically coated with monolayer graphene has also been demonstrated (**Figure 17(e)**) [274]. By using a single CW pump to drive the over-modal resonant cavity, multiple Stokes solitons corresponding to different cavity modes with different center frequencies, at the same repetition rate locked to the co-generated Kerr soliton, were generated. This produced sensitive beat notes that allowed for sub-Hz spectral resolution for single molecule detection of $NH_3$ and the identification of $NH_3$, $CO_2$, and $H_2O$ in a mixture.

**5.2 Ranging**

Light detection and ranging (LiDAR) technology plays a significant role in advanced metrology, with wide applications to autonomous vehicles, satellite formation flying, gravitational waves detection, and many others [48, 49]. Recently, microcomb-based systems have been developed to realize fast, precision, and long-distance ranging [46-49], enabling significant advances towards miniature photonic integrated ranging systems.

The time-of-flight (ToF) method, which measures the time it takes for a wave to travel from a source to a target and back, is commonly used for ranging systems [275, 276]. For microcomb-based ranging systems, ToF measurements can be realized by combining optical interferometry and multi-heterodyne detection of two Vernier microcombs similar to that in DCS. Precision and ultrafast ranging using ToF method has been demonstrated based on two soliton microcombs generated by $Si_3N_4$ MRRs (**Figure 18(a)**) [46], achieving a lateral spatial resolution of ~2 μm for moving targets at a speed of ~150 m/s. The large optical bandwidth (>11 THz) allowed for ultra-precision ranging with an Allan deviation of ~12 nm at 13 μs, and the high repetition rates of microcombs (~100 GHz) enabled ultrafast measurement with an acquisition rate of ~100 MHz.

Similarly, ToF-based distance measurement has been demonstrated using two counterpropagating soliton microcombs generated by a single silica wedge resonator (**Figure**

**18(b)**) [47]. The generation of two microcombs from the same resonator device allowed for both an improvement in their relative coherence and the simplification of the measurement system. In the demonstration, a precision of ~200 nm at an averaging time of ~500 ms was achieved over a range ambiguity of ~16 mm.

ToF-based ranging with the precision of a few nanometers has also been achieved based on a single soliton microcomb generated by a tapered $Si_3N_4$ MRR (**Figure 18(c)**) [266]. Spectrally resolved interferometry between the reference and measurement pulses was utilized to map information onto the optical ToF. The large repetition rate of the soliton microcomb (~88.5 GHz) enables a comb-tooth-resolved interferogram that could be directly read out via an optical spectrum analyzer. A precision of ~3 nm over an unambiguous range of 23 mm and stable distance measurement over 1000 s were achieved.

Other methods have also been employed for microcomb-based ranging, including the frequency-modulated continuous-wave (FMCW) method [277, 278] and the dispersive interferometry (DPI) method [49].

The FMCW method has been used for massively parallel ranging based on a soliton microcomb generated from a $Si_3N_4$ MRR (**Figure 18(d)**) [48], where the distance information was mapped to frequency by sending a linearly chirped laser to the target and the reflected signal was measured by delayed homodyne detection. Compared to ToF, FMCW can measure both distance and velocity at the same time, which is ideal for autonomous driving. In the demonstration, the pump laser was fast chirped at a speed of $>10^{17}$ $Hz^2$ in the soliton region, which was transferred to all comb lines without changing the pulse-to-pulse repetition rate, enabling parallel ranging measurement for 30 channels at an equivalent sampling rate of 3 megapixels per second.

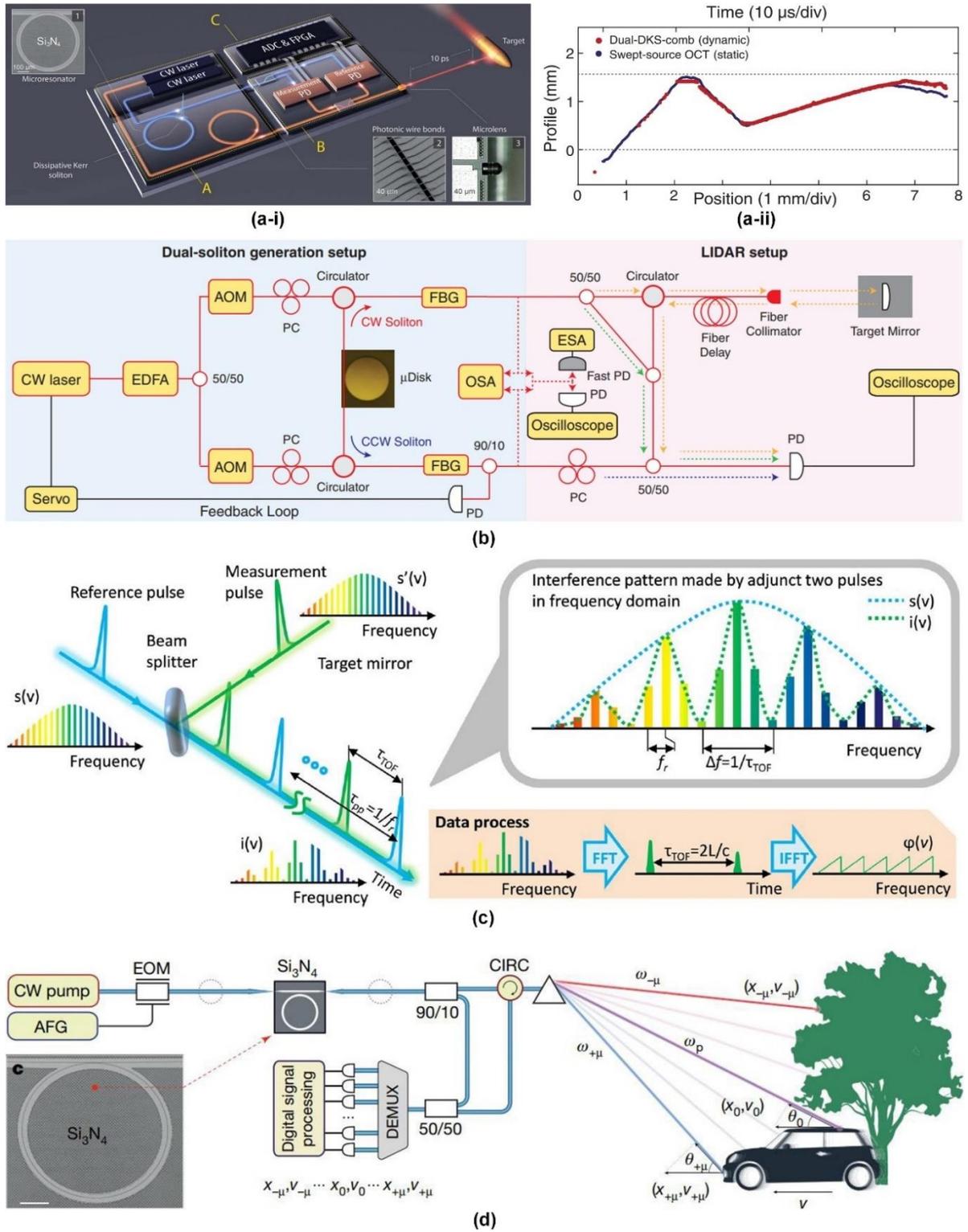

**Figure 18.** Ranging based on optical microcombs. (a) Ultrafast ranging of moving targets based on soliton microcombs generated by two $Si_3N_4$ MRRs. (b) Distance measurement based on counterpropagating soliton microcombs generated by a single silica wedge resonator. (c) Spectrally resolved laser ranging with nanometric precision based on a soliton microcomb generated by a tapered $Si_3N_4$ MRR. (d) Massively parallel ranging using a soliton microcomb generated by a $Si_3N_4$ MRR pumped by chirped laser. (a) Reprinted with permission from [*Science.*, 359, 887 (2018)].[46] (b) Reprinted with permission from [*Science.*, 359, 884 (2018)].[47] (c) Reprinted with permission from [*Phys. Rev. Lett.*, 126, 023903 (2021)].[266] (d) Reprinted with permission from [*Nature.*, 581, 164 (2020)].[48]

The DPI method is an LFC-based ranging technique that is more suited to long-distance measurements due to its high tolerance to interference and long coherent lengths [279, 280]. However, in DPI a measurement dead zone appears when there is a mismatch between the optical spectrum analyzer's resolution and the LFC's repetition rate. Optical microcombs with high repetition rates can bridge this gap, thus enabling DPI-based long-distance ranging without a dead zone. Field trial ranging over a distance of ~1179 m has been demonstrated based on a 49-GHz soliton microcomb generated by a Hydex MRR [49], achieving an Allan deviation of ~$5.6 \times 10^{-6}$ m at an average time of ~$2 \times 10^{-4}$ s.

**5.3 Astrocombs**

Astrocombs, which are LFCs used for calibrating astronomical spectrographs, represent a key approach to astronomical spectroscopy that can enable the detection of exoplanets and cosmological phenomena [52, 53]. A main technical challenge confronting this field is the ability to match the comb spacing to the resolution of spectrographs, which is critical for the accurate characterization of Doppler shifts of measured light from astronomical objects [50, 281], for example. However, for conventional astrocombs implemented with mode-locked fiber lasers at typical repetition rates < 1 GHz [282-284], additional filtering steps are needed to coarsen the comb spacing in order to match the resolution of astronomical spectrometers. In contrast, optical microcombs with both high repetition rates (typically > 10 GHz) and compact device footprint show better compatibility with the astronomical spectrometers, providing an attractive way to implement astrocomb systems with significantly reduced SWaP and complexity.

A spectrograph calibrator used for measuring periodic Doppler shifts in astronomical spectra has been realized based on a soliton microcomb generated by a $SiO_2$ microdisk (**Figure 19(a)**) [51]. The large comb spacing of the microcomb (~20 GHz) matched the astronomical spectrometer resolution well (10 – 30 GHz). In the drift measurement, the repetition rate of the soliton microcomb was locked to a Rb atomic clock, achieving a high precision of 3 – 5 m/s in wavelength solution that enabled calibration of near-infrared spectrometers at a few m/s.

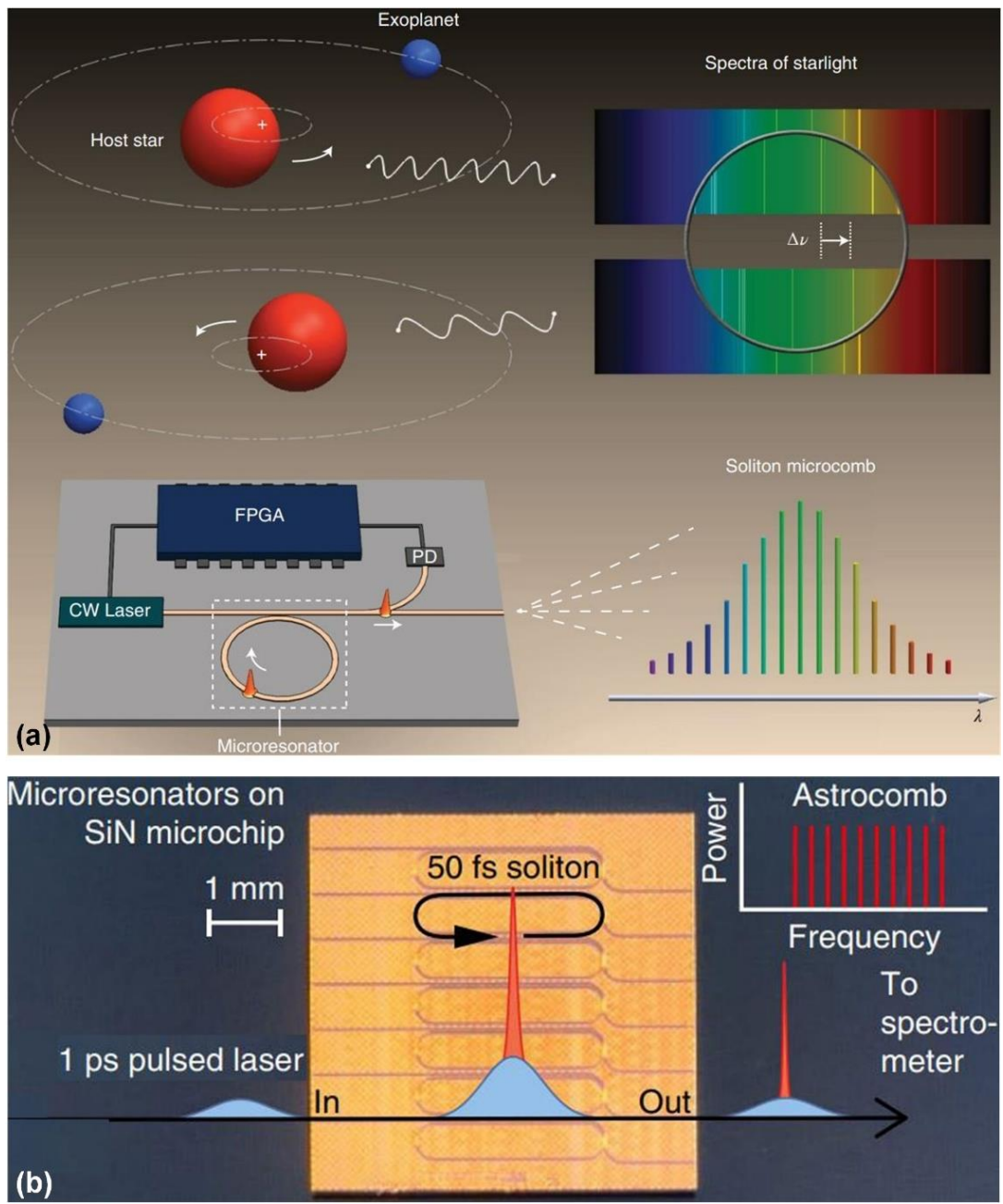

**Figure 19.** Astrocombs based on optical microcombs. (a) A spectrograph calibrator based on a soliton microcomb generated by a $SiO_2$ microdisk resonator. (b) A microphotonic astrocomb based on soliton microcomb generated by a $Si_3N_4$ MRR. (a) Reprinted with permission from [*Nat. Photonics.*, 13, 25 (2019)].[51] (b) Reprinted with permission from [*Nat. Photonics.*, 13, 31 (2019)].[50]

An astrocomb based on soliton microcombs generated by a $Si_3N_4$ MRR with an FSR of ~23.7 GHz and a Q factor of $6.4 \times 10^5$ (**Figure 19(b)**) was recently reported [50]. By using a sub-harmonic driven system, the MRR was pumped by periodically coupled picosecond optical pulses at half the repetition rate of the microcomb, which enabled all-optical control of the microcombs and the stabilization of its repetition rate. A calibration precision of ~25 cm/s was achieved, together with the verification of the consistency of the astrocomb frequency with the U–Ne standard.

**5.4 Frequency measurements**

Frequency measurements are of fundamental importance in many fields such as spectroscopy, communications, and sensing [143]. Compared with benchtop spectrometers based on gratings and spatial-light interferometers, LFCs offer stable measurement grids at microwave frequencies enabled by their equally spaced comb lines [17]. Paralleling the development of optical microcomb technologies, many microcomb-based frequency measurement systems have been demonstrated, successfully measuring not only frequencies spanning from microwave to optical region [143, 268, 269], but also frequency related parameters such as the frequency drift and frequency spectra [267]. Some also provide useful methods for the characterization of optical microcombs [268, 269].

A Vernier frequency division approach has been used to measure the repetition rate of an optical microcomb at a microwave frequency (~197 GHz) well beyond the bandwidth of electronic devices (**Figure 20(a)**) [268]. The measurement was realized by beating the target comb and another comb that have periodically aligned comb lines and then processing the generated microwave beat notes at low frequencies. Frequency division from ~197 GHz (*i.e.*, the repletion rate of the targe microcomb) to ~995 MHz was demonstrated by using another microcomb with a line spacing of ~216 GHz.

Rapid optical frequency measurement across a 60-GHz frequency range near 192.78 THz has been demonstrated based on a microcomb-based Vernier spectrometer (**Figure 20(b)**) [143], where two counterpropagating soliton microcombs with slightly different repetition rates generated by a single silica wedge resonator were simultaneously phase-locked [285], enabling the mapping of the differences between the two combs' frequencies to a stable microwave frequency grid. The tracking of rapidly tuning lasers undergoing fast chirping and discontinuous tuning has been demonstrated using the Vernier spectrometer, achieving a high laser tuning rate up to ~10 THz/s.

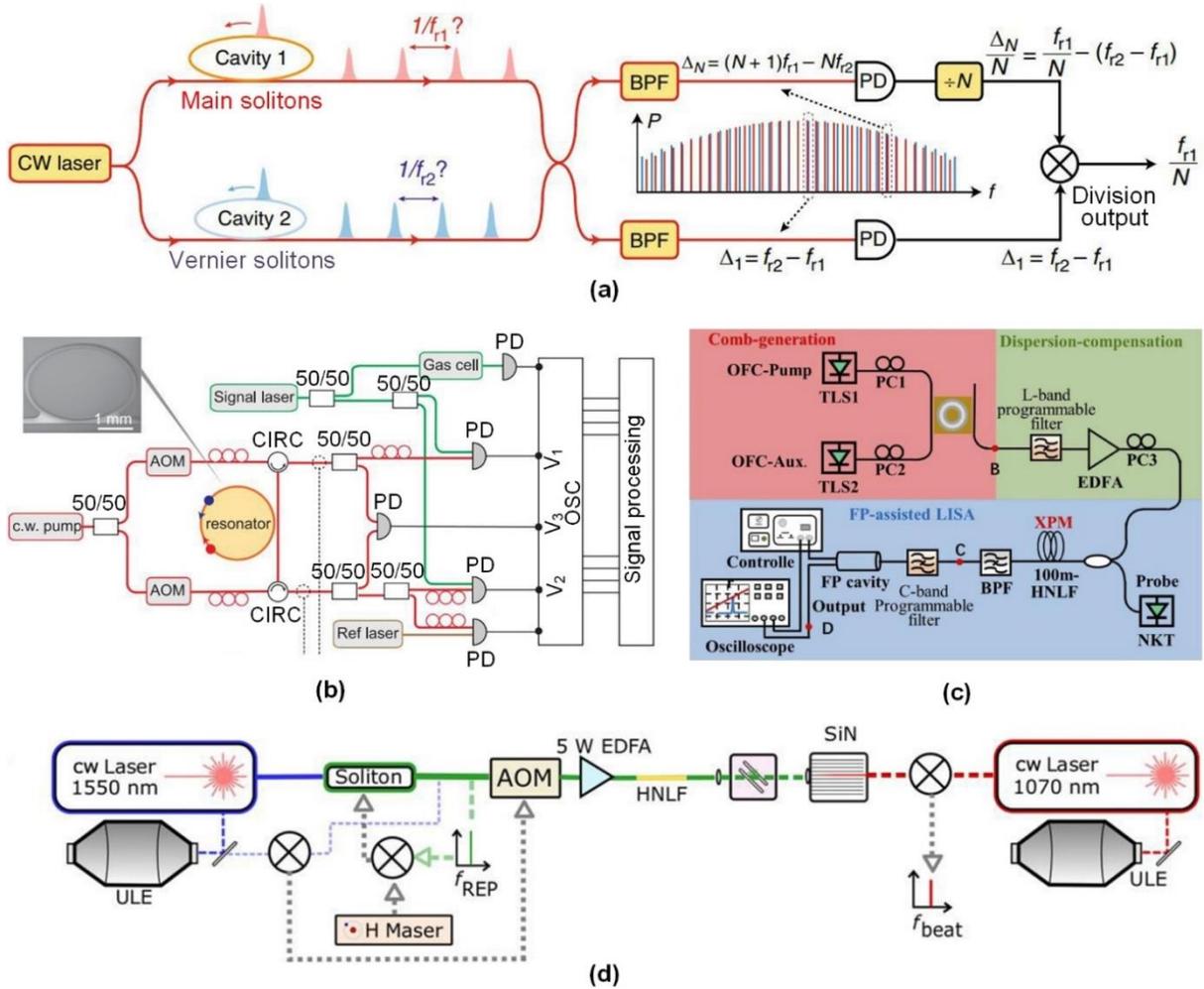

**Figure 20.** Frequency measurements based on optical microcombs. (a) Detecting repetition rates of microcombs based on Vernier frequency division. (b) Rapid and broadband optical frequency measurement based on a Vernier spectrometer with two counterpropagating soliton microcombs generated by a single silica wedge resonator. (c) Measuring the RF spectra of microcombs from the microwave to terahertz bands based on an all-optical RF spectrum analyzer. (d) Measuring the relative frequency drift between two optical cavities based on a soliton microcomb generated by a silica microdisk resonator and subsequently broadened via coherent supercontinuum generation after passing a $Si_3N_4$ waveguide. (a) Reprinted with permission from [*Nat. Commun.*, 11, 3975 (2020)].[268] (b) Reprinted with permission from [*Science.*, 363, 965 (2019)].[143] (c) Reprinted with permission from [*Opt. Express.*, 29, 2153 (2021)].[269] (d) Reprinted with permission from [*Phys. Rev. Appl.*, 9, 024030 (2018)].[267]

All-optical RF spectrum analyzers based on optical microcombs have been demonstrated [286, 287], which can measure RF spectra by detecting the optical signals directly in the optical domain via the Kerr nonlinear effect, thus allowing the operation bandwidth to extend into the terahertz regime. Recently, a microcomb-based RF spectrum analyzer has been used to monitor the comb dynamics (**Figure 20(c)**) [269], where the RF spectra of microcombs generated by a Hydex MRR were mapped onto the optical spectrum of a probe beam via XPM and subsequently observed using a Fabry-Perot spectrometer, achieving a resolution of ~7.5 MHz and a frame rate of ~100 Hz in a bandwidth from 200 GHz to 2 THz. Successful measurements of the RF spectra of microcombs in both modulation instability and soliton states were demonstrated within a time window of 2 s.

The precision measurement of the relative frequency drift between two stabilized optical resonant cavities has been demonstrated (**Figure 20(d)**) [267], where a silica microdisk resonator was employed to generate a soliton micro-comb with a repetition rate of ~15 GHz. The generated optical microcomb was broadened via coherent supercontinuum generation after going through an integrated $Si_3N_4$ waveguide, which enabled the frequency comb to be fully self-referenced for $f$ - $2f$ referencing. The generated supercontinuum comb line at ~1070 nm was beaten against a stabilized 1070-nm CW laser cavity, generating a beat note down to ~50 MHz that enabled measuring a frequency drift at 180 mHz/s.

## 5.5 Spectrum channelizers

In precision measurements, spectrum channelizers are often employed to slice the spectra of wideband signals with instantaneous bandwidths that are beyond the capability of available instruments, thereby enabling high-resolution measurement to be performed in multiple frequency channels with smaller bandwidths [288]. Recently, microcomb-based microwave spectrum channelizers have been demonstrated [270, 271], where the microcombs served as compact sources providing numerous wavelength channels, yielding improved slicing resolution as well as reduced SWaP and complexity compared to microwave spectrum channelizers based on fiber gratings [289], nonlinear fibers [290], acousto-optic crystals [291], and discrete laser arrays [292].

For spectrum channelizers, channelized frequency step and slicing resolution are two important performance parameters. The former is used for characterizing the frequency difference between channelized RF spectral segments, and the latter for the bandwidth of channelized RF spectral segments. Considering $N$ comb lines with a comb spacing of $\Delta f_{MC}$, the optical frequency of the $k_{th}$ comb line can be given by [271]

$$f(k) = f(1) + (k-1)\Delta f_{MC} \qquad (26)$$

where $f(1)$ is the frequency of the first comb line on the red side. After modulation, the input RF signal is multicast onto each comb line, and the spectrum of the RF signal can be spectrally sampled, or sliced by using another passive microresonator with a slightly different FSR of $\Delta f_{MR}$. Therefore, the $k_{th}$ channelized RF spectrum can be expressed as [271]

$$f_{MW}(k) = f_{MR}(k) - f(k) = \left[f_{MR}(1) - f(1)\right] + (k-1)(\Delta f_{MR} - \Delta f_{MC}) \qquad (27)$$

where $f_{MR}(k)$ is the $k$th center frequency of the passive microresonator and $[f_{MR}(1) - f(1)]$ is the relative spacing between the first comb line and the adjacent resonance of the passive microresonator (*i.e.*, the offset of the channelized RF frequency). According to **Eq. (27)**, the channelized frequency step between adjacent comb lines is $\Delta f_{MR} - \Delta f_{MC}$, *i.e.*, the difference between the FSRs of the two resonators. On the other hand, the slice resolution is determined by the filtering bandwidth of the passive filter, *i.e.*, the resonance bandwidth of the passive resonator.

A microwave spectrum channelizer with a broad operation bandwidth of ~90 GHz has been demonstrated based on microcombs generated by a Hydex MRR, with an FSR of ~200 GHz, resulting in 20 channels in the C band (**Figure 21(a)**) [270]. An additional Hydex MRR with an FSR of ~49 GHz was employed as a passive Vernier filter to slice microwave signals modulated on the microcomb at every 4th resonance, achieving a channelized frequency step of ~4.43 GHz and a slice resolution of ~1.04 GHz.

To further improve the performance, a microwave spectrum channelizer using two MRRs with similar FSRs has been demonstrated [271], where soliton crystal microcombs generated by a 49-GHz-FSR Hydex MRR provided 92 channels in the C band (**Figure 21(b)**), yielding significantly improved channelized frequency step of ~87.5 MHz and slice resolution of ~121.4 MHz.

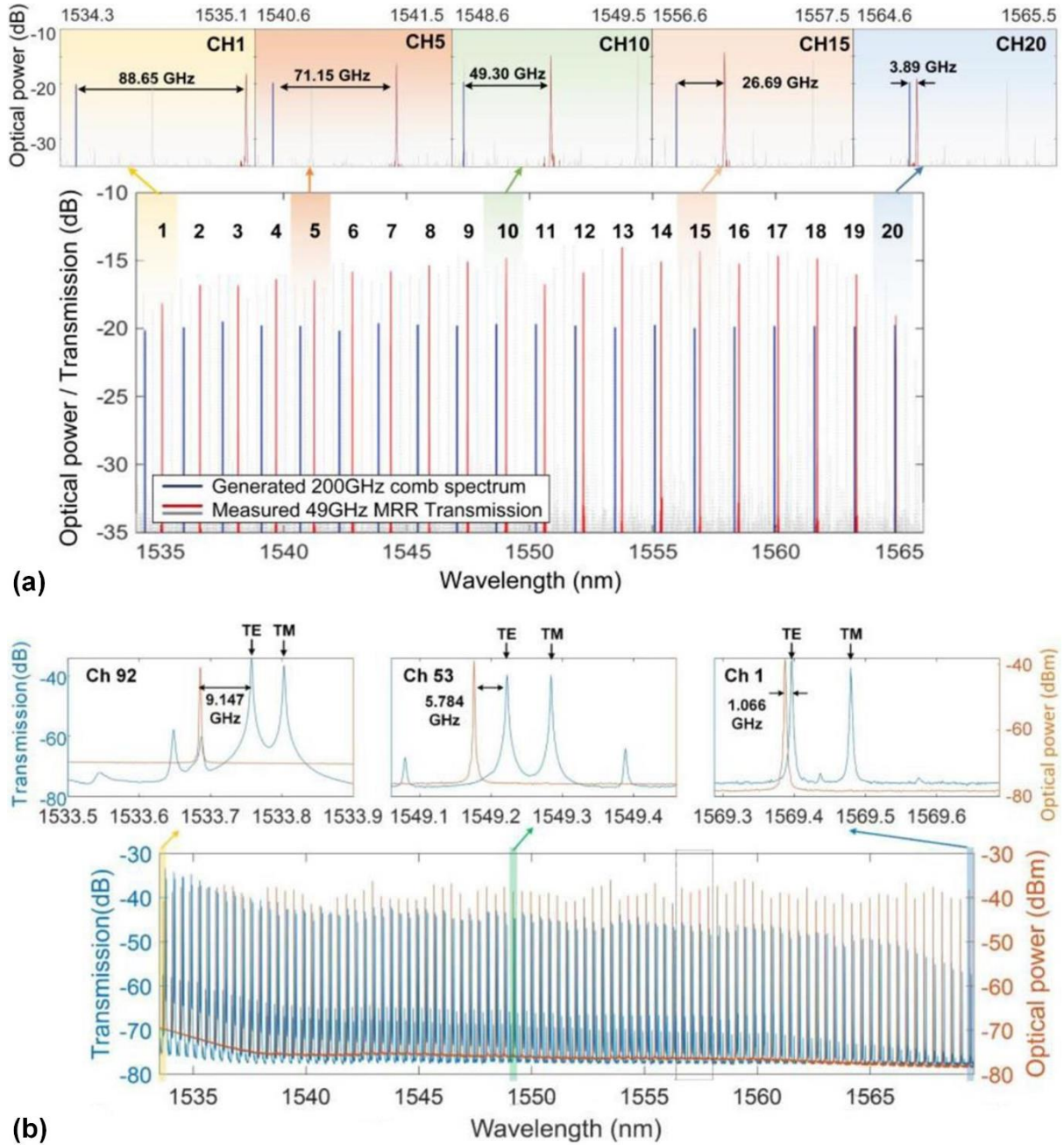

**Figure 21**. Microwave spectrum channelizers based on optical microcombs. (a) A microwave spectrum channelizer with 20 channels in the C band based on microcombs generated by a 200-GHz-FSR Hydex MRR. (b) A microwave spectrum channelizer with 92 channels in the C band based on soliton crystal microcombs generated by a 49-GHz-FSR Hydex MRR. (a) Reprinted with permission from [*J. Lightwave Technol.*, 36, 4519 (2018)].[270]

(b) Reprinted with permission from [*J. Lightwave Technol.*, 38, 5116 (2020)].[271]

## 6. Neuromorphic computing based on optical microcombs

Neuromorphic computing systems create high-speed and power-efficient hardware for information processing by emulating the distributed topology of the brain [19]. The term 'neuromorphic computing', first used by *Dr. Carver Mead* in 1990, [293] has expanded in recent years to include artificial neural networks (ANNs) using non-von Neumann architectures [294]. Compared to traditional von Neumann architectures, neuromorphic computing has the potential to handle complex calculations faster and more energy-efficiently with a smaller footprint. Neuromorphic computing also underpins the performance of machine learning for sophisticated tasks, making it possible to develop algorithms that are capable for online and real-time learning.

The implementation of ANNs based on photonic hardware can overcome the bandwidth bottleneck of their electrical counterparts [5]. The overall network capacity can also be significantly improved by making use of the parallel processing capability of light [55, 56]. All of these make optical neural networks (ONNs) promising candidates for ultrahigh-speed neuromorphic computing in this era of big data. Compared to bulky and power-hungry discrete laser sources, the compact device footprint of optical microcombs enables the provision of a large number of wavelength channels that can be exploited as carriers to implement vector computing. This allows for power-efficient massively parallel processing in neuromorphic computing systems and provides an attractive way for implementing low-SWaP neuromorphic hardware to perform sophisticated tasks. In this section, we review and discuss the emerging applications of optical microcombs to neuromorphic computing, including single neurons and neural networks.

### 6.1 Single neurons

Single neurons, or neuron perceptrons, which are fundamental building blocks in the ANNs,

determine whether the input belongs to some specific classes [54, 295]. For a single neuron in an ANN, the output can be regarded as the dot product of an input vector and a synaptic weight vector being biased and mapped by a nonlinear function onto a desired range [296, 297]. As the basic computing function for single neurons, vector dot product (VDP) has a general form given by

$$\boldsymbol{A} \bullet \boldsymbol{B} = \sum_{i=1}^{k} a_i b_i \qquad (28)$$

where $\boldsymbol{A}$ and $\boldsymbol{B}$ are vectors including $k$ elements, $a_i$ and $b_i$ are the $i^{\text{th}}$ elements in $\boldsymbol{A}$ and $\boldsymbol{B}$, respectively.

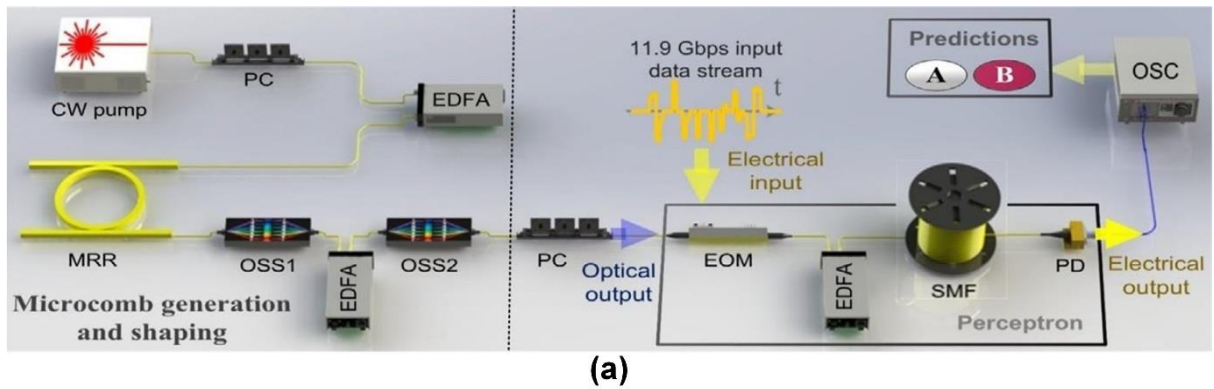

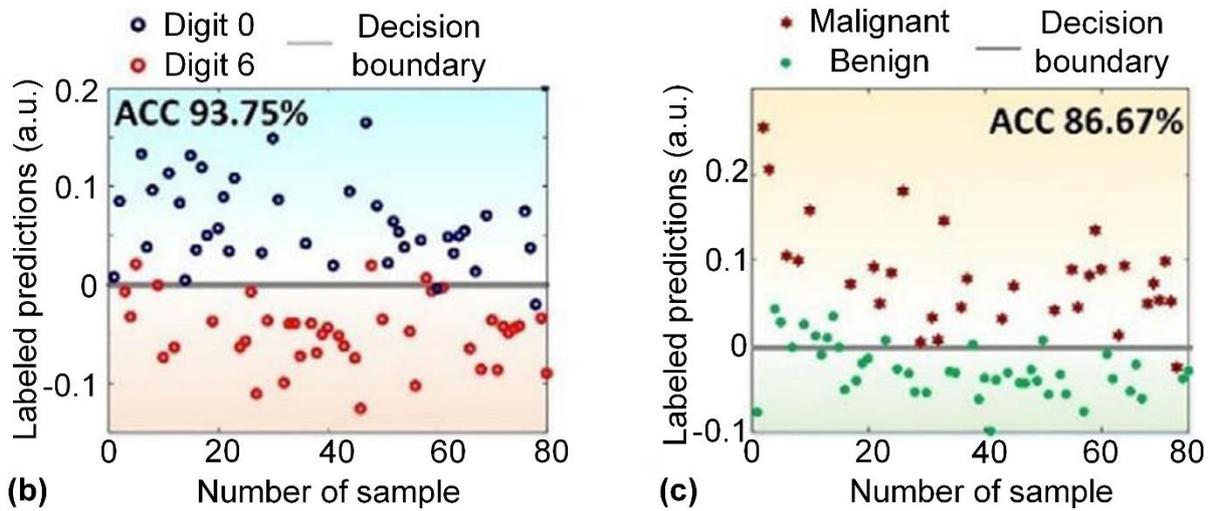

**Figure 22**. A reconfigurable single neuron based on optical microcombs. (a) Experimental setup. (b) Experimental results for classification of handwritten digits and cancer cells. Reprinted with permission from [*Laser Photon. Rev.*, 14, 10 (2020)].[54]

Recently, a reconfigurable single optical neuron has been realized on the basis of the microcomb-based transversal filter system [54], which unlocks the applications of microcombs for neuromorphic computing [298-300]. Optical microcombs can provide many wavelength channels that serve as synapses, with different values of synaptic weights being obtained by shaping the power of the comb lines. Moreover, the transversal filter system enables the delay of different wavelength channels and the sum of the delayed replicas in the same time slot. These form the basis of VDP computing and hence neuron perceptrons based on optical microcombs. In contrast to processing analog signals for differentiation, integration, or Hilbert transformation in **Section 3**, the realization of VDP enables processing digital signals such as digital figures by converting them into vectors, which greatly broadens the application scope of optical microcombs.

F**igure 22(a)** shows the experimental setup of the reconfigurable single neuron based on optical microcombs [54], where the synapses are mapped onto 49 wavelength channels of a soliton crystal microcomb generated by a Hydex MRR. The input data were first flattened and converted to vectors, and then the vectors were time-multiplexed via a digital-to-analog converter (DAC), with each symbol occupying a single time slot. The transversal filter system simultaneously performed synapse weighting in the spectral domain and vector scaling in the temporal domain, yielding a high processing speed of up to ~95.2 Gbit/s. The performance of the perceptron was tested by classifying handwritten digits and cancer cells, achieving accuracies of ~93.75% and ~86.67%, respectively (**Figure 22(b)**).

## 6.2 Neural networks

By connecting multiple single neurons to implement neural networks, complex neuromorphic computing functions can be realized. To date, many different architectures of ONNs have been reported, mainly including feed-forward neural networks [299, 301], spiking neural networks (SNNs) [297], recurrent neural networks (RNNs) [302], reservoir computing [303], and convolutional neural networks (CNNs) [55, 56, 304]. Optical microcombs, which offer a large number of wavelength channels that can be used for vector computing, have been mainly employed in CNN architectures based on vector convolution [55, 56].

**Table 8. Performance comparison of optical CNNs based on different photonic hardware.**

| Photonic hardware | Input data dimension | Computing speed (OPS $^{a)}$) | Scalability & reconfigurability $^{b)}$ | Integrated components | Year | Refs. |
|---|---|---|---|---|---|---|
| Angle-sensitive pixels | $2.3 \times 10^6$ | – | Level 1 | ASP image sensor | 2016 | [304] |
| Mach-Zenhder interferometer arrays | 4 | CW $^{c)}$ | Level 2 | Weight & sum circuits | 2017 | [299] |
| Diffractive optical surfaces | 784 | CW | Level 2 | None | 2018 | [305] |
| Digital micromirrors | $2 \times 10^6$ | 1000 $^{d)}$ | Level 2 | None | 2020 | [306] |
| Spatial light modulators | 65536 | CW | Level 2 | None | 2021 | [307] |
| Optical microcombs | 16 | ~4 trillions | Level 2 | Light source, and weight & sum circuits | 2021 | [55] |
| Optical microcombs | $2.5 \times 10^5$ | ~11.3 trillions | Level 3 | Light source | 2021 | [56] |

$^{a)}$ OPS: operations per second, *i.e.*, floating-point operations per second.
$^{b)}$ Scalability & reconfigurability: Level 1 – the synaptic weights can hardly be reconfigured; Level 2 – the synaptic weights can be reconfigured, but the network structure (*i.e.*, the number of layers and neurons in each layer) can hardly be reconfigured; Level 3 – the synaptic weights and network structure can be reconfigured.
$^{c)}$ CW: continuous-wave light source is used in the architecture as the input data signal, and high-speed updating of the input data is not demonstrated to achieve a high computing speed.
$^{d)}$ Convolution operations per second here.

CNNs are an important class of ANNs that have been widely used for visual imagery analysis such as image recognition, classification, and segmentation [308, 309]. The implementation of optical CNNs can greatly increase the processing speed and reduce the power consumption over their electronic counterparts [55, 56]. In **Table 8**, we compare the performance of optical CNNs based on different photonic hardware. In 2016, an optical CNN was experimentally demonstrated [304], where optical computing was performed in the first layer using angle-sensitive pixels, yielding up to a ~90% reduction of power consumption in the image sensor. Subsequently, optical CNNs based on Mach-Zenhder interferometer arrays [299], diffractive optical surfaces [305], digital micromirrors [306], and spatial light modulators [307] have been investigated. Recently, microcomb-based optical CNNs have also been demonstrated [55, 56], providing a promising path towards compact on-chip CNNs with high processing speed and low power consumption for data-heavy artificial intelligence applications.

In CNNs, the diverse computing functions are realized based on vector convolution [309, 310], which can be achieved on the basis of VDP. For two vectors $A = [a_1\ a_2\ \ldots\ a_m]$ and $B = [b_1\ b_2\ \ldots\ b_n]$, their convolution $A * B$ is a new vector $C = [c_1\ c_2\ \ldots\ c_{m+n-1}]$, where $c_i$ is the $i^{th}$ element in $C$ that can be expressed as

$$c_i = \sum_u a_u b_{i-u+1} \tag{29}$$

In **Eq. (29)**, $u$ ranges over all legal subscripts for $a_u$ and $b_{i-u+1}$. For vector convolution, the processing speed is normally dictated by the number of accomplished floating-point operations (FLOPs) per second, or operations per second (OPS). For microcomb-based transversal filter systems, the OPS can be expressed as [56]

$$OPS = \frac{2R}{\Delta t} \cdot \frac{K-R+1}{K+R-1} \approx \frac{2R}{\Delta t} \tag{30}$$

where $\Delta t$ is the time delay between adjacent taps (the same as that in **Eq. (18)**), $R$ is the number of multiply-accumulate (MAC) operations accomplished in $\Delta t$, and $K$ is the length of the input vector. The effect of the term $(K−R+1)/(K+R−1)$ can be omitted considering $K$ is generally much larger than $R$ in practical neuromorphic processing. Since there are two FLOPs in each MAC operation, *i.e.*, a multiply operation and an accumulate operation, there is a factor of 2 in **Eq. (30)**. Note that the OPS can be scaled up by dividing the overall available comb lines into several groups, with **Eq. (30)** working for each group and the overall OPS being scaled by a factor equaling to the number of the divided groups.

**Figure 23(a)** shows a tensor core capable of performing parallel convolutional computing for optical CNNs based on a soliton microcomb generated by a $Si_3N_4$ MRR [55]. The vector convolution was realized by transmitting shaped microcombs through a passive waveguide network patched onto phase-change materials, where the input vector was encoded on the amplitudes of different comb lines and the kernel vector was encoded on the phase states (*i.e.*, amorphous or crystalline) of the phase-change materials. An experimental demonstration of a 9 × 4 waveguide network with four multiplexed input vectors was reported at a modulation speed of 14 GHz, achieving a maximum overall processing speed of 2 trillion MAC operations per second. The classification of handwritten digits (0 to 9) was performed to test the performance of the tensor core, achieving an accuracy of ~95.3% agreeing well with theoretical predictions (96.1%).

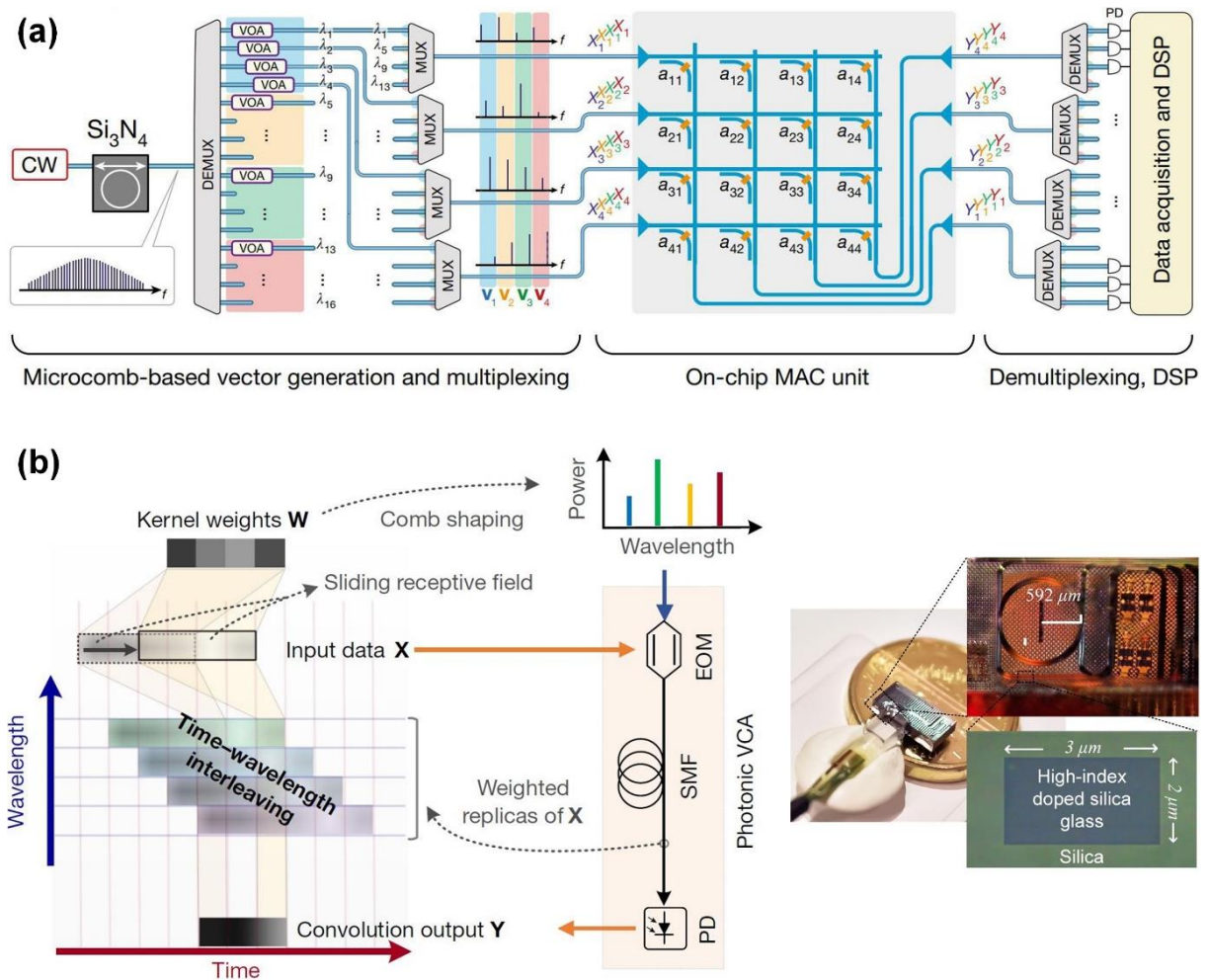

**Figure 23**. Convolutional neural networks (CNNs) based on optical microcombs. (a) Schematic of a convolution computation accelerator. Parallel operations are achieved by using multiple wavelengths derived from a soliton microcomb generated by a Si$_3$N$_4$ MRR. (b) Operation principle of an optical CNN based on time-wavelength interleaving using a soliton crystal microcomb generated by a Hydex MRR. (a) Reprinted with permission from

[*Nature.*, 589, 52 (2021)].[55] (b) Reprinted with permission from [*Nature.*, 589, 44 (2021)].[56]

**Figure 23(b)** shows another optical CNN realized by a microcomb-based transversal filter system [56], where the same hardware was used in both the convolutional layer and the fully connected layer. In the convolutional layer, 90 wavelength channels of a soliton crystal microcomb generated by a Hydex MRR were divided into 10 groups for parallel processing, each included 9 channels for a kernel and was modulated with flattened vectors converted from images to be processed. Convolution of the flattened input vectors and the shaped comb lines as kernel vectors was realized by computing the VDP in each time slot, which formed one of the elements of the output convolution vector. A computing speed of ~1.13 trillion operations per second (TOPS) was achieved for each kernel, representing a total of ~11.3 TOPS for 10 kernels. The output data of the convolutional layer were then pooled electronically and flattened into vectors to form the input data for the fully connected layer, where VDP computing was performed for predictions. The performance of the optical CNN was tested by classifying the full set of handwritten digits (0 to 9), achieving an accuracy of ~88%, which was very close to the theoretical accuracy for the system of 90%.

**7. Quantum optics based on optical microcombs**

Although classical applications have been of predominant interest for the microcomb community to date, the use of optical microcombs for quantum applications has been a fast-growing field in recent years [57, 311]. Despite the fact that solid-state and atom-based quantum systems have shown advantages in efficient multiqubit generation as well as the ease of control [312, 313], they generally have a low system scalability [57], which makes it challenging to scale up the system for information processing. With the capability of providing hundreds of quantum channels from a single microresonator, optical microcombs represent a novel generation of non-classical light sources for implementing compact and large-scale quantum optical systems with applications to quantum communication, quantum computation, and quantum metrology [57, 314, 315]. Since there have been excellent reviews on quantum applications based on optical microcombs [11, 57, 58, 311], here we just provide a brief overview of this field, including the generation of single / entangled photons and squeezed light.

## 7.1 Generation of single / entangled photons

Single and entangled photons are of fundamental importance for quantum photonic devices, both of which can be produced by spontaneous nonlinear parametric processes [6]. The generation of quantum microcombs involves interactions among individual photons, which features distinctive quantum characteristics [316]. The spontaneous nonlinear parametric processes can produce entangled photons via either second- ($\chi^{(2)}$) or third-order ($\chi^{(3)}$) optical nonlinearities. The former is based on spontaneous parametric down-conversion (SPDC, *i.e.*, quantum counterpart of classical difference-frequency generation), whereas the latter is enabled by spontaneous four-wave mixing (SFWM, *i.e.*, quantum counterpart of classical FWM). For Kerr microcombs, the SFWM process can be triggered when the resonator is pumped below the optical parametric oscillation threshold, where two pump photons annihilate, resulting in simultaneous and spontaneous generation of the signal and idler photons. The photonic interaction can be expressed as [11]

$$2\hbar\omega_p \to \hbar\omega_s + \hbar\omega_i \tag{31}$$

where $\hbar$ is the reduced Planck constant, $\omega_p$, $\omega_s$, and $\omega_i$ are the pump, signal, and idler angular frequencies, respectively. Since SFWM intrinsically arises from the coupling between the pump photons and the vacuum fluctuation of the side modes [11], it has a purely quantum nature and can be analyzed from a fully quantum perspective [316].

In 2014, the generation of heralded single photons at different wavelengths of a microcomb was demonstrated based on a Hydex MRR embedded in an external fiber cavity (**Figure 24(a)**) [317]. Subsequently, a SFWM scheme employing two orthogonally-polarized excitation pumps was introduced to produce complex quantum states (**Figure 24(b)**) [318], achieving direct generation of orthogonally-polarized photon-pairs. In 2016, the generation of bi- and multiphoton-entangled qubits based on a microcomb generated by a Hydex MRR was demonstrated (**Figure 24(c)**) [59], showing attractive features such as high scalability enabled by operation over multiple modes, high-purity for the generated entangled photons, and high compatibility with fiber technology.

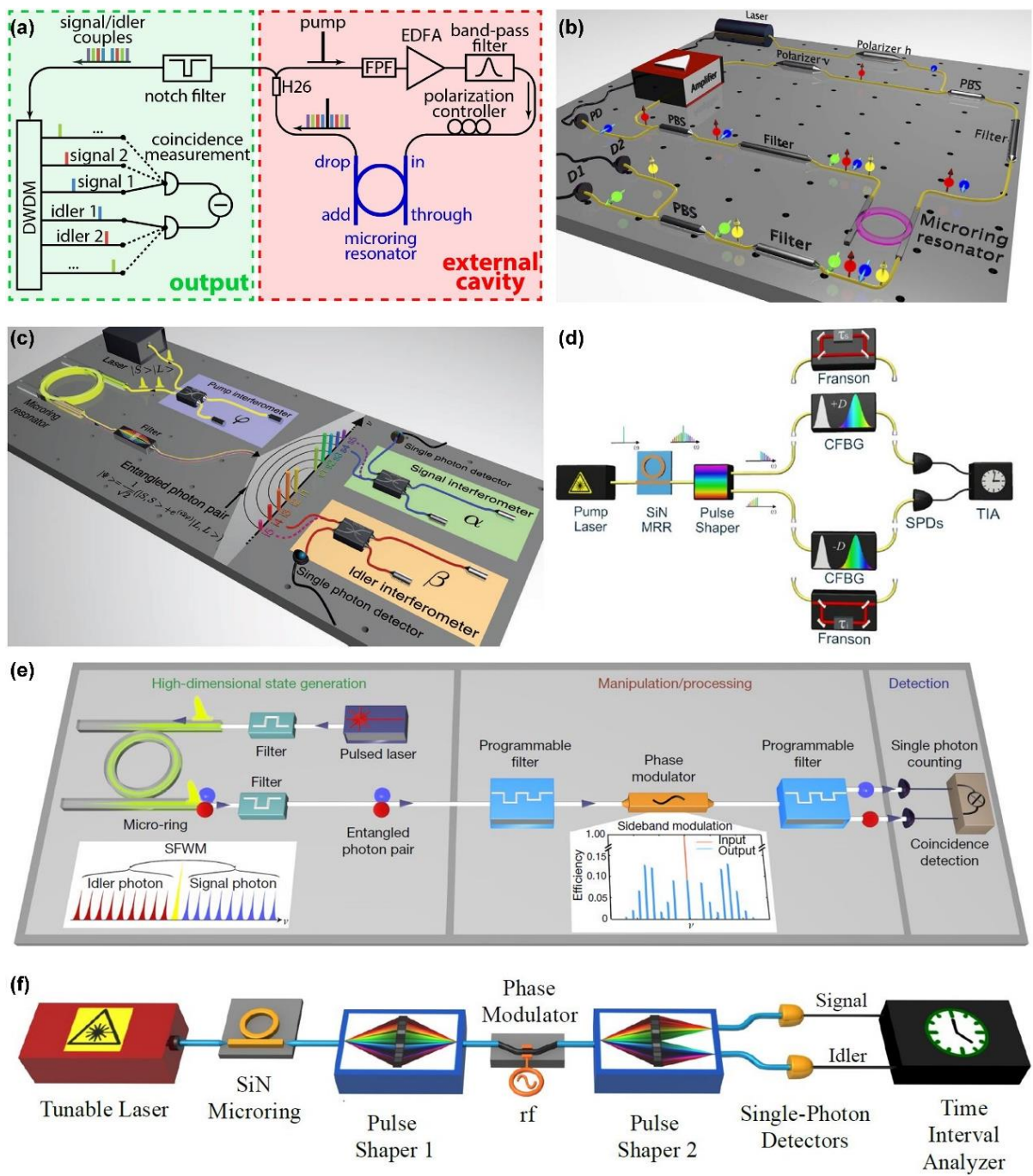

**Figure 24**. Generation of single / entangled photons based on optical microcombs. (a) Generation of heralded

single photons based on a Hydex MRR embedded in an external fiber cavity. (b) Generation of orthogonally-polarized photon-pairs based on spontaneous four-wave mixing (SFWM) in a bi-chromatically pumped Hydex MRR. (c) Generation of bi- and multiphoton-entangled qubits based on a microcomb generated by a Hydex MRR. (d) Generation of energy-time entangled states based on a biphoton microcomb generated by a $Si_3N_4$ MRR. (e) Generation of high-dimensional entangled quantum states based on SFWM in a Hydex MRR. (f) Generation of qubit and qutrit frequency-bin entanglement based on a microcomb generated by a $Si_3N_4$ MRR. (a) Reprinted with permission from [*Opt. Express.*, 22, 6535 (2014)].[317] (b) Reprinted with permission from [*Nat. Commun.*, 6, 8236 (2015)].[318] (c) Reprinted with permission from [*Science.*, 351, 1176 (2016)].[59] (d) Reprinted with permission from [*Optica.*, 4, 655 (2017)].[319] (e) Reprinted with permission from [*Nature.*, 546, 622 (2017)].[320] (f) Reprinted with permission from [*Opt. Express.*, 26, 1825 (2018)].[321]

In 2017, the generation of energy-time entangled states covering multiple resonances was demonstrated based on biphoton microcombs generated by a $Si_3N_4$ MRR (**Figure 24(d)**) [319]. By using these quantum frequency combs, nonlocal dispersion cancellation has also been demonstrated. Subsequently, the successful generation of high-dimensional entangled quantum states was achieved based on SFMW in a Hydex MRR (**Figure 24(e)**) [320], where a coherent manipulation scheme was introduced to control the entangled states, allowing for deterministic high-dimensional gate operations.

In 2018, qubit and qutrit frequency-bin entanglement was demonstrated based on a microcomb with 40 mode pairs generated by a $Si_3N_4$ MRR with an FSR of ~50 GHz (**Figure 24(f)**) [321]. In 2019, the generation of cluster states, *i.e.*, multipartite entangled states where each particle is entangled with more than one other particle, was achieved based on SFWM in a Hydex MRR [322]. Experimental characterization of the noise sensitivity of the generated three-level, four-partite cluster states was performed, together with demonstrations of high-dimensional one-way quantum operations.

**7.2 Generation of squeezed light**

In quantum theory, the energy exchange between physical systems is quantized. As a result, the sensitivity of a measurement system is intrinsically limited by the quantum noise [323]. Squeezed states of light, which show 'squeezed' fluctuations with less uncertainty than coherent states [324], open up new avenues to improve the measurement sensitivity beyond the quantum noise limit [325]. A notable example is their application to GEO 600 – an optical interferometer employed for gravitational wave detection [323].

In a Kerr microcomb, the out-coupled photon flux is the same for the two symmetric sidemodes $\pm l$ from the pump in the semi-classical limit. As a result, the difference between the photon numbers of the two sidemodes given by

$$N_{out,\Delta} = N_{out,l} - N_{out,-l} \qquad (32)$$

is anticipated to be null [11]. The photon number difference in **Eq. (32)** can be experimentally characterized by measuring the power difference of the two sidemodes. From a quantum viewpoint, the operator corresponding to **Eq. (32)** can be expressed as [11]

$$\hat{N}_{out,\Delta} = \hat{N}_{out,l} - \hat{N}_{out,-l} \qquad (33)$$

which is not null and results in fluctuations since there is a shot-noise floor level after photodetection [61]. When specific conditions are satisfied, squeezed light with a noise level below the photocurrent shot-noise level can be generated [11], and the noise reduction (NR) relative to the shot-noise level is usually employed to characterize the squeezing performance. **Table 9** compares the performance of state-of-the-art squeezed light sources based on optical microcombs.

**Table 9. Comparison of squeezed light generation based on optical microcombs.**

| Microresonator | Pump power (mW) | Measured NR (dB) | Squeezing after loss correction(dB) | Year | Refs. |
|---|---|---|---|---|---|
| $Si_3N_4$ MRR | 90 | 1.7 | 5 | 2015 | [326] |
| $LiNbO_3$ WGMR | 0.3 | 1.4 | − | 2019 | [327] |
| $Si_3N_4$ MRRs | 48 at ~1543 nm<br>54 at ~1559 nm | 1.34 | 3.09 | 2020 | [325] |
| $Si_3N_4$ MRR | 104.9 | 1.0 [a)]<br>1.5 [b)] | 4 [c)]<br>7 [d)] | 2020 | [328] |
| $Si_3N_4$ MRRs | 70 | 1.65 | 8 | 2021 | [329] |
| $SiO_2$ wedge resonator | 120 | 1.6 | 3.1 | 2021 | [330] |

a) for quadrature squeezing
b) for photon number difference squeezing

In 2015, all-optical squeezing was experimentally realized based on a $Si_3N_4$ MRR pumped at ~90 mW (**Figure 25(a)**) [326], achieving a NR of ~1.7 dB measured by the intensity-difference squeezing method that corresponded to ~5-dB squeezing after correcting system losses. Subsequently, the generation of squeezed vacuum states has been demonstrated based on a $LiNbO_3$ WGM resonator with high second-order optical nonlinearity (**Figure 25(b)**) [327], achieving a measured NR of ~1.4 dB at a low pump power of ~300 μW in degenerate single-mode operation.

In 2020, the generation of quadrature-phase squeezed states in the RF carrier sideband was demonstrated based on a $Si_3N_4$ MRR with dual pumps [325], achieving a NR of ~1.34 dB characterized by homodyne measurements that corresponded to ~3.09-dB squeezing after loss correction. Subsequently, both quadrature squeezing and photon number difference squeezing were demonstrated based on SFWM in a $Si_3N_4$ MRR pumped at ~104.9 mW [328], achieving NRs of ~1.0 dB and ~1.5 dB, respectively (**Figure 25(c)**).

In 2021, strongly squeezed light was generated based on a photonic molecule resonator consisting of two coupled $Si_3N_4$ MRRs pumped at ~70 mW (**Figure 25(d)**) [329]. By using micro-heaters to tune the photonic molecule resonator, the unwanted parametric processes were suppressed, yielding a measured NR of ~1.65 dB that corresponded to ~8-dB squeezing after loss correction. Subsequently, the generation of a deterministic, two-mode-squeezed quantum microcomb was demonstrated based on a silica wedge resonator pumped at ~120 mW (**Figure 25(e)**) [330], where 40 frequency multiplexed qumodes were obtained, achieving a measured NR of ~1.6 dB that corresponded to ~3.1-dB squeezing after correcting system losses.

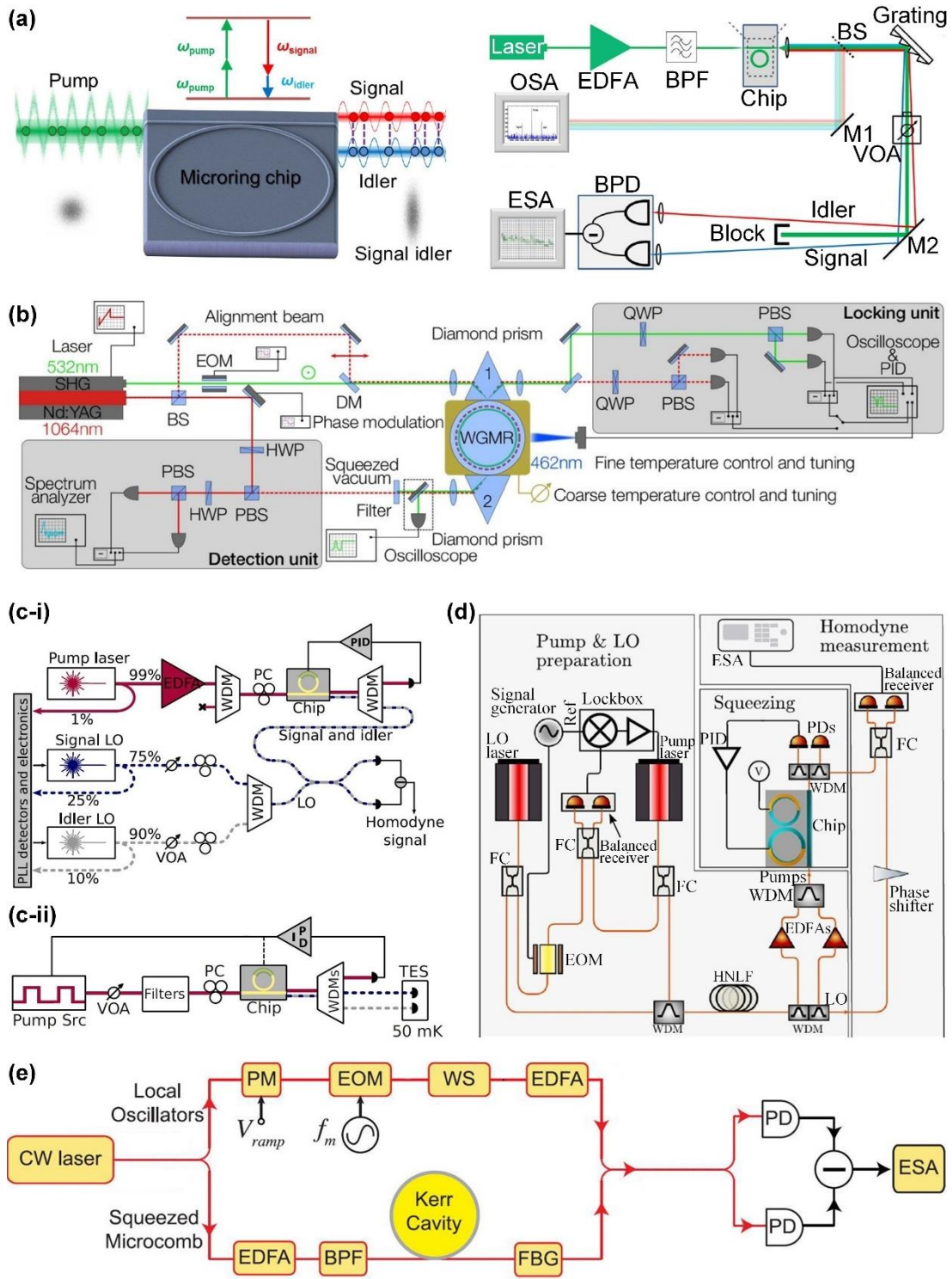

**Figure 25**. Generation of squeezed light based on optical microcombs. (a) All-optical squeezing based on a $Si_3N_4$ MRR pumped above the power threshold of optical parametric oscillation. (b) Generation of squeezed vacuum states based on parametric down-conversion in a $LiNbO_3$ WGM resonator. (c) Generation of (i) quadrature squeezing and (ii) photon number difference squeezing based on SFWM in a $Si_3N_4$ MRR. (d) Generation of strongly squeezed light based on SFWM in a photonic molecule resonator consisting of two coupled $Si_3N_4$ MRRs. (e) Generation of squeezed quantum microcomb in a silica wedge resonator. (a) Reprinted with permission from [*Phys. Rev. Appl.*, 3, 044005 (2015)].[326] (b) Reprinted with permission from [*Optica.*, 6, 1375 (2019)].[327] (c) Reprinted with permission from [*Sci. Adv.*, 6, eaba9186 (2020)].[328] (d) Reprinted with permission from [*Nat. Commun.*, 12, 2233 (2021)][329] (e) Reprinted with permission from [*Nat. Commun.*, 12, 4781 (2021)].[330]

## 8. Challenges and perspectives

The past decade has witnessed rapid growth in research on optical microcombs, including the improvement of their performance as well as expansion of their range of applications. As evidenced by the substantial body of work reviewed in previous sections, the applications of optical microcombs have rapidly expanded and achieved considerable progress – in both traditional and emerging new applications. Despite this remarkable success, though, there are still open challenges and limitations that need to be addressed. In this section, we discuss these in terms of both research and industrial applications.

Paralleling the development of micro / nano fabrication technologies, the family of material platforms for generating optical microcombs has grown substantially. **Figure 26** compares the state-of-the-art of the different material platforms. For optical microcomb generation, a high Kerr nonlinearity to improve the conversion efficiency, a large bandgap to reduce the TPA, and a high Q factor with reduced linear loss are key goals. As shown in **Figure 26(a)**, materials with high Kerr nonlinearities (*i.e.*, $n_2$) tend to have lower bandgaps [83, 91, 92, 95]. Similarly, there is also a trade-off between achieving a high Q factor versus a high Kerr nonlinearity, as shown in **Figure 26(b),** since higher refractive index materials yield larger nonlinearities, but on the other hand also tend to produce higher waveguide loss, and hence lower Q factors. These trade-offs need to be considered and properly balanced in developing new material platforms for optical microcomb generation. Amongst the various optical microcomb platforms, CMOS-compatible platforms such as $Si_3N_4$, Hydex, SiC, $SiO_2$, and $Ta_2O_5$, are promising to facilitate large-scale manufacturing of commercial devices. Some platforms, such as $Si_3N_4$, $LiNbO_3$, GaP, and AlGaAs, that can be co-integrated with other functional components on chips such as lasers, EO modulators, and PDs, are attractive for implementing monolithically integrated systems.

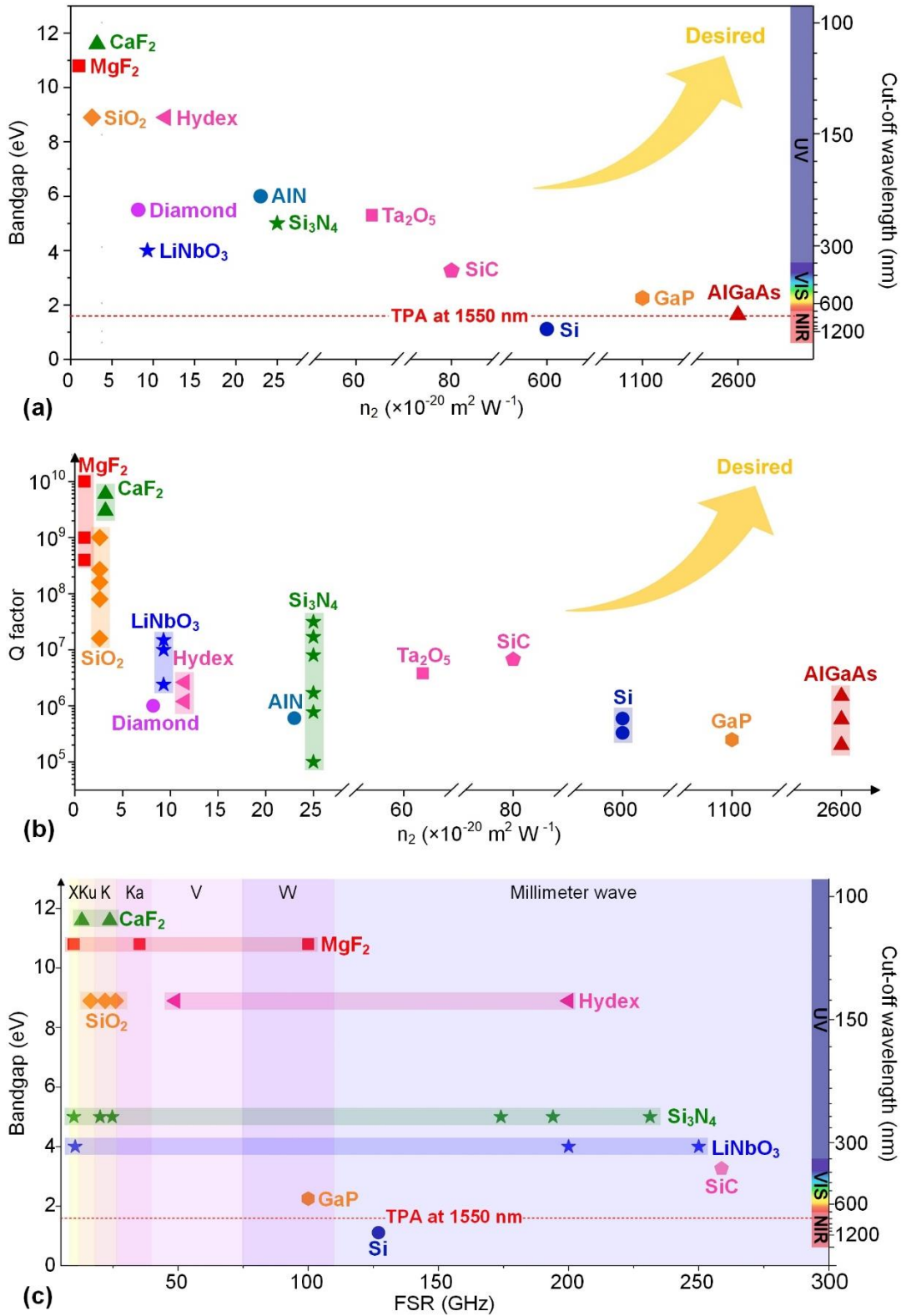

**Figure 26.** Comparison of the state-of-the-art material platforms for optical microcomb generation. (a) Materials' bandgaps and cut-off wavelengths versus their Kerr nonlinear coefficients ($n_2$). (b) Microresonators' quality (Q) factors versus the materials' $n_2$. (c) Materials' bandgaps and cut-off wavelengths versus microresonators' free-spectral ranges (FSRs) in the microwave frequency band (0.3 GHz – 300 GHz). Values of $n_2$, bandgap and cut-off wavelength are taken from Ref. [83] for $MgF_2$ and $CaF_2$, Ref. [91] for $Ta_2O_5$, Ref. [92] for $SiO_2$, diamond, Hydex, AlN, $Si_3N_4$, SiC, Si, GaP, and AlGaAs, and Ref. [95] for $LiNbO_3$. Values of Q factors are taken from Refs. [65, 73, 74] for $MgF_2$, Refs. [21, 63, 102, 109, 331] for $SiO_2$, Refs. [75, 76] for $CaF_2$, Ref. [85] for diamond, Refs. [93,

95, 108] for LiNbO$_3$, Refs. [68, 112] for Hydex, Ref. [98] for AlN, Refs. [81, 82, 99, 114, 139] for Si$_3$N$_4$, Ref. [91] for Ta$_2$O$_5$, Ref. [90] for SiC, Refs. [79, 332] for Si, Ref. [92] for GaP, and Refs. [86-88] for AlGaAs. Values of FSRs are taken from Refs. [65, 73, 74] for MgF$_2$, Refs. [21, 63, 102] for SiO$_2$, Refs. [75, 76] for CaF$_2$, Ref. [85] for diamond, Refs. [93, 95, 108] for LiNbO$_3$, Refs. [30, 68, 333] for Hydex, Ref. [98] for AlN, Refs. [26, 69, 81, 82, 99, 114, 139] for Si$_3$N$_4$, Ref. [90] for SiC, Refs. [79] for Si, Ref. [92] for GaP, and Ref. [87] for AlGaAs.

**Figure 26(c)** compares the FSRs of microresonators based on different material platforms, which are plotted in the microwave frequency band (0.3 GHz – 300 GHz) given that it has close relations with the majority of state-of-the-art applications of optical microcombs. Before 2010, microcombs with small FSRs of less than 100 GHz were mainly generated by discrete and often bulk based WGM cavities such as toroids, wedged disks, and rod resonators [65, 71, 142]. By virtue of the advances in integration fabrication technologies in the past decade, many fully integrated MRRs with much smaller FSRs can now generate microcombs. This is highlighted by micro-combs generated by Si$_3$N$_4$ MRRs with FSRs down to 10 GHz [26], Hydex MRRs with FSRs down to 49 GHz [30], and LiNbO$_3$ MRRs with FSRs down to 10 GHz [93]. In addition, recent progress in the on-chip integration of WGM microcavities [138] opens up a new avenue for generating small-FSR optical microcombs based on photonic integrated chips.

Achieving reliable and self-starting mode locking are both critical capabilities needed to enable stable and coherent optical microcombs for many applications such as frequency synthesizers, optical communications, and precision measurements. As reviewed in **Section 2.4**, in recent years, many new mode-locking approaches such as cryogenic cooling [88], auxiliary laser heating [159], and nonlinear dynamics engineering [110] have been proposed, and conventional mode-locking approaches such as self-injection locking and filter-driven four-wave mixing have also experienced new developments [106, 334]. For practical applications, there are two main directions for improvement. One is achieving easy-to-operate mode-locking without complex startup processes, recently demonstrated with turnkey soliton microcomb generation [110], offering the promise of further improving the system stability and control of the feedback phases. The other direction involves realizing on-chip mode-locking systems, where, for example, the piezoelectric control of microcombs via integrated actuators and the portable battery-operated soliton microcomb generator have represented significant advances. Alternative approaches may involve the use of 2D materials with strong saturable absorption to enable simple passive mode locking, as has been demonstrated in mode-locked fiber lasers [335, 336].

Microcomb-based frequency synthesizers have been used for synthesizing not only microwave but also optical frequencies – well beyond that offered by LFCs generated from solid-state lasers and mode-locked fiber lasers [16, 206, 207]. Although the frequency stability and spectral purity of microcomb-based frequency synthesizers are still not as good as their bulky counterparts, the gap between them is continuously narrowing owing to the advances in technologies for generating soliton microcombs with high coherence and stability [88, 159, 192]. By further reducing the linewidths and noise of lasers for pumping microcombs as well as improving the mechanical and thermal stability of microcomb-based frequency synthesizing systems, there is still room for future improvement. In addition, compared with direct generation of microwave frequencies via photodetection, frequency synthesizers based on optical-to-microwave frequency division normally show higher purity and better frequency stability. As a result, they are preferable for demanding applications with more stringent requirements for frequency deviation and noise performance.

For microcomb-based microwave photonic filters, the filtering resolution has been significantly improved by using microcomb sources with smaller FSRs to provide increased numbers of wavelength channels or taps, *e.g.*, a Hydex MRR with an FSR of ~49 GHz that provides 80 wavelength channels in the C-band [30]. To increase the resolution further, more available wavelength channels can be obtained by using MRRs with even smaller FSRs down to ~10 GHz [26], but this comes at the expense of reduced FSRs in the RF response spectra. Another option is to extend the operation band, for example, into the L band (1570 – 1620 nm).

This requires some specialized L-band components in the transversal filter system such as EO modulators, wave shapers, and PDs to be upgraded with broader operation bandwidths. Out-of-band rejection is another critical parameter for microwave photonic filters, which can also be improved by increasing the tap numbers in the microwave photonic transversal filters. In addition, although Gaussian apodization has been applied to improve the out-of-band rejections [30], it also results in degraded filter resolution, reflecting the trade-off between these two parameters for a fixed tap number.

For microcomb-based microwave photonic signal processors, basic analog signal computing functions, such as (either integral or fractional-order) differentiation [37, 224], integration [38], and (either integral or fractional-order) Hilbert transform [34, 35], have already been realized based on transversal filter systems. More complex functions, such as high-order integration and solving differential equations [235, 337-339], are expected to be achieved by designing the corresponding tap coefficients. State-of-the-art microwave photonic signal processors based on microcombs show higher processing accuracy than what can be achieved by directly processing optical analog signals using passive optical devices [232, 233, 340], although their accuracy is still not as high as their electronic counterparts. In practical systems, there are a number of factors that lead to processing inaccuracy. **Figure 27(a)** shows a typical microcomb-based transversal filter system, where the error sources mainly include (I) a limited number of taps, (II) phase noise of the microcombs, (III) uneven gain and noise of the optical amplifiers, (IV) time delay and shaping inaccuracy caused by optical spectral shapers, (V) chirp of the electro-optic modulators (EOM), (VI) high-order dispersion of the dispersive fiber, and (VII) noise of PDs. Amongst them, the errors induced by (I) can be reduced by using resonators with smaller FSRs or optical amplifiers with broader operation bandwidths. (II) can be lowered by using a range of mode-locking technologies as mentioned previously. The errors arising from (III) – (VI) can be compensated for by using feedback control circuits to calibrate the system impulse response. By using a two-stage comb shaping strategy [38], the inaccuracy induced by (III) and (IV) can also be improved. (VII) can be reduced by employing balanced PD to eliminate the common mode noise.

Monolithic integration is a key issue that has experienced significant recent development. It is critical for future microwave photonic filters and signal processors based on optical microcombs. Although just using integrated microcomb sources to replace discrete laser arrays already yields significant benefits in terms of SWaP, cost, and complexity, there is much more to be gained by increasing the level of integration for the overall system. **Figure 27(b)** shows the concept of a monolithically integrated transversal filter system with a microcomb source, which includes a CW laser, an MRR, an optical amplifier, an EOM, an optical spectral shaping array, a delay element array, and a PD. Compared with the system in **Figure 27(a)**, the SWaP are greatly reduced, reaching the chip scale. In principle, all the components in **Figure 27(b)** can be integrated on the same chip. For instance, InP/Si semiconductor lasers and $Si_3N_4$ MRRs have been heterogeneously integrated to generate soliton microcombs [135]. Semiconductor optical amplifiers (SOAs) [341], Si or InP spectral shapers [342, 343], $LiNbO_3$ modulators [344], delay lines [345], and Ge PDs [346] have also been demonstrated in integrated forms. Recently, some submodules of the system have already been realized, such as the microcomb source chip consists of heterogeneously integrated pump lasers and microresonators [136], and the processing chip contains modulators, shaping arrays, and delay lines [33]. All of these pave the way for implementing the whole system on a single chip in the future.

For coherent optical communications based on microcombs, the state-of-the-art data rate and SE have reached over 50 Tbit/s [42] and 10 bits/s/Hz [43], respectively, using high-order modulation format up to 256-QAM [43, 44]. To improve the data rate and SE, even higher-order modulation formats as well as other multiplexing methods such as poly polarization or orbital angular momentum multiplexing can be employed [347, 348], although these would also bring more stringent requirements for the OSNR and the BER. Emerging mode-locking technologies, such as auxiliary laser heating [159] and turnkey soliton microcomb generation [110], can be applied to improve the comb stability and simplify the operation of these systems. In addition, in demonstrations of coherent communications reported to date based on microcombs, most of the components in the transmitters and receivers other than the microcombs have been discrete devices. Similar to the microcomb-based signal processors in **Figure 27(b)**, monolithically integrated transmitters and receivers with significantly reduced SWaP will be the subject of future research. For IM-DD optical communications, PAM8 or even higher modulation formats can be used to further increase the data rate on the basis of state-of-the-art PAM4 systems. This also requires a higher OSNR of the comb lines, which can be achieved through modifying the device architecture or employing soliton crystals, dark solitons, or laser cavity solitons that have high conversion efficiencies. In addition, the quantum characteristics of optical microcombs could be exploited for realizing secured communications with unprecedented capability [57].

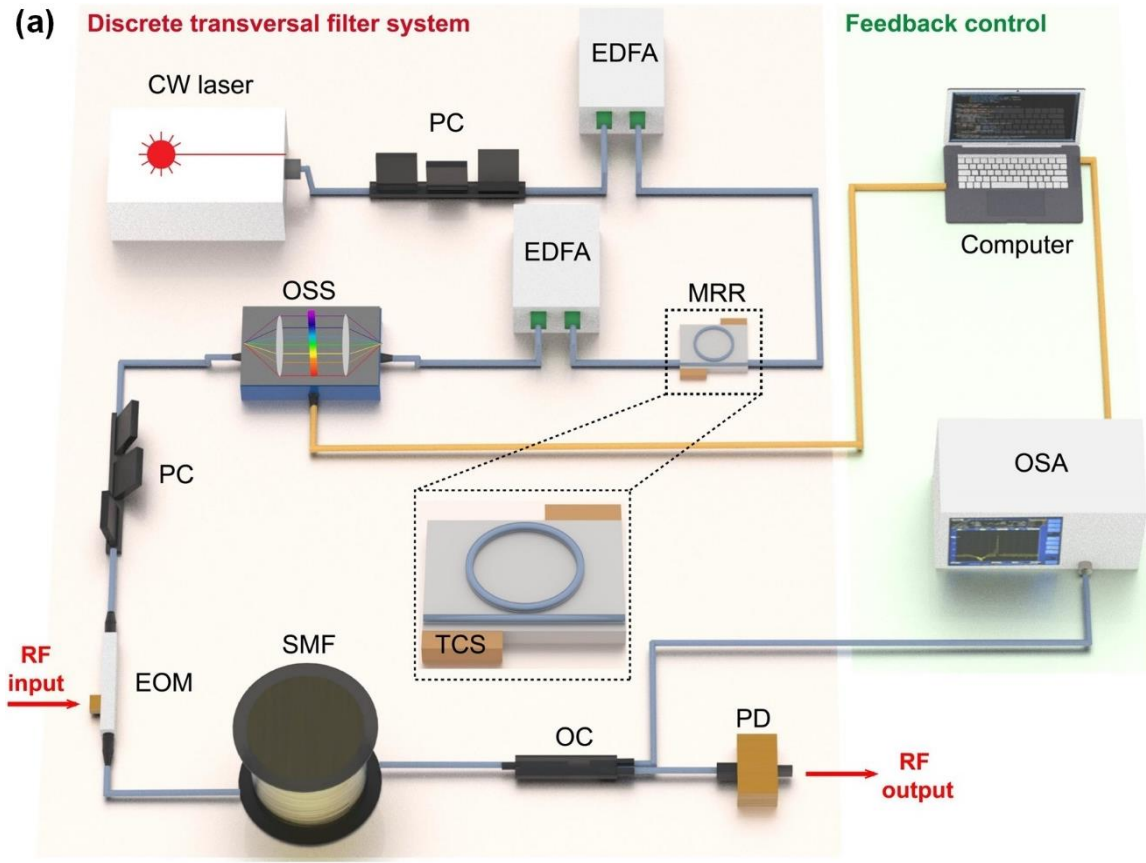

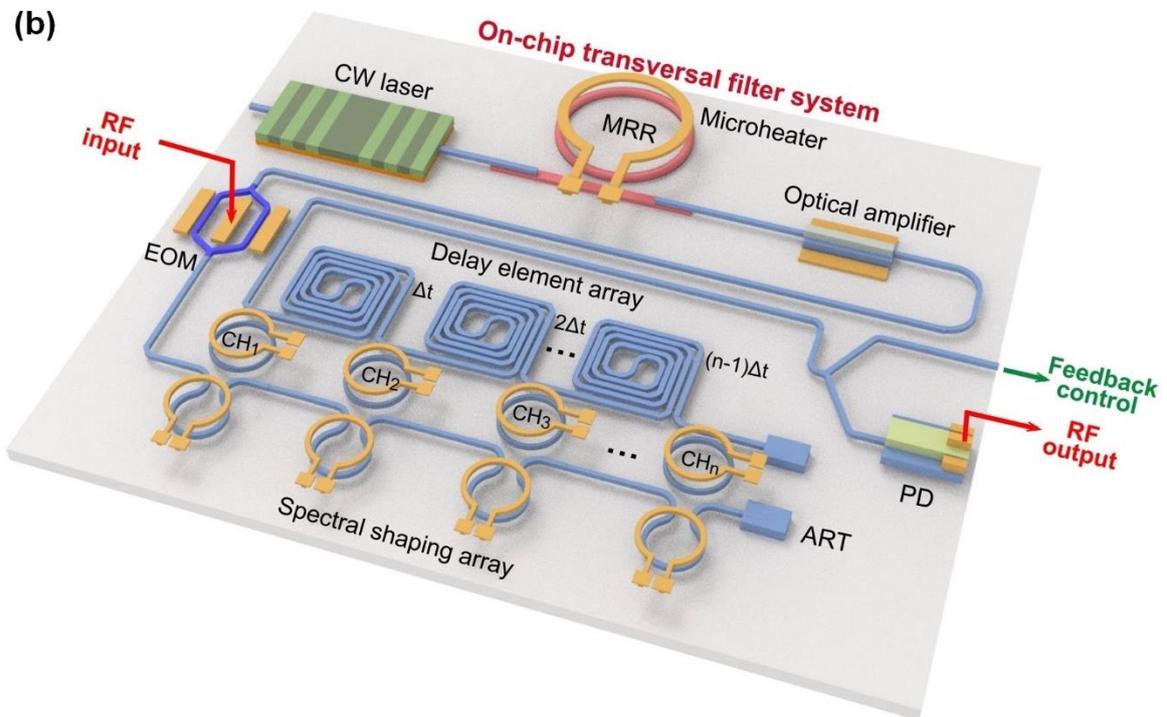

**Figure 27**. (a) A schematic of the state-of-the-art microcomb-based microwave photonic transversal filter system. (b) A schematic showing the concept of a monolithically integrated microcomb-based microwave photonic transversal filter system. CW laser: continuous-wave laser. PC: polarization controller. EDFA: erbium-doped fiber amplifier. MRR: microring resonator. TCS: temperature control stage. OSS: optical spectral shaper. EOM: electro-optic Mach-Zehnder modulator. SMF: single-mode fiber. OC: optical coupler. PD: photodetector. OSA: optical spectrum analyzer. ART: anti-reflection termination.

For precision measurements based on optical microcombs, in addition to conventional spectroscopy, many other functions have been demonstrated such as ranging [46, 48], measurements of Doppler shift and frequency [52, 143]. Compared with LFCs generated by mode-locked fiber lasers, optical microcombs feature larger repetition rates, which enables low acquisition time and broadband sampling of optical spectra, but at the same time results in relatively low resolution. To achieve higher resolution, channelizers can be employed to slice the spectra of broadband microwave signals into small-bandwidth segments. As reviewed in **Section 5.5**, the state-of-the-art microcomb-based microwave spectrum channelizers already achieve a 100-MHz-level slice resolution [271], which is compatible with the real-time bandwidths of electronic devices. In the future, portable or even monolithically integrated sensors based on DCSs are expected to be achieved, with expanded applications in fields such as medical diagnosis, exoplanet analysis, combustion studies, biological tests, meteorology observation, and bio-chemical threat detection. New measurement functions are also expected to enable the characterization of other parameters, such as time-frequency distribution, angle-to-arrival, and noise. In addition, microcomb-based precision measurements can be applied to more complex microwave photonic radar systems, where there has already been encouraging preliminary work [212, 217].

For applications to neuromorphic computing, single neurons and simple CNNs with a single hidden layer have been realized based on vector operations such as dot product and convolution. The microcomb-based CNNs have achieved processing speeds that are far superior to their electronic counterparts [56], but the accuracies have not reached the same level. Aside from the previously mentioned methods for improving the computing accuracy of microwave photonic signal processors, the recognition accuracy can be improved further by increasing the scale of the fully connected layers [56]. On the other hand, the existing CNN systems can be scaled to realize deep CNNs with multiple hidden layers, which can accomplish more complicated tasks. Issues that need to be addressed include the performance of input / output interfaces between adjacent layers, deteriorated processing accuracy caused by increased system complexity, and synergy control among different layers. In addition, through flattening digital images into vectors, and then converting them to analog signals, digital image processing or even video processing can be realized based on optical microcombs. This offers new possibilities for realizing a wide range of image processing functions, including traditional functions such as edge enhancement and motion blur [228, 229], as well as new computer vision applications such as autonomous vehicles, face recognition, remote drones, and medical diagnosis [50, 51, 349].

The scale of state-of-the-art quantum systems based on optical microcombs can be further expanded by using microresonators with smaller FSRs as well as integrated EO modulation and spectral phase manipulation schemes [342, 350]. To achieve this, broadband detection techniques for frequency-encoded photon states will also be needed [351]. Deterministic sources are critical for implementing efficient quantum systems. Similar to other sources based on the spontaneous nonlinear optical processes, optical microcombs are stochastic in nature. Therefore, the challenges for implementing deterministic microcomb-based quantum systems are similar to those of deterministic single-photon sources. Since optical microcombs are capable of providing a large quantum resource per photon, they offer new possibilities to compensate for the drawbacks of non-determinism. The miniaturization of quantum microcombs, which is crucial for practical applications, remains a technological challenge. A pulsed quantum-microcomb excitation scheme has been demonstrated recently [352], paving the way for the realization of monolithic quantum microcombs. The continuous improvement of microcombs' performance as well as progress in their classical applications will also naturally benefit the future advances in quantum applications of optical microcombs.

Achieving microcomb generation in the visible region, where several atomic transitions

(*e.g.*, Rb, Hg$^+$, and Yb) that are being used as time references for atomic clocks are located [189, 353], offers new possibilities for realizing miniatured optical atomic clocks, optical coherence tomography systems [354], and visible-band mode-locked systems [355]. The challenges here not only revolve around having to work with platforms that are low loss in these shorter wavelength regimes (*e.g.*, ruling out silicon) but the challenge is that it is much more difficult to engineer anomalous dispersion. Furthermore, the waveguide dimensions become significantly smaller to achieve single-mode operation. Many experimental demonstrations of visible microcomb generation have already been reported, such as those based on engineering the mode hybridization [331], the second-order nonlinear interactions [101, 113, 356], high-order modes [111], and the Cherenkov-like radiation [357]. However, most have only managed to generate microcombs above 700 nm, at the boundary between the visible and near IR wavelengths, hinting at more breakthroughs in generating visible microcombs at shorter visible wavelengths (< 600 nm).

Finally, there is a high degree of synergy between advances in optical microcombs and their fast-expanding areas of applications. On the one hand, optical microcombs with compact footprint, low power consumption, and large comb spacing offer enormous new possibilities for a variety of applications in both conventional and new interdisciplinary fields. On the other hand, driven by the growing requirements in practical applications, controlling and optimizing microcombs' performance, as well as investigating the underlying physics (*e.g.*, imaging the DKS dynamics with a high resolution < 1 ps [160]) become increasingly demanding, which will facilitate the rapid development and wide deployment of optical microcombs. This synergy will have a long-lasting positive impact, which will be a strong driving force for the continual improvement in performance and the broadening of its applications to deliver the promised benefits of LFCs based on photonic integrated chips. Research on the applications of optical microcombs will undoubtedly continue to thrive, in parallel with the development of relevant commercial products that will eventually enable the bridging of the gap between laboratory-based research and practical industrial applications.

## 9. Conclusion

Optical microcombs have opened up promising avenues towards implementing LFCs in a compact, energy-efficient, and cost-effective manner. Since the ground-breaking discovery of optical microcombs in 2007 [10], these revolutionary light sources have experienced remarkable advances, facilitating a wide range of applications in both conventional and emerging areas. In this paper, we provide a comprehensive review of the applications of optical microcombs. We provide an overview of the devices and methods for generating optical microcombs. We systematically review the use of optical microcombs for a variety of applications in microwave photonics, optical communications, precision measurements, neuromorphic computing, and quantum optics. We also discuss the current challenges and future perspectives. Over the next decade, we foresee that the synergy between microcombs and their rapid expanded applications will facilitate the wide deployment of optical microcombs and fast transition of this technology into commercial production, bringing further exciting breakthroughs for improving the microcombs' performance and underpinning their seamless and versatile applications in practical systems.

**Acknowledgments**

This work was supported by the Australian Research Council Discovery Projects Program (grant numbers DP150104327, DP190102773 and DP190101576) and the Swinburne ECR-SUPRA program. R.M. acknowledges support by the Natural Sciences and Engineering Research Council of Canada (NSERC) through the Strategic, Discovery and Acceleration Grants Schemes, by the MESI PSR-SIIRI Initiative in Quebec, and by the Canada Research Chair Program.

**Conflict of interest**

The authors declare no conflicts of interest.

**Author contributions**

Y. S. and J. W. contributed equally to the manuscript preparation. J. W. conceived of the idea and designed the outline. Y. S. and J. W. did the literature review and prepared the figures and tables. J. W., Y. S., and D. J. M. prepared the text. J. W., A. M., and D. J. M. jointly supervised the project. All authors participated in the review and discussion of the manuscript.